\documentclass[a4paper,11pt]{article} 
\pdfoutput=1 
\usepackage{jheppub}

\usepackage{amsmath, amssymb}
\usepackage{mathrsfs}
\usepackage{graphicx}
\usepackage{dcolumn}
\usepackage{bm}

\usepackage{comment}
\usepackage{braket}
\usepackage{appendix}
\usepackage{color,soul}
\usepackage{amsthm}
\usepackage{bbm}
\newtheorem{claim}{Claim}

\newcommand{\diff}{\mathrm{d}}
\newcommand{\p}{\partial}
\newcommand{\ve}{\varepsilon}
\newcommand{\Diff}{{\mathcal{D}}}

\newcommand{\tr}{\mathrm{tr}}
\newcommand{\im}{\mathrm{i}}

\newcommand{\calB}{\mathcal{B}}

\newcommand{\calZ}{\mathcal{Z}}

\newcommand{\rme}{\mathrm{e}}

\newcommand{\CR}{\mathcal{CR}}
\newcommand{\Spin}{\mathrm{Spin}}

\title{Non-invertible self-duality defects of Cardy-Rabinovici model and mixed gravitational anomaly}

\author[a,b]{Yui Hayashi}
\emailAdd{yui.hayashi@yukawa.kyoto-u.ac.jp}
\affiliation[a]{
Department of Physics, Graduate School of Science and Engineering, Chiba University, Chiba 263-8522, Japan
}

\author[b]{and Yuya Tanizaki}
\affiliation[b]{Yukawa Institute for Theoretical Physics, Kyoto University, Kyoto 606-8502, Japan
}
\emailAdd{yuya.tanizaki@yukawa.kyoto-u.ac.jp}

\preprint{YITP-22-12}

\abstract{
We study properties of self-duality symmetry in the Cardy-Rabinovici model. The Cardy-Rabinovici model is the $4$d $U(1)$ gauge theory with electric and magnetic matters, and it enjoys the $SL(2,\mathbb{Z})$ self-duality at low-energies. 
$SL(2,\mathbb{Z})$ self-duality does not realize in a naive way, but we notice that the $ST^{p}$ duality transformation becomes the legitimate duality operation by performing the gauging of $\mathbb{Z}_N$ $1$-form symmetry with including the level-$p$ discrete topological term. 
Due to such complications in its realization, the fusion rule of duality defects becomes a non-group-like structure, and thus the self-duality symmetry is realized as a non-invertible symmetry. 
Moreover, for some fixed points of the self-duality, the duality symmetry turns out to have a mixed gravitational anomaly detected on a $K3$ surface, and we can rule out the trivially gapped phase as a consequence of anomaly matching. 
We also uncover how the conjectured phase diagram of the Cardy-Rabinovici model satisfies this new anomaly matching condition. 
}

\begin{document}

\maketitle

\section{Introduction}

Symmetry provides a powerful tool to study nonperturbative aspects of quantum field theories (QFTs). 
Its spontaneous breaking can characterize the phase structures, and we can obtain various associated low-energy theorem thanks to its universal feature. 
Furthermore, gauging the symmetry often gives more detailed data, and anomaly matching condition severely constrains the possible low-energy dynamics of QFT. 

Recently, the notion of symmetry in QFT has been generalized, and we recognize that the essential ingredient of symmetry turns out to be the topological defect operators~\cite{Gaiotto:2014kfa}. 
These defect operators are extended objects defined on submanifolds in the Euclidean spacetime. The topological property means that we can continuously deform the submanifold of the defect without affecting the expectation value, and this property gives the generalized version of the Ward-Takahashi identity. 
From this perspective, we can describe usual symmetry by codimension-$1$ topological defects with a group-like fusion rule. 
An important example of the generalized symmetry given in Ref.~\cite{Gaiotto:2014kfa} is the higher-form symmetry, where the generators are topological defects with higher codimension, and we can understand the center symmetry of gauge theories as a special case of $1$-form symmetry. 
Thanks to such generalizations, we can apply conventional techniques related to symmetry in much broader areas of QFTs. 

In this paper, we study the self-duality symmetry of the Cardy-Rabinovici model, which is the $4$d $U(1)$ gauge theories with electric and magnetic matters~\cite{Cardy:1981qy, Cardy:1981fd}. 
This model enjoys the $\mathbb{Z}_N$ $1$-form symmetry as the dynamical electric charge is quantized in $N$, and let us denote it as $\mathbb{Z}_{N}^{[1]}$. 
As a result, the Higgs phase and confinement phases must be separated by quantum phase transitions. 
The Cardy-Rabinovici model at strong couplings also realizes oblique confinement phases proposed by 't~Hooft~\cite{tHooft:1981bkw}, and this theory provides a playground to understand the dynamics of $4$d gauge theories. 
Moreover, as recently discussed in Ref.~\cite{Honda:2020txe}, this theory at $\theta=\pi$ has the mixed anomaly between the $\mathbb{Z}_{N}^{[1]}$ and $\mathsf{CP}$ symmetries, and this is analogous to the case of $SU(N)$ Yang-Mills theory at $\theta=\pi$~\cite{Gaiotto:2017yup, Gaiotto:2017tne}. 

The most notable feature of the Cardy-Rabinovici model is the $SL(2,\mathbb{Z})$ invariance of its phase diagram~\cite{Cardy:1981fd}  generated by the $S$ and $T$ transformations. The $S$ transformation exchanges the electric and magnetic charges, and the $T$ transformation shifts the $\theta$ angle by $2\pi$. 
Naively, these operations seem to act as the self-duality of the model, but it is not the case~\cite{Honda:2020txe}. 
The Cardy-Rabinovici model has the electric $\mathbb{Z}_N^{[1]}$ symmetry, while it does not have the magnetic one. Due to this imbalance between the electric and magnetic $1$-form symmetries, the $S$ transformation does not realize self-duality. 
In Ref.~\cite{Honda:2020txe}, the $\mathbb{Z}_{M}^{[1]}$-gauged Cardy-Rabinovici model was considered when $N=M^2$. 
In this case, the $SL(2,\mathbb{Z})$ transformation realizes the self-duality as both the electric and magnetic sectors enjoy the $\mathbb{Z}_M^{[1]}$ symmetry. 
In the $\mathbb{Z}_{M}^{[1]}$-gauged model, some of $SL(2,\mathbb{Z})$ duality transformations have a mixed gravitational anomaly, and we can rule out the trivially gapped phase at the fixed point in the space of couplings~\cite{Honda:2020txe, Seiberg:2018ntt}. 
However, it was unclear if we could obtain a similar constraint on the original model with general values of $N$. 
It is one of our motivations to resolve this problem in the original Cardy-Rabinovici model. 

Here, another extension of the symmetry comes in. As long as there are nontrivial topological defects, we can regard it as symmetry even if their fusion rule does not form a group-like structure. Such a class of symmetries is dubbed non-invertible symmetries, and it has been extensively studied basically in the context of $2$d QFTs and higher-dimensional topological theories~\cite{Aasen:2016dop, Bhardwaj:2017xup, Buican:2017rxc, Freed:2018cec,  Chang:2018iay, Thorngren:2019iar, Thorngren:2021yso, Ji:2019jhk, Rudelius:2020orz, Gaiotto:2020iye, Komargodski:2020mxz, Aasen:2020jwb, Inamura:2021wuo, Inamura:2021szw, Burbano:2021loy}. 
Even in higher-dimensional and non-topological QFTs, its usefulness has been found out first in the context of the classification of confining string spectra~\cite{Nguyen:2021yld, Nguyen:2021naa}. 
Recently, Koide, Nagoya, and Yamaguchi explicitly constructed the $S$-duality defect as a non-invertible symmetry for a lattice $\mathbb{Z}_2$ gauge theory~\cite{Koide:2021zxj} by extending the construction of Kramers-Wannier duality defects on the $2$d lattice Ising model~\cite{Aasen:2016dop}. 
A few months later, Refs.~\cite{Choi:2021kmx, Kaidi:2021xfk} proposed more systematic constructions of duality symmetry defects. In particular, we can have Kramers-Wannier-type duality defects by the half-space gauging procedure.

Motivated by these breakthroughs, we construct the $ST^{p}$ duality transformation as a non-invertible topological defect. 
We first perform the $\mathbb{Z}_{N}^{[1]}$ gauging of the Cardy-Rabinovici model with including the level-$p$ discrete $\theta$ term, and show that it is dual to the original Cardy-Rabinovici model via the $ST^{p}$ transformation. 
When we consider the fixed point of the $ST^{p}$ transformation, this duality indicates that we can obtain the codimension-$1$ topological defect of $ST^{p}$ by gauging $\mathbb{Z}_{N}^{[1]}$ with the level-$p$ discrete topological term on the half-space, following the idea of Refs.~\cite{Choi:2021kmx, Kaidi:2021xfk}. 
For $p=0$, our topological defect is the same with the $S$ transformation defect constructed in Ref.~\cite{Choi:2021kmx}, so our result gives its generalization to other values of $p$. 
We shall compute the fusion rule for $p=-1$ in detail. Since $(ST^{-1})^3=\mathsf{C}$ even at the naive level, we have to merge three $ST^{-1}$ defects to relate it with invertible symmetries.\footnote{Sometimes, such a defect is called a triality defect to emphasize that it relates three different theories instead of two. In this paper, we simply call duality defects including those cases. }

Remarkably, we also find that the $ST^{-1}$ transformation turns out to have a mixed gravitational anomaly. 
By evaluating this anomaly on a $K3$ surface, we show that the symmetry-protected topological (SPT) states with $\mathbb{Z}_N^{[1]}$ symmetry can be ruled out from the possible ground states at the fixed point of $ST^{-1}$. 
In the conjectured phase diagram, three first-order phase transition lines merge at the $ST^{-1}$ fixed point, and $ST^{-1}$ cyclically exchanges the Higgs phase, monopole-induced confinement phase, and dyon-induced confinement phase. 
From this observation, we uncover how the Cardy-Rabinovici model satisfies the new anomaly-matching constraint obtained by the non-invertible $ST^{-1}$ duality.

This paper is organized as follows.
In Sec.~\ref{Sec:CR_model}, a short review of the Cardy-Rabinovici model is provided.
We first introduce the original construction of the model on the lattice and then give its formal continuum description.
We also review the conjectured phase diagram obtained by the free-energy argument.
In Sec.~\ref{Sec:duality_defect_fusion_rule}, we realize the electromagnetic duality by gauging $\mathbb{Z}_N^{[1]}$.
In Claim~\ref{claim:self-duality-1}, we explain how the $ST^{p}$ duality is realized via $\mathbb{Z}_{N}^{[1]}$ gauging with the level-$p$ discrete topological term.
As the simplest nontrivial example, we focus on the $ST^{-1}$ self-duality thereafter, which is expressed in Claim \ref{claim:self-dual-stinv}.
Then, we construct the duality defect as a half-space gauging operation and derive its fusion rule in Claim \ref{claim:fusion_rule}.
In Sec.~\ref{sec:constraints_dynamics}, we discuss constraints on low-energy dynamics of the Cardy-Rabinovici model from the $ST^{-1}$ self-duality realized in the previous section.
We also see how the constraints are consistent with the conjectured phase diagram shown in Sec.~\ref{Sec:CR_model}.
Lastly, in Sec.~\ref{sec:conclusion} we summarize our findings and outline possible future directions.
Appendices \ref{app:proofs}, \ref{sec:proof_claim_fusion_rule} and \ref{App:computation_bordism} gives technical details of the main claims of the main text.
Appendix \ref{App:other_parameters} includes discussions on other self-dualities $S$ and $ST^{-1}ST^2S$.

\section{Cardy-Rabinovici model} \label{Sec:CR_model}

In this section, we give a brief review on the lattice $U(1)$ gauge theory with the $\theta$ term proposed by Cardy and Rabinovici~\cite{Cardy:1981qy, Cardy:1981fd}, which we call  Cardy-Rabinovici model. 
Because of the presence of magnetic monopoles, the $\theta$ term affects the local dynamics of the $(3+1)$d $U(1)$ gauge theory. 
We also give a review on the formal continuum description given in Ref.~\cite{Honda:2020txe}. We then explain the conjectured phase diagram obtained by the energy-versus-entropy argument of loop excitations. 

\subsection{Lattice description of the model}

Cardy-Rabinovici model~\cite{Cardy:1981qy, Cardy:1981fd} is defined as the lattice $U(1)$ gauge theory coupled to the charge-$N$ Higgs field and the magnetic monopole. 
To have a good control of magnetic monopoles, we use the Villain form of the lattice $U(1)$ gauge theory~\cite{Villain:1974ir}, and the $U(1)$ gauge field $a$ on the lattice is given by the pair $(\widetilde{a}_\mu,s_{\mu\nu})$ of the $\mathbb{R}$-valued link variable $\widetilde{a}_\mu$ and the $\mathbb{Z}$-valued plaquette variable $s_{\mu\nu}$. 
The $U(1)$ gauge transformation is realized as the combination of the $\mathbb{R}$-valued $0$-form gauge transformation and the $\mathbb{Z}$-valued $1$-form gauge transformation, 
\begin{align}
    & \widetilde{a}_\mu\to \widetilde{a}_\mu+\partial_\mu \lambda^{(0)}+2\pi \lambda^{(1)}_\mu,\nonumber\\
    & s_{\mu\nu}\to s_{\mu\nu}+\partial_\mu \lambda^{(1)}_\nu - \partial_\nu \lambda^{(1)}_\mu, 
    \label{eq:gauge_trans_Villain}
\end{align}
where $\lambda^{(0)}$ is the $\mathbb{R}$-valued site variable and $\mathbb{\lambda}^{(1)}_{\mu}$ is the $\mathbb{Z}$-valued link variable. 
Here, the lattice derivative is defined by $\partial_{\mu} \lambda^{(0)}(x)=\lambda^{(0)}(x+\hat{\mu})-\lambda^{0}(x)$, where $\hat{\mu}$ denotes the unit vector along $x_\mu$ axis. 
The gauge-invariant field strength $f=\diff a$ is defined as 
\begin{equation}
    f_{\mu\nu}:=\partial_\mu \widetilde{a}_\nu - \partial_\nu \widetilde{a}_\mu -2\pi s_{\mu\nu}, 
    \label{eq:Villain_f}
\end{equation}
and the lattice Maxwell kinetic term is defined with this field strength:
\begin{equation}
    S_{\mathrm{kin}}:=\frac{1}{2g^2}\sum_{(x,\mu,\nu)}f_{\mu\nu}(x)^2. 
\end{equation}
The monopole current in the Villain-type lattice can be defined as 
\begin{equation}
    m_\mu(\tilde{x}):=\frac{1}{2}\ve_{\nu\mu\lambda \sigma}\partial_\nu s_{\lambda \sigma}(x),
    \label{eq:Villain_monopole}
\end{equation}
where $\tilde{x}=x+\frac{1}{2}(\hat{1}+\hat{2}+\hat{3}+\hat{4})$ denotes the site on the dual lattice.
The monopole current $m_{\mu}$ is thus given as the $\mathbb{Z}$-valued link variable that satisfies the conservation law, 
\begin{equation}
    \partial_\mu m_\mu=0, 
\end{equation}
and it describes the closed worldline of magnetic monopoles.

To treat the electric and magnetic matters on equal footing, the dynamical electric charges are introduced as the closed worldlines in the Cardy-Rabinovici model. 
The corresponding electric current $n_\mu(x)$ is defined as the $\mathbb{Z}$-valued link variable on the original lattice, and it satisfies the charge conservation,  
\begin{equation}
    \partial_\mu n_\mu=0. 
    \label{eq:charge_conservation}
\end{equation}
To define the confinement as the area law of the Wilson loops, we would like to have the $\mathbb{Z}_N$ $1$-form symmetry, denoted as $\mathbb{Z}_N^{[1]}$. 
To this end, we assume that the dynamical electric matter has the charge $N$, and thus the minimal coupling term in the Lagrangian is given by 
\begin{equation}
    \im N n_\mu(x)\widetilde{a}_\mu(x). 
\end{equation}
We note that this satisfies the gauge invariance under \eqref{eq:gauge_trans_Villain}: The $\mathbb{R}$-valued $0$-form gauge invariance obeys from the charge conservation, and the $\mathbb{Z}$-valued $1$-form gauge transformation changes the action only by $2\pi \im N \mathbb{Z}$ so the path-integral weight is not affected.

Let us introduce the $\theta$ term to this model. 
Following Refs.~\cite{Cardy:1981qy, Cardy:1981fd}, we note that the magnetic monopole acquires the electric charge $\frac{\theta}{2\pi}$ by the Witten effect~\cite{Witten:1979ey}, and thus the electric current should be modified as\footnote{We note, however, that this introduction of the $\theta$ term has some problems, and the most serious one would be lack of the $\mathbb{Z}$-valued $1$-form gauge invariance.
Topological aspects of the Villain-type lattice $U(1)$ gauge theories are scrutinized in Refs.~\cite{Sulejmanpasic:2019ytl, Anosova:2022cjm}, where the well-defined lattice $\theta$ term for $4$d $U(1)$ lattice gauge theory has been discussed.
Since our arguments are based on the formal continuum description, we do not go far into the construction here.}
\begin{equation}
    \widetilde{n}_\mu(x):=n_{\mu}(x)+\frac{\theta}{2\pi} \sum_{\tilde{x}}F(x-\tilde{x}) m_\mu(\tilde{x}), 
\end{equation}
where $F(x-\tilde{x})$ is a short-range function that relates the original and dual lattices with the normalization $\sum_{\tilde{x}}F(x-\tilde{x}) = 1$. 
The minimal coupling term is then replaced by 
\begin{equation}
    S_{\mathrm{mat}}:=\im N \sum_x \widetilde{n}_\mu(x)\widetilde{a}_\mu(x). 
\end{equation}
In the long wavelength limit, we expect that the original and dual lattices are almost identical, and the effective electric current $\widetilde{n}_\mu$ may be simply written as
\begin{equation}
    \widetilde{n}_\mu(x) = n_\mu(x) + \frac{\theta}{2\pi} m_\mu(x). 
\end{equation}
Under this approximation, we can observe the $2\pi$ periodicity of the $\theta$ angle. Since both $n_\mu$ and $m_\mu$ are $\mathbb{Z}$-valued link fields, $\tilde{n}_\mu$ stays the same under $\theta\mapsto \theta+2\pi$ associated with $n_\mu\mapsto n_\mu-m_\mu$. 
The partition function of the Cardy-Rabinovici model is given by 
\begin{equation}
    \calZ_{\CR (\mathrm{lattice})}:=\sum_{\{s_{\mu\nu}\},\{n_\mu\}}\int \Diff \widetilde{a}_\mu \exp\Bigl(-S_{\mathrm{kin}}[\widetilde{a}_\mu, s_{\mu\nu}]-S_{\mathrm{mat}}[\widetilde{a}_\mu,s_{\mu\nu},n_{\mu}]\Bigr), 
\end{equation}
where the integration of $\widetilde{a}_\mu(x)$ is restricted to $[-\pi,\pi)$ to fix the $\mathbb{Z}$-valued $1$-form gauge redundancy. 

We should keep in mind that the $\theta$ angle periodicity is an emergent feature in the Cardy-Rabinovici model, and the lattice-scale fluctuations do not enjoy this property. 
Recently, topological features of the Villain-type $U(1)$ gauge theory were scrutinized in Refs.~\cite{Sulejmanpasic:2019ytl, Anosova:2022cjm} (see also Refs.~\cite{Gattringer:2018dlw, Sulejmanpasic:2020lyq, Sulejmanpasic:2020ubo, Anosova:2022yqx}). Especially in Ref.~\cite{Anosova:2022cjm}, it is found that both the exact $2\pi$ periodicity of $\theta$ and the self-duality can be established within the exponentially local lattice action, while one has to give up with the ultra-local action. 
In this work, we keep using the original formulation of the Cardy-Rabinovici model, since our computations are done with the formal continuum description that we shall discuss soon later, and various details at the lattice scale are neglected. 
It would be quite interesting if our calculations can be performed with the explicit lattice regularization based on the proposal of Ref.~\cite{Anosova:2022cjm}.

\subsection{Formal continuum description of the model}

Here, let us present the formal continuum description of the Cardy-Rabinovici model following Ref.~\cite{Honda:2020txe} as it is useful to understand the topological aspect of the model. 
However, this is the $U(1)$ gauge theory coupled to both electric and magnetic particles, and thus we do not know its Lagrangian formulation to have the path-integral expression with the manifest locality and Euclidean (or Lorentz) invariance.
Therefore, we use the worldline representation for the matter fields, which spoils the manifest locality of the theory.

Introducing the complex coupling,
\begin{align}
    \tau := \frac{\theta}{2 \pi} + \im \frac{2 \pi}{Ng^2},
\end{align}
we denote the Maxwell action with the $\theta$ term as 
\begin{align}
    S^\tau_{U(1)}[\diff a ] 
    :=& -\frac{\im N}{16\pi}\int \left(\tau (\diff a+*\diff a)^2+\overline{\tau}(\diff a-*\diff a)^2\right) \nonumber\\
    =&\, \frac{1}{2g^2} \int \diff a \wedge *\diff a - \frac{\im N \theta}{8 \pi^2} \int \diff a \wedge \diff a.
    \label{eq:Maxwell_action}
\end{align}
We note that the $\theta$ term in \eqref{eq:Maxwell_action} has an extra factor $N$ compared with the usual definition, so one might wonder if the $\theta$ angle periodicity is given by $2\pi/N$ instead of $2\pi$. 
This is not the case for the Cardy-Rabinovici model because of dynamical magnetic monopoles. 
We should notice that the $\theta$ angle periodicity cannot be determined just by the Maxwell action in the case of the $U(1)$ gauge theory, and we must specify the electric and magnetic matter contents and their Boltzmann weights to determine it. 

We introduce the Wilson line as 
\begin{equation}
    W(C):=\exp\left(\im \int_C a\right), 
    \label{eq:WilsonLoop}
\end{equation}
where $C$ is a closed loop in the spacetime. In the lattice description, the worldline of electric charge is specified by the $\mathbb{Z}$-valued link field $n_\mu$ satisfying the conservation law~\eqref{eq:charge_conservation}, and we may relate them by $\delta(C)=*(n_\mu \diff x^\mu)$. 
To treat magnetic monopoles, we also introduce the 't~Hooft line,
\begin{equation}
    H(C').
    \label{eq:tHooftLoop}
\end{equation}
This is defined as a defect operator, which determines the boundary condition of the $U(1)$ gauge field $a$ near the closed loop $C'$: For sufficiently small two-spheres $S^2$ linking to the loop $C'$, $a$ must satisfy
\begin{equation}
    \int_{S^2} \diff a=2\pi. 
\end{equation}
The path integral of matter fields is given by 
\begin{equation}
    \calZ_{\mathrm{mat}}[a]=\sum_{C,C':\,\mathrm{loops}} \exp(-S_{\mathrm{mat}}[C,C']) W^N(C) H(C'), 
\end{equation}
where $\sum_{C,C'}$ represents the summation over all possible closed loops $C, C'$ of electric and magnetic worldlines, and $S_{\mathrm{mat}}[C,C']$ controls the Boltzmann weight for configurations of the loops before taking into account the $U(1)$ gauge interaction. 
We note that the Wilson loop appears in the form of $W^N(C)$, and thus dynamical electric charge is quantized in the integer multiples of $N$. 

Formal continuum description of the Cardy-Rabinovici model is then given by 
\begin{equation}
    \calZ_{\CR}^{\tau}:=\int \Diff a \exp(-S^\tau_{U(1)}[\diff a]) \calZ_{\mathrm{mat}}[a]. 
\end{equation}
For later purpose, we make the $\tau$ dependence explicit.
Let us discuss the periodicity of the $\theta$ angle~\cite{Honda:2020txe}. When we assume that the spacetime manifold $X$ is a closed spin $4$-manifold, the topological charge is quantized as 
\begin{equation}
    \frac{1}{8\pi^2}\int_X \diff a\wedge \diff a\in \mathbb{Z}. 
\end{equation}
At the first sight, this suggests that the partition function is invariant under $\theta \mapsto \theta+\frac{2\pi}{N}$, or $\tau\mapsto \tau+\frac{1}{N}$, according to the definition~\eqref{eq:Maxwell_action}. 
However, the story is not so simple under the presence of the 't~Hooft line, because the Witten effect~\cite{Witten:1979ey} suggests 
\begin{equation}
    \left\langle H(C')\right\rangle_{S^{\tau+1/N}_{U(1)}}=\left\langle H(C') W(C')\right\rangle_{S^{\tau}_{U(1)}}. 
\end{equation}
Here, the expectation value is understood in terms of the path integral of pure Maxwell theory, where we set the complex coupling as $\tau+\frac{1}{N}$ on the left-hand side while we set it as $\tau$ on the right-hand side. 
Since the dynamical electric charge is assumed to have the charge $N$, the matter part explicitly breaks the $2\pi/N$ periodicity even though it is the minimal possible periodicity in the pure Maxwell theory. 

Let us then try the $2\pi$ shift of the $\theta$ term. We take the expectation value of the matter partition function at the complex coupling $\tau+1$, and then we can rewrite it as follows:
\begin{align}
    \left\langle \calZ_{\mathrm{mat}}[a]\right\rangle_{S^{\tau+1}_{U(1)}}
    &=\sum_{C,C':\,\mathrm{loops}} \exp(-S_{\mathrm{mat}}[C,C']) 
    \left\langle W^N(C) H(C')\right\rangle_{S^{\tau+1}_{U(1)}}\nonumber\\
    &=\sum_{C,C':\,\mathrm{loops}} \exp(-S_{\mathrm{mat}}[C,C']) 
    \left\langle W^N(C) \{H(C')W^{N}(C')\}\right\rangle_{S^{\tau}_{U(1)}}\nonumber\\
    &=\sum_{C,C':\,\mathrm{loops}} \exp(-S_{\mathrm{mat}}[C-C',C']) 
    \left\langle W^N(C) H(C')\right\rangle_{S^{\tau}_{U(1)}}. 
    \label{eq:check_2pi_theta}
\end{align}
To find the last expression, we replace $C$ by $C+C'$. If the matter action satisfies 
\begin{equation}
    S_{\mathrm{mat}}[C-C', C']=S_{\mathrm{mat}}[C,C'], 
    \label{eq:matter_T}
\end{equation}
then the right hand side of \eqref{eq:check_2pi_theta} becomes the expectation value of the matter partition function at $\tau$. 
As a result, we have established the $2\pi$ periodicity of $\theta$ by assuming a property~\eqref{eq:matter_T} of the matter action. 
In our computations in the next section, we assume that $S_{\mathrm{mat}}$ has a sufficiently nice property under various transformations.\footnote{In addition to \eqref{eq:matter_T}, we should also require that $S_{\mathrm{mat}}[C,C']=S_{\mathrm{mat}}[-C,-C']$ for the $\mathsf{C}$ symmetry, $S_{\mathrm{mat}}[C,C']=S_{\mathrm{mat}}[-C,C']$ for the $\mathsf{CP}$ symmetry, and $S_{\mathrm{mat}}[C,C']=S_{\mathrm{mat}}[-C', C]$ for the electromagnetic duality. For details, see Ref.~\cite{Honda:2020txe}.} 
It would be convenient to think that we simply set
\begin{equation}
    S_{\mathrm{mat}}\equiv 0. 
\end{equation}
This drastic assumption is legitimate for our purpose, since we just try to understand the algebraic property of the topological defects generating symmetries. Since it is robust under continuous symmetric deformations, we may start with the easiest situation. 
As a result, our formal continuum description becomes
\begin{align}
    \calZ_{\CR}^\tau 
    &= \int \mathcal{D}a \exp\left(-S^{\tau}_{U(1)}[\diff a]\right) \sum_{C, C':\,\mathrm{loops}} \hspace{-0.8em} W^N (C) H(C'). 
    \label{eq:CR_def}
\end{align}
In the following, we regard \eqref{eq:CR_def} as the definition of the Cardy-Rabinovici model in this paper. 

\subsection{Conjectured phase diagram from the free-energy argument}
\label{sec:phase_diagram}

In this subsection, we give a review on the conjectured phase diagram obtained from the energy versus entropy argument~\cite{Cardy:1981qy, Cardy:1981fd} (see also \cite{Banks:1977cc, Savit:1977fw}). 

\begin{figure}
    \centering
    \includegraphics[scale=0.8]{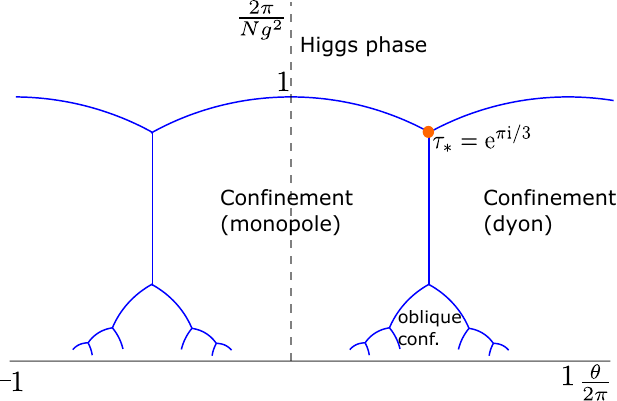}
    \caption{Conjectured phase diagram of the Cardy-Rabinovici model. In the weak-coupling regime, electric charge condensation occurs and the system is in the Higgs phase. In the strong-coupling regime, magnetic charge condensation occurs. Depending on the values of $\theta$, different types of dyon start to condense, and the Cardy-Rabinovici model is expected to have the rich phase structure. }
    \label{fig:phase_diagram}
\end{figure}

For the line operator $W^{Nn}(C)H^m(C)$, the coefficient of the Coulomb energy is given by 
\begin{align}
    \ve_{n,m}(\tau,\overline{\tau})&=N^2 g^2\left(n + \frac{\theta}{2\pi}m\right)^2+\left(\frac{2\pi}{g}\right)^2 m^2 \nonumber\\
    &=\frac{2\pi N}{\mathrm{Im}(\tau)}\left|n + m\tau\right|^2. 
    \label{eq:Coulomb}
\end{align}
The long-range part of the Coulomb interaction can be screened by the presence of other lines, so let us only take into account the short-range part. Then, its contribution to the action density is proportional to 
\begin{equation}
    \ve_{n,m}(\tau,\overline{\tau}) G(0) L(C), 
\end{equation}
where $G(0)$ is the short-range part of lattice massless Green function and $L(C)$ is the length of the loop $C$. 
This should be compared with the entropy of this line configuration with the given length, and it is given by  
\begin{equation}
    \ln(2\cdot 4-1) L(C)
\end{equation}
for the $4$d cubic lattice. 
When the entropy factor overcomes the energy suppression, 
\begin{equation}
    \ve_{n,m}(\tau,\overline{\tau})<\mathrm{const.}\sim \frac{\ln 7}{G(0)},
    \label{eq:criterion}
\end{equation}
then we regard that the charge $(n,m)$ can condense in the vacuum. 
When there are several candidates for condensation, we choose $(n,m)$ that gives the minimal energy. 
If there are no $(n,m)$ satisfying the inequality, the system is in the Coulomb phase with massless photon. 
We show the phase diagram in Fig.~\ref{fig:phase_diagram} when there is always some condensation at any values of $\tau$ (For detailed explanations on the figure, see Sec.~2 of Ref.~\cite{Honda:2020txe}).

The conjectured phase diagram in Fig.~\ref{fig:phase_diagram} has a rich structure because of the large invariance of $\ve_{n,m}(\tau,\overline{\tau})$. 
Notably, it has the electromagnetic $SL(2,\mathbb{Z})$ invariance, generated by $S$ and $T$ transformations~\cite{Cardy:1981fd}:
\begin{align}
    &S:\tau\mapsto -\frac{1}{\tau},\quad
    \begin{pmatrix}
    n\\
    m
    \end{pmatrix}\mapsto 
    \begin{pmatrix}
    0&-1\\
    1&0
    \end{pmatrix}\begin{pmatrix}
    n\\
    m
    \end{pmatrix}
    =\begin{pmatrix}
    -m\\
    n
    \end{pmatrix},
    \label{eq:S_transformation}\\
    &T:\tau\mapsto \tau+1,\quad 
    \begin{pmatrix}
    n\\
    m
    \end{pmatrix}\mapsto\begin{pmatrix}
    1&-1\\
    0&~1
    \end{pmatrix} \begin{pmatrix}
    n\\
    m
    \end{pmatrix}=\begin{pmatrix}
    n-m\\
    m
    \end{pmatrix},
    \label{eq:T_transformation}
\end{align}
and then the group structure can be written as 
\begin{equation}
    SL(2,\mathbb{Z})=\bigl\langle S,T\, \bigl|\, S^2=(ST^{-1})^3,\, S^4=1\bigr\rangle. 
\end{equation}
We can identify $\mathsf{C}=S^2=(ST^{-1})^3$ as the charge conjugation symmetry, and we also have the $\mathsf{CP}$ invariance, $\tau\to -\overline{\tau}$ with $(n,m)\to (-n,m)$. 
The phase diagram has to be symmetric under these transformations according to the above criterion~\eqref{eq:criterion} for condensation, 
and the conjectured phase diagram shown in Fig.~\ref{fig:phase_diagram} satisfies these requirements. 

In Fig.~\ref{fig:phase_diagram}, we put an orange dot at the special point of the complex complex coupling,
\begin{equation}
    \tau_*=\rme^{\pi\im/3}=\frac{1+\sqrt{3}\,\im}{2}. 
\end{equation}
This is the unique point in the upper half-plane, $\mathrm{Im}(\tau)>0$, that is fixed under the $ST^{-1}$ transformation:
\begin{equation}
    ST^{-1}:\tau\mapsto \frac{1}{-\tau+1},\quad 
    \begin{pmatrix}
    n\\
    m
    \end{pmatrix}\mapsto\begin{pmatrix}
    0&-1\\
    1&~1
    \end{pmatrix} \begin{pmatrix}
    n\\
    m
    \end{pmatrix}=\begin{pmatrix}
    -m\\
    n+m
    \end{pmatrix}. 
\end{equation}
We can check that $ST^{-1}$ generates the $\frac{2\pi}{3}$ anticlockwise rotation around the fixed point $\tau_*$. For $|\delta \tau|\ll 1$, we have 
\begin{equation}
    ST^{-1}:\tau_*+\delta\tau\to \tau_*+\rme^{2\pi\im/3}\delta \tau.
\end{equation}
This generates the $(\mathbb{Z}_6)_{ST^{-1}}$ transformation, and we should note that $(ST^{-1})^3=\mathsf{C}$ acts on $(n,m)\to(-n,-m)$ while $\tau=\tau_*+\delta \tau$ is fixed. 
As shown in Fig.~\ref{fig:phase_diagram}, the $ST^{-1}$ transformation exchanges the Higgs phase, confinement phase via monopole condensation, and confinement phase via dyon condensation in the cyclic manner.

\section{Non-invertible \texorpdfstring{$ST^{p}$}{ST[p]} duality defect and fusion rules} \label{Sec:duality_defect_fusion_rule}

In this section, we construct the $ST^p$ self-duality defect and discuss its properties. 
After having general formula, we especially focus on the $ST^{-1}$-self-dual point $\tau_* = \rme^{\pi \im /3}$ for concreteness. Self-dualities on other parameters are discussed in Appendix \ref{App:other_parameters}.

In Sec.~\ref{Sec:self-duality}, we describe how the self-duality is realized with the $\mathbb{Z}_N^{[1]}$ gauging.
Then, in Sec.~\ref{Sec:fusion-rule}, we introduce its duality defect as a half-space gauging and derive the fusion rule for this defect.

\subsection{\texorpdfstring{$ST^{p}$}{ST[p]} self-duality, \texorpdfstring{$\mathbb{Z}_N^{[1]}$}{ZN[1]} gauging, and mixed gravitational anomaly} \label{Sec:self-duality}

To realize the $ST^{p}$ transformation as a topological defect, we first need to understand how the `self-duality' is realized. 
If two quantum field theories $\mathcal{T}_1$ and $\mathcal{T}_2$ are related by a duality transformation, let us denote it symbolically as 
\begin{equation}
    \mathcal{T}_1\simeq \mathcal{T}_2. 
\end{equation}
As we have reviewed in Sec.~\ref{sec:phase_diagram}, the conjectured phase diagram of Cardy-Rabinovici model is understood from the $SL(2,\mathbb{Z})$ duality transformations. 
Let us denote the Cardy-Rabinovici model with the complex coupling $\tau$ as $\CR^\tau$. Does the $SL(2,\mathbb{Z})$ duality imply
\begin{equation}
    \CR^{\tau}\simeq \CR^{ST^{p}(\tau)}=\CR^{-(\tau+p)^{-1}}\,\,?
    \label{eq:self_duality_naive}
\end{equation}
We can immediately see that this cannot be the case. 
Under the $S$ transformation, the electric and magnetic charges are exchanged. Since the dynamical electric charges are quantized in $N$, the original theory enjoys the electric $\mathbb{Z}_N^{[1]}$ symmetry, while the transformed theory has the magnetic, or dyonic, $\mathbb{Z}_N^{[1]}$ symmetry. 
As a result, the self-duality of the Cardy-Rabinovici model cannot be understood in the naive way when we take into account the global aspects of field theories~\cite{Honda:2020txe}.

Even though the naive duality relation~\eqref{eq:self_duality_naive} does not hold, we find that the duality of Cardy-Rabinovici model is realized as 
\begin{align}
    \CR^\tau/(\mathbb{Z}_N^{[1]})_p \simeq \CR^{ST^{p}(\tau)}=\CR^{-(\tau+p)^{-1}}. 
    \label{eq:self-duality-rough}
\end{align}
On the left-hand-side of the duality relation, we consider the $\mathbb{Z}_N^{[1]}$-gauged Cardy-Rabinovici model with the discrete $\theta$ parameter $p$, and we claim that it is dual to the original Cardy-Rabinovici model at $\tau'=ST^p(\tau)$. 
By gauging the $\mathbb{Z}_N^{[1]}$ symmetry with the discrete topological term, the genuine line operator with the nontrivial dual $\mathbb{Z}_N^{[1]}$ transformation has the electromagnetic charge $(e,m)=(p,1)$. The $T^{p}$ transformation makes it to have $(e,m)=(0,1)$ due to the Witten effect, and the $S$ transformation brings it back to the electric line $(e,m)=(-1,0)$, which is a $(e,m)=(1,0)$ line with the inverse orientation.
Since all loops are summed up in \eqref{eq:CR_def}, the orientation does not matter for the evaluation of the partition function.
This suggests that the level-$p$ $\mathbb{Z}_N^{[1]}$-gauged model can be reduced to the original Cardy-Rabinovici model via the $ST^p$ transformation.
This is the basic idea behind the duality relation~\eqref{eq:self-duality-rough}. 
More precisely, 
\begin{claim}\label{claim:self-duality-1}
Cardy-Rabinovici model and the level-$p$ $\mathbb{Z}_N^{[1]}$-gauged model are dual in the following sense:
\begin{enumerate}
    \item[\textsc{(i)}] The partition functions are identical up to a gravitational counterterm,
    \begin{align}
        \calZ_{\CR/(\mathbb{Z}_N^{[1]})_p}^{\tau} [B] 
        &= N^{\frac{\chi(X)}{2} }{\left( ST^p(\tau)\right)^{\frac{\chi(X)+\sigma(X)}{4} } \left( ST^p(\overline{\tau})\right)^{\frac{\chi(X)-\sigma(X)}{4} } } \, \calZ_{\CR}^{ST^p(\tau)} [B]\notag\\
        &=N^{\frac{\chi(X)}{2} }{\left( \tau+p\right)^{-\frac{\chi(X)+\sigma(X)}{4} } \left( \overline{\tau}+p\right)^{-\frac{\chi(X)-\sigma(X)}{4} } } \, \calZ_{\CR}^{-(\tau+p)^{-1}} [B],
        \label{eq:self-duality-1}
    \end{align}
    where $\chi(X)$ is the Euler number and $\sigma(X)$ is the signature of the $4$d spacetime $X$. 

    \item[\textsc{(ii)}] Genuine line operators of the level-$p$ $\mathbb{Z}_N^{[1]}$ gauged model are generated by $W^N_{\diff a+b}(C)$ and $H_{N(\diff a + b)}W^{p/N}_{N(\diff a+b)}(C')$. 
    Under the duality, these line operators correspond to $W(C)$ and $H(C)$ of the Cardy-Rabinovici model as 
    \begin{align}
        \begin{pmatrix}
        H_{N(\diff a+ b)}W^{p/N}_{N(\diff a +b)}(C')\\
        W^N_{\diff a + b}(C)
        \end{pmatrix}\mapsto 
        \begin{pmatrix}
        W^{-1}(C')\\
        H(C)
        \end{pmatrix}. \label{eq:Claim-1-operators}
    \end{align}
\end{enumerate}
\end{claim}

Here, let us introduce several notations used in Claim~\ref{claim:self-duality-1}:
\begin{itemize}

    \item We denote the Wilson and 't~Hooft loops as $W_f(C)$ and $H_f(C)$ when they are defined with the $2$-form field strength $f$. 
    The 't~Hooft line $H_f(C)$ is defined by $\int_{S^2} f=2\pi$ for small $S^2$ linking to $C$ whether or not this is the minimal quantized value for $f$. 

    \item Under the presence of the background $\mathbb{Z}_N$ two-form gauge field $B$, the partition function of Cardy-Rabinovici model is defined as 
    \begin{equation}
        \calZ_{\CR}^\tau[B]=\int \Diff a \exp(-S^\tau_{U(1)}[\diff a+B])\sum_{C,C':\,\mathrm{loops}} \hspace{-0.8em} W_{\diff a+B}^N(C) H_{\diff a+B}(C'). \label{eq:CR-partition-func-w-background}
    \end{equation}
    Here, we note that  $W_{\diff a + B}^N(C)$ is a genuine line operator, while the charge-$1$ Wilson loop $W_{\diff a + B}(C, \Sigma) = \rme^{\im \int_{\Sigma: \partial \Sigma = C} (\diff a + B)}$ depends on the choice of the surface $\Sigma$ with nontrivial $B$.
    
    \item We define the $\mathbb{Z}_N^{[1]}$-gauged partition function with the discrete $\theta$ term $\frac{N p}{4\pi}\int b\wedge b$ under the presence of the background field $B$ as\footnote{Precisely speaking, the the form of the topological action depends on whether $N$ is even or odd. For odd $N$, we may use the cup product $B\cup B$, but for even $N$, we need to use the Pontryagin square, $\mathfrak{P}(B)=B\cup B+B\cup_1\delta B$ to define the SPT action. Fortunately, we do not have to care about this subtlety as long as $H^*(X;\mathbb{Z})$ is torsion free. In such cases, $B\in H^2(X;\mathbb{Z}_N)$ has an integer lift $\tilde{B}\in H^2(X;\mathbb{Z})$, and $\tilde{B}\cup\tilde{B} \mod 2N$ is well defined. For more details, see Ref.~\cite{Kapustin:2013qsa}. }  
    \begin{equation}
        \calZ_{\CR/(\mathbb{Z}_N^{[1]})_p}^\tau[B]
        =\int \Diff b\,\, \calZ_{\CR}^\tau[b]\exp\left(\frac{\im N p}{4\pi}\int_X b\wedge b+\frac{\im N}{2\pi}\int_X b\wedge B\right). 
        \label{eq:Z_N-gauging}
    \end{equation}
    Here, $X$ is the $4$d spacetime realized as the closed spin $4$-manifold, and we further assume that the integer-valued cohomology $H^*(X;\mathbb{Z})$ is torsion free. 

    \item  When we should care about the normalization of the path integral, we replace $\int \Diff b$ by $\frac{|H^0(X;\mathbb{Z}_N)|}{|H^1(X;\mathbb{Z}_N)|}\sum_{b\in H^2(X;\mathbb{Z}_N)}$. The prefactor $\frac{|H^0 (X; \mathbb{Z}_N)|}{|H^1 (X; \mathbb{Z}_N)|}$ represents the inverse of the gauge volume. 
    For discrete gauge fields, we follow the convention of Ref.~\cite{Kapustin:2014gua}, and we write the wedge product in place of the cup product. 

    \item We denote the Betti numbers as $\beta_k=\dim H^k(X;\mathbb{R})$, and we decompose $\beta_2=\beta_2^+ + \beta_2^-$ into the (anti-)self-dual parts. They are related to the Euler number and the signature as $\beta_{2}^{\pm}-\beta_1+\beta_0=(\chi\pm\sigma)/2$. 
    Using the curvature $2$-form $R$, $\chi$ and $\sigma$ are given as
    \begin{align}
        \chi(X)=\int_{X}\frac{1}{2(4\pi)^2}\ve^{ijk\ell}R_{ij}\wedge R_{k\ell},\quad 
        \sigma(X)=\int_{X} \frac{-1}{6(2\pi)^2}\tr(R\wedge R),
    \end{align}
    due to the Gauss-Bonnet theorem and the Hirzebruch signature theorem.
    This is why we can regard the overall constant in \eqref{eq:self-duality-1} as a gravitational counterterm. 
\end{itemize}

Here, we only give an outline for the derivation of  Claim~\ref{claim:self-duality-1} in the case of pure Maxwell theory, and we postpone the detailed proof to Appendix~\ref{sec:proof_claim1} as it is a bit lengthy. 
In the case of pure Maxwell theory, the $(\mathbb{Z}_N^{[1]})_p$ gauging procedure is given by
\begin{align}
    \calZ^\tau_{U(1)/(\mathbb{Z}_N^{[1]})_p}[B]&=\int \Diff b\, \calZ^\tau_{U(1)}[b]\rme^{\im \frac{Np}{4\pi}\int b\wedge b+\im \frac{N}{2\pi}\int b\wedge B}\nonumber\\
    &=\int \Diff b \Diff a\,\rme^{-S^\tau_{U(1)}[\diff a+b]}\rme^{\im \frac{Np}{4\pi}\int b\wedge b+\im \frac{N}{2\pi}\int b\wedge B}. \label{eq:integral-repr-Euler-signature}
\end{align}
Thanks to the Dirac quantization, we have $\int_{M_2}\diff a\in 2\pi \mathbb{Z}$ for any closed $2$-manifolds $M_2$, and thus we can replace the $\mathbb{Z}_N$ $2$-form gauge field $b$ by $\diff a+b$ in the topological factor without affecting its value. Therefore, we get 
\begin{align}
    \calZ^\tau_{U(1)/(\mathbb{Z}_N^{[1]})_p}[B]&=
    \int \Diff a \Diff b\, \rme^{-S^\tau_{U(1)}[\diff a+b]}\rme^{\im \frac{Np}{4\pi}\int (\diff a+b)^2+\im \frac{N}{2\pi}\int (\diff a+b)\wedge B}\nonumber\\
    &=\int \Diff a \Diff b\, \rme^{-S^{\tau+p}_{U(1)}[\diff a+b]}\rme^{\im \frac{N}{2\pi}\int (\diff a+b)\wedge B}\nonumber\\
    &=\int \Diff a'\, \rme^{-S^{(\tau+p)/N^2}_{U(1)}[\diff a']}\rme^{\frac{\im}{2\pi}\int \diff a'\wedge B}. 
\end{align}
In the above derivation, we can see that the discrete $\theta$ term of level $p$ introduces the $T^p$ transformation. 
In order to obtain the last line, we note that $N(\diff a+b)$ satisfies the proper Dirac quantization as the $U(1)$ gauge field, and thus we introduce the new $U(1)$ gauge field $a'$ by $\diff a'=N(\diff a+b)$. 
Because of this rescaling, the charge $N$ Wilson lines becomes the unit charge $1$, and instead, the charge $1$ 't~Hooft line becomes charge $N$. 
We then perform the Abelian duality transformation, 
\begin{align}
    \calZ^\tau_{U(1)/(\mathbb{Z}_N^{[1]})_p}[B]
    &=\int \Diff \widetilde{h} \Diff \tilde{a}\rme^{-S^{(\tau+p)/N^2}_{U(1)}[\widetilde{h}]}\rme^{\frac{\im}{2\pi}\int \widetilde{h}\wedge (\diff \tilde{a}+B)} \nonumber\\
    &\propto \calZ^{-(\tau+p)^{-1}}_{U(1)}[B]. 
\end{align}
In the intermediate step, we have introduced the $\mathbb{R}$-valued $2$-form field $\widetilde{h}$ and another $U(1)$ gauge field $\tilde{a}$. 
Integrating out $\tilde{a}$, $\widetilde{h}$ satisfies the Dirac quantization condition so it becomes the $U(1)$ gauge field strength $\widetilde{h}=\diff a$, and we obtain the original expression. 
Integrating out $\widetilde{h}$ instead, we find the $ST^p$-duality relation. 
To obtain the gravitational counterterm in the proportionality coefficient, we have to extend the discussion of Refs.~\cite{Witten:1995gf, Verlinde:1995mz} by including the $\mathbb{Z}_N^{[1]}$ gauging with the discrete topological term. In Appendix~\ref{sec:proof_claim1}, we determine the gravitational counter term by introducing the UV cutoff and the renormalization prescription, and also discuss the behaviors of line operators to include the effect of the matter contribution.  

As a corollary of Claim \ref{claim:self-duality-1}, we obtain the important relation at the $ST^{-1}$ self-dual point $\tau=\tau_*$:

\begin{claim}[Corollary of Claim \ref{claim:self-duality-1}] \label{claim:self-dual-stinv}
Let us set $\tau=\tau_*=\rme^{\pi\im/3}$, which is the fixed point of $ST^{-1}$. 
Then, the relation~\eqref{eq:self-duality-1} of partition functions with $p=-1$ becomes
\begin{align}
    \calZ_{\CR/(\mathbb{Z}_N^{[1]})_{-1}}^{\tau_*} [B] = N^{\frac{\chi(X)}{2}} \rme^{-\frac{\pi \im}{3} \sigma(X)} \,\calZ_{\CR}^{\tau_*} [B]. 
    \label{eq:self-duality-fixed-param}
\end{align}
\end{claim}

This follows immediately from \eqref{eq:self-duality-1} by substituting $\tau=\tau_*=\rme^{\pi\im/3}$ and $p=-1$. 
This prefactor $ N^{\frac{\chi(X)}{2}} \rme^{-\frac{\pi \im}{3} \sigma(X)}$ can be regarded as a mixed gravitational anomaly, and especially the signature part can be used to constrain the dynamics as we shall discuss in Sec.~\ref{sec:constraints_dynamics}.

Here, we would like to emphasize that Claim~\ref{claim:self-duality-1} can be used to derive all the possible duality relations of Cardy-Rabinovici model. 
To see how it works, let us first consider the duality relation,
\begin{equation}
    \CR^\tau/(\mathbb{Z}_N^{[1]})_{p_1}\simeq \CR^{ST^{p_1}(\tau)}. 
\end{equation}
We can again gauge the $\mathbb{Z}_N^{[1]}$ symmetry of both sides with the discrete level-$p_2$, and then we have 
\begin{equation}
    (\CR^\tau/(\mathbb{Z}^{[1]}_N)_{p_1})/(\mathbb{Z}^{[1]}_N)_{p_2}\simeq \CR^{ST^{p_1}(\tau)}/(\mathbb{Z}^{[1]}_N)_{p_2}\simeq \CR^{ST^{p_2} ST^{p_1}(\tau)}. 
\end{equation}
By applying the formula~\eqref{eq:self-duality-1} for each step, we obtain the relation between the partition functions for $ST^{p_2}ST^{p_1}$. 
As an example, let us set $\tau_{**}=\frac{\sqrt{3}+\im}{2\sqrt{3}}$, which is the fixed point of $ST^{-1}ST^2S$, and then we obtain (see Appendix~\ref{sec:triple_point_oblique} for more details)
\begin{equation}
    \calZ^{\tau_{**}}_{((\CR/(\mathbb{Z}_N^{[1]})_0)/(\mathbb{Z}^{[1]}_{N})_{2})/(\mathbb{Z}^{[1]}_N)_{-1}}[B]
    =N^{\frac{3\chi(X)}{2}}\rme^{-\frac{\pi\im}{6}\sigma(X)} \calZ^{\tau_{**}}_{\CR}[B].
    \label{eq:self_duality_ST-1ST2S}
\end{equation}
In this way, we can find various self-duality relations from Claim~\ref{claim:self-duality-1}.

\subsection{Non-invertible topological self-duality defect and fusion rule} \label{Sec:fusion-rule}

So far, we have seen that the self-duality of the Cardy-Rabinovici model involves the gauging of $\mathbb{Z}_N^{[1]}$ symmetry with the appropriate choice of the discrete $\theta$ angle. 
The duality is promoted to the symmetry at the fixed point, and there exists the topological defect for the duality symmetry.  
In the following, we construct the topological defect for the $ST^{-1}$ transformation at $\tau=\tau_*$. 

When the duality relation is realized via the gauging of the symmetry, the duality symmetry cannot be realized as the ordinary symmetry operation that gives the unitary transformation. 
Instead, it should be realized as the non-invertible topological defects~\cite{Koide:2021zxj, Choi:2021kmx, Kaidi:2021xfk}. 
Such an example in $4$d QFT was first realized in Ref.~\cite{Koide:2021zxj}, and later it is noticed that we can systematically define the non-invertible duality symmetry by performing the gauging in the half spacetime~\cite{Choi:2021kmx, Kaidi:2021xfk}. 

We note that the fusion rule will not be used in the derivation of low-energy theorems, so one may directly go to Sec.~\ref{sec:constraints_dynamics} before reading the following of this section. 

\subsubsection{Construction of the defect}

Let us construct the duality defect $\mathscr{D} (M^{(3)})$ on a closed codimension 1 submanifold $M^{(3)}$.
We assume that $M^{(3)}$ is a compact orientable manifold.
We divide the spacetime manifold $X$ into two parts $X^+$ and $X^-$ with the same boundary with an opposite orientation $\partial X^+ = -\partial X^- = M^{(3)}$.
 We then consider the half-space gauging in $X^+$ with a Dirichlet boundary condition for $b$, and we call this defect operator as $\mathscr{D}(M^{(3)})$:
\begin{align}
  \langle \mathscr{D}(M^{(3)})\cdots \rangle =N^{-\frac{\chi(X^+)}{2}}\rme^{\frac{\pi\im}{3}\sigma(X^+)}\int_{X^+} \Diff b~  \rme^{-\frac{\im N}{4 \pi} \int b \wedge b}  \langle \cdots \rangle_b.
  \label{eq:repr-defect-half-space-gauging}
\end{align}
Here, $\langle \cdots \rangle_b$ stands for the expectation value of the Cardy-Rabinovici model with the $\mathbb{Z}_N^{[1]}$ background $b$ insertion. The normalization of the path integral for $b$ is given by
\begin{align}
    \int_{X^+} \Diff b~ (\cdots) := \frac{|H^0 (X^+, \partial X^+; \mathbb{Z}_N)|}{|H^1 (X^+, \partial X^+; \mathbb{Z}_N)|} \sum_{b \in H^2 (X^+, \partial X^+; \mathbb{Z}_N)}  (\cdots),
\end{align}
where the relative cohomology represents the space of $\mathbb{Z}_N$-valued gauge fields obeying the Dirichlet boundary condition on $\partial X_+$. 

We have to check if the definition~\eqref{eq:repr-defect-half-space-gauging} is well-defined as the codim-$1$ defect, since it uses the information of the half-space $X^+$. 
Indeed, this does not define the codim-$1$ defects for generic theories with $\mathbb{Z}_N^{[1]}$ symmetry, but it does for the Cardy-Rabinovici model at $\tau=\tau_*$ thanks to the duality relation~\eqref{eq:self-duality-fixed-param}. 
As a consequence of the locality, the half-space gauging on the bulk $X^+$ goes back to the original Cardy-Rabinovici model via the $ST^{-1}$ duality (up to the gravitational counter term), and the nontrivial effect takes place only at the boundary $M^{(3)}$. 
As discussed in Ref.~\cite{Choi:2021kmx} for the case of $S$ transformation, the duality defect $\mathscr{D}(M^{(3)})$ can be seen as the Chern-Simons coupling on $M^{(3)}$, 
\begin{align}
&\int_{X^+} \Diff b~ \calZ_{\CR}^{\tau_*} [b] \rme^{-\frac{\im N}{4 \pi} \int b \wedge b}\notag \\
  &\propto \int_{X^+} \Diff \tilde{a}_+  \int_{X^-} \Diff a_- ~ \left[ \rme^{- \frac{\im N}{2 \pi} \int_{M^{(3)}} a_- \wedge \diff \tilde{a}_+ } \rme^{ \frac{\im N}{4 \pi} \int_{M^{(3)}} a_- \wedge \diff a_- } \right] \notag \\
  &~~ \times \rme^{-S_{U(1)}^{\tau_*} [\diff \tilde{a}_+ ]-S_{U(1)}^{\tau_*} [\diff a_- ]} \calZ_{\mathrm{mat}}^{X^+}[\diff \tilde{a}_+ ] \calZ_{\mathrm{mat}}^{X^-} [\diff a_- ].  \label{eq:half-space-S-transformation}
\end{align}
Here, we put a technical assumption that the matter partition functions can be separated as $\calZ_{\mathrm{mat}}^{X^-} [\diff a]$ and $\calZ_{\mathrm{mat}}^{X^-} [\diff a]$ for $X^+$ and $X^-$, respectively.

Since the derivation of (\ref{eq:half-space-S-transformation}) is almost the same as that of of Claim \ref{claim:self-duality-1}, let us explain how the boundary terms $\left[ \rme^{- \frac{\im N}{2 \pi} \int_{M^{(3)}} a_- \wedge \diff \tilde{a}_+ } \rme^{ \frac{\im N}{4 \pi} \int_{M^{(3)}} a_- \wedge \diff a_- } \right]$ appear.
In the derivation of duality relation, we replaced $\rme^{\frac{\im p N}{4 \pi} \int b \wedge b}$ by $\rme^{ \frac{\im p N}{4 \pi} \int (\diff a + b) \wedge (\diff a + b) }$. In the presence of the boundary, we need subtract the boundary Chern-Simons term $\rme^{- \frac{\im p N}{4 \pi} \int_{M^{(3)}} a \wedge \diff a}$, and the appearance of the diagonal Chern-Simons term is a new ingredient of the $ST^p$ duality defect compared with the $S$ duality defect in Ref.~\cite{Choi:2021kmx}. 
Note that, on the boundary $M^{(3)}$, the $U(1)$ fields (before $S$ transformation) on both sides have to match: $\left. a_+ \right|_{M^{(3)}} = \left. a_-  \right|_{M^{(3)}}$, so we can write this Chern-Simons term using the $U(1)$ gauge field $a_-$ on $X^-$. 
When we apply the $S$ transformation on $X^+$, we replace $N(\diff a_+ + b)$ by the $\mathbb{R}$-valued $2$-form field $\widetilde{h}$ with the topological coupling $\frac{1}{2\pi}\widetilde{h}\wedge \diff \widetilde{a}_+$, and we gauge away $\diff a_+$. In this step, we used $e^{ \frac{\im }{2 \pi} \int \diff a \wedge \diff \tilde{a}} = 1$ on closed manifolds, but it produces the off-diagonal Chern-Simons term $ \rme^{- \frac{\im N}{2 \pi} \int_{M^{(3)}} a_- \wedge \diff \tilde{a}_+ }$ when the boundary $M^{(3)}$ exists.

One may worry that a line operator across $M^{(3)}$ could invalidate the above half-space gauging and $ST^{-1}$ transformation.
However, such a line operator, say a Wilson line operator $W(C)$, can be rewritten as $W(\tilde{C}_+)W(\tilde{C}_-)$ with closed loops $\tilde{C}_+ \subset X^+,~\tilde{C}_- \subset X^-$.
In fact, when the loop $C$ consists of open paths $C_+ \subset X^+$ and $C_- \subset X^-$, i.e., $C = C_+ + C_-$, we can attach paths on $M^{(3)}$ with the opposite direction, $+\gamma$ and $- \gamma$, closing $C_+$ and $C_-$:
\begin{align}
    C = C_+ + C_- = (C_+ + \gamma) + ( - \gamma + C_-) =: \tilde{C}_+ + \tilde{C}_-.
\end{align}
Therefore, the correspondence of the line operators (\ref{eq:Claim-1-operators}) can be used here with this decomposition.
For example, the Wilson loop is transformed as $W(C) = W(\tilde{C}_+)W(\tilde{C}_-) \rightarrow  H^{-1/N}(\tilde{C}_+)W(\tilde{C}_-)$.
We understand the matter sector in this sense\footnote{
Note that the well-definedness of $H^{-1/N}(C_+)W(C_-)$ can be seen from the boundary off-diagonal Chern-Simons term $ \rme^{- \frac{\im N}{2 \pi} \int_{M^{(3)}} a_- \wedge \diff \tilde{a}_+ }$.
Indeed, the 't Hooft line $H^{-1/N}(C_+)$ introduces a magnetic defect surface $\diff \tilde{a}_+ \rightarrow \diff \tilde{a}_+ + \frac{2 \pi}{N} \delta(\Sigma)$ attached to $C_+$.
On the boundary $M^{(3)}$, this defect affects $\tilde{a}_+$ as $\diff \tilde{a}_+ \rightarrow \diff \tilde{a}_+ + \frac{2 \pi}{N} \delta(\gamma)$.
Therefore, the additional boundary term arising from this defect is $ \rme^{- \int_\gamma a_-}$, which closes the Wilson loop $W(C_-)$ in the other side.
Therefore, the off-diagonal Chern-Simons term $ \rme^{- \frac{\im N}{2 \pi} \int_{M^{(3)}} a_- \wedge \diff \tilde{a}_+ }$ makes $H^{-1/N}(C_+)W(C_-)$ well-defined.
}.



Based on (\ref{eq:Claim-1-operators}), the correspondence of the line operators between $X^+$ theory and $X^-$ theory is as follows.
\begin{align}
\mathrm{CR~model~in~}X^+~&\leftrightarrow~\mathrm{CR~model~in~}X^- \notag \\
        \begin{pmatrix}
        H_{\diff a_+ + b}(C') \rightarrow  W^{-N}_{\diff \tilde{a}_+} H_{\diff \tilde{a}_+} (C')  \\
        W_{\diff a_+ + b}(C, \Sigma) \rightarrow H_{\diff \tilde{a}_+}^{1/N}(C, \Sigma)
        \end{pmatrix}&\leftrightarrow
        \begin{pmatrix}
        H_{\diff a_-}(C') \\
        W_{\diff a_-}(C) 
        \end{pmatrix}, \label{eq:defect-line-operators}
\end{align}
where $a_+$ is the U(1) field on $X^+$ before $ST^{-1}$ transformation.
A genuine line operator can become a non-genuine line operator, e.g., $W_{\diff a_-}(C) \mapsto H_{\diff \tilde{a}_+}^{1/N}(C, \Sigma)$, in this correspondence.
Schematically, this can be expressed as the phenomenon that the sweeping of the topological defect $\mathscr{D} (M^{(3)})$ leaves the topological two-surface $\Sigma$ attached to the line operator ($W_{\diff a_-}(C) \rightarrow W_{\diff a_+ + b}(C, \Sigma) = W_{\diff a_+}(C) e^{i \int_\Sigma b}$) \cite{Choi:2021kmx}. 
We note that the appearance of the topological two-surface does not matter in the computation of the partition function, since all the electric matter has charge $N$ and the choice of the topological surface becomes irrelevant for those lines. 

\subsubsection{Fusion rules}
In this section, we discuss the fusion rule related to the self-duality topological defect, and we observe its non-invertible nature.

A naive expectation from the $SL(2,\mathbb{Z})$ group structure would suggest $\mathscr{D} (M^{(3)})^3 \overset{?}{=} \mathsf{C}(M^{(3)})$ and $\mathscr{D} (M^{(3)}) \times \Bar{\mathscr{D}} (M^{(3)}) \overset{?}{=} (\mathrm{const.}) \mathbbm{1}$.
However, due to the half-space gauging construction, the defect $\mathscr{D} (M^{(3)})$ does not obey these naive fusion rules of the group structure.
Let us determine the fusion rule of the defects more precisely\footnote{
In the original preprint version, the authors were implicitly assuming that the algebra was commutative and computed the fusion rule only for one of the orderings. 
In Ref.~\cite{Choi:2022zal}, it is pointed out that the actual fusion rule becomes noncommutative for even $N$. 
In the following, we show the correct result of the fusion rule taking into account the ordering of the symmetry defects. 

We thank the anonymous referee for pointing out the discrepancy on this point and drawing our attention. We also thank Shu-Heng Shao for the discussion on this issue during the conference, ``Continuous Advances in QCD'' at the University of Minnesota. 
}.

\begin{claim}
\label{claim:fusion_rule}
Let $\mathscr{D}(M^{(3)})$ be the $ST^{-1}$ self-duality defect, $\Bar{\mathscr{D}}(M^{(3)})$ be its orientation inverse, and $\eta(\Sigma)$ be a $\mathbb{Z}_N^{[1]}$ symmetry operator on a two-cycle $\Sigma\subset {M^{(3)}}$.
Here, $\Bar{\mathscr{D}} (M^{(3)})$ can be obtained by the half-space gauging of the other side $X^-$.
We also introduce the charge conjugation operator $\mathsf{C}(M^{(3)})$. 

We have the following fusion rules that involve the duality defects and $1$-form symmetry defects:
\begin{align}
    \mathscr{D} (M^{(3)}) \times \eta(\Sigma) &= \mathscr{D} (M^{(3)}),~~ \eta(\Sigma) \times \mathscr{D} (M^{(3)}) = (-1)^{Q(\Sigma)} \mathscr{D} (M^{(3)}), \label{eq:Deta-fusion-rule}\\
    \mathscr{D} (M^{(3)}) \times \Bar{\mathscr{D}} (M^{(3)}) &=  \frac{1}{N}\sum_{\Sigma \in H_2(M^{(3)}; \mathbb{Z}_N)} (-1)^{Q(\Sigma)} \eta(\Sigma), \notag \\
    \Bar{\mathscr{D}} (M^{(3)}) \times \mathscr{D} (M^{(3)}) &=  \frac{1}{N}\sum_{\Sigma \in H_2(M^{(3)}; \mathbb{Z}_N)}  \eta(\Sigma)    \label{eq:orientation_inv_fusion_rule}\\
    \mathscr{D} (M^{(3)})^3 
    &= \mathcal{N}_{ \mathscr{D}^3 }(M^{(3)}) ~ \mathsf{C}(M^{(3)})
    \sum_{\Sigma \in H_2(M^{(3)}; \mathbb{Z}_N)} (-1)^{Q(\Sigma)} \eta(\Sigma).
    \label{eq:ST-1_fusion_rule}
\end{align}
Here, $(-1)^{Q(\Sigma)}$ is a sign defined by
\begin{align}
    (-1)^{Q(\Sigma)} := \rme^{\frac{\im N}{4\pi}\int a_1(\Sigma) \wedge d_M a_1(\Sigma)}, 
\end{align}
where $a_1(\Sigma) \in H^1 (M^{(3)}; \mathbb{Z}_N)$ is the Poincar\'e dual of $\Sigma$. 
When $N$ is odd or the homology $H^*(M^{(3)}; \mathbb{Z})$ is torsion free, the sign $(-1)^{Q(\Sigma)}$ is trivial: $(-1)^{Q(\Sigma)} = 1$.
The normalization constant $\mathcal{N}_{ \mathscr{D}^3 }$ for $\mathscr{D}(M_3)^3$ is given by
\begin{align}
    \mathcal{N}_{ \mathscr{D}^3 }(M^{(3)}) :=& N^{-3\chi(X^+)/2} e^{\im \pi \sigma(X^+)} \frac{|H^0 (X^+, \partial X^+; \mathbb{Z}_N)|^3 |H^2 (X^+, \partial X^+; \mathbb{Z}_N)|}{|H^1 (X^+, \partial X^+; \mathbb{Z}_N)|^3} \notag \\
    &~~~ \times \left( \sum_{b \in H^2 (X^+, \partial X^+; \mathbb{Z}_N)} \rme^{-\frac{\im N}{4 \pi} \int b \wedge b}\right). \label{eq:fusion-rule-normalization}
\end{align}
\end{claim}

An important message is that some of the fusion rules involve the summation over the $1$-form symmetry generators, which is quite unconventional compared with the usual group-like symmetries. 
Such a non-group-like fusion rules have been observed in the fusion category symmetry~\cite{Aasen:2016dop, Bhardwaj:2017xup, Buican:2017rxc, Freed:2018cec,  Chang:2018iay, Thorngren:2019iar, Thorngren:2021yso, Ji:2019jhk, Rudelius:2020orz, Komargodski:2020mxz, Aasen:2020jwb, Inamura:2021wuo, Inamura:2021szw}, and also in a recent generalization of the duality symmetry in $4$d gauge theories~\cite{Koide:2021zxj,Choi:2021kmx, Kaidi:2021xfk}. 

We postpone the detailed discussion and derivation of these fusion rules in Appendix~\ref{sec:proof_claim_fusion_rule}. 
Here, we give a brief sketch of the derivation of the fusion rules.

The first fusion rule follows from the definition of $\mathscr{D} (M^{(3)})$ as the gauging procedure.
The $\mathbb{Z}_N^{[1]}$ symmetry operator $\eta(\Sigma)$ can be absorbed into $\mathscr{D} (M^{(3)})$ and the sign factor, $(-1)^{Q(\Sigma)}$, arises from the discrete $\theta$-term in this process depending on the ordering of the defects.
At a first glance, the fusion rules~\eqref{eq:Deta-fusion-rule} may seem to contradict with the other fusion rule~\eqref{eq:ST-1_fusion_rule} by violating the associativity. According to the first rule of \eqref{eq:Deta-fusion-rule}, we have 
\begin{equation}
    \eta\times \mathscr{D}^3=(\eta\times \mathscr{D})\times \mathscr{D}^2=\mathscr{D}\times \mathscr{D}^2=\mathscr{D}^3, 
\end{equation}
but we obtain $\eta \times \mathscr{D}^3=(-1)^Q \mathscr{D}^3$ when we first apply \eqref{eq:ST-1_fusion_rule}, and they give different answers.
However, as firstly revealed in Ref.~\cite{Choi:2022zal}, the normalization factor $\mathcal{N}_{ \mathscr{D}^3 }(M^{(3)})$ contains the partition function of the $U(1)$ level-$N$ Chern-Simons theory, and thus $\mathscr{D}^3(M^{(3)})$ vanishes whenever $M^{(3)}$ contains a surface $\Sigma$ such that $(-1)^{Q(\Sigma)}$ becomes nontrivial.
Hence, the fusion rules~\eqref{eq:Deta-fusion-rule} are consistent with the associativity and the other fusion rules.

The second rule can be derived from a parallel discussion with Refs.~\cite{Choi:2021kmx, Kaidi:2021xfk}.
Naively, from Claim \ref{claim:self-dual-stinv}, the gauging in $X^+$ and $X^-$ would be an identity transformation.
A more careful consideration with slightly separating these defects gives an extra gauging on the interval $M^{(3)}\times (-\ve,\ve)$.
This extra gauging on the interval leads to the right-hand side of (\ref{eq:orientation_inv_fusion_rule}).
There might be a possible local counterterm depending on the topology of $M$, but this can be absorbed by the local redefinition of $\Bar{\mathscr{D}} (M^{(3)})$.

The third fusion rule (\ref{eq:ST-1_fusion_rule}) consists of (1) charge conjugation, (2) sum over $\mathbb{Z}_{N}^{[1]}$ generators on $M^{(3)}$, (3) the normalization factor.
Let us look into each factor.

First, the action of the triple defect $\mathscr{D} (M^{(3)})^3:= \mathscr{D} (M^{(3)}) \times \mathscr{D} (M^{(3)}) \times \mathscr{D} (M^{(3)}) $ on the line operators can be determined from (\ref{eq:defect-line-operators}) as follows.
\begin{align}
        \begin{pmatrix}
        W (C)  \\
        H (C)
        \end{pmatrix}
        \mapsto
        \begin{pmatrix}
        H^{1/N}(C, \Sigma) \\
        W^{-N} H (C) 
        \end{pmatrix}
        \mapsto
        \begin{pmatrix}
        W^{-1} H^{1/N} (C, \Sigma) \\
        W^{-N} (C) 
        \end{pmatrix}
        \mapsto
        \begin{pmatrix}
        W^{-1}  (C) \\
        H^{-1} (C) 
        \end{pmatrix}, \label{eq:triple-gauging-charge-conjugation}
\end{align}
which is exactly the charge conjugation, as expected from $(ST^{-1})^3 = \mathsf{C}$.

Incidentally, the appearance of the charge conjugation can be also observed from the partition function with $\mathbb{Z}_N^{[1]}$ background field as follows.
Let $\mathcal{T}$ be a QFT with $\mathbb{Z}_{N}^{[1]}$ symmetry.
Let us compute the partition function of $((\mathcal{T}/(\mathbb{Z}_{N}^{[1]})_{-1})/(\mathbb{Z}_{N}^{[1]})_{-1})/(\mathbb{Z}_{N}^{[1]})_{-1}$ under the presence of background $B$ field. 
It is given by
\begin{align}
    &\quad \calZ_{((\mathcal{T}/(\mathbb{Z}_{N}^{[1]})_{-1})/(\mathbb{Z}_{N}^{[1]})_{-1})/(\mathbb{Z}_{N}^{[1]})_{-1}}[B] \notag\\
    &=\int \Diff b_1\Diff b_2 \Diff b_3\, \calZ_{\mathcal{T}}[b_1]\rme^{\frac{\im N}{4\pi}\int (-b_1^2-b_2^2-b_3^2+2b_1\wedge b_2+2b_2\wedge b_3+2b_3\wedge B)} \notag\\
    &=\int \Diff b_1\, \calZ_{\mathcal{T}}[b_1]\rme^{-\frac{\im N}{4\pi}\int(b_1^2-B^2)} \int \Diff b_2\, \rme^{\frac{\im N}{2\pi}\int b_2\wedge (b_1+B)} \int \Diff b_3\,\rme^{-\frac{\im N}{4\pi}\int (b_3-b_2-B)^2}\notag\\
    &\propto \int \Diff b_1\, \calZ_{\mathcal{T}}[b_1]\rme^{-\frac{\im N}{4\pi}\int(b_1^2-B^2)}\delta(b_1+B)
    =\calZ_{\mathcal{T}}[-B]. \label{eq:charge-conjugation-B}
\end{align}
By completing the square in terms of $b_3$, we can perform the $b_3$ integration explicitly, which just gives an overall constant. 
Then, the $b_2$ integration does not have the quadratic term, and thus it gives the delta-functional constraint, $b_1=-B$, and we obtain the result, which shows the sign flipping of the $\mathbb{Z}_{N}^{[1]}$ background field.
As shown in (\ref{eq:half-space-S-transformation}), the background field $B$ coupled to $\mathbb{Z}_{N}^{[1]}$ gauge field $b$ is equivalent to the background field of ``electric'' $\mathbb{Z}_{N}^{[1]}$ symmetry in terms of $ST^{-1}$-transformed gauge field $\tilde{a}_+$.
Therefore, the equality (\ref{eq:charge-conjugation-B}) suggests that the triple-gauged side theory can be regarded as the charge-conjugated ungauged one.

Second, when we discuss the fusion rule of the duality defects, we must take into account the boundary condition of the half-space gauging. 
By slightly separating three $ST^{-1}$ defects, we notice that the delta-functional constraint for $b_1$ appears on the bulk due to the $b_2$ and $b_3$ path integrals, but $b_1$ can freely fluctuate on the tiny interval $M^{(3)}\times (-\ve,\ve)$. 
As a result, for each nontrivial $2$-cycle of $M^{(3)}$, we must sum up all the possible $\mathbb{Z}_{N}^{[1]}$ generators. 
This is the basic idea behind the formula~\eqref{eq:ST-1_fusion_rule} up to the normalization. 

Lastly, the normalization (\ref{eq:fusion-rule-normalization}) can be understood as follows\footnote{We check the $X^+$ independence of $\mathcal{N}_{ \mathscr{D}^3 }(M^{(3)})$ in Appendix \ref{sec:proof_claim_fusion_rule}.
Incidentally, we note that the normalization constant becomes trivial $\mathcal{N}_{ \mathscr{D}^3 }(M^{(3)}) = 1$, if $H^* (X^+, \partial X^+; \mathbb{Z}_N)$ is torsion free and the intersection form is in the same class as those of closed spin manifolds, i.e., the form is symmetric bilinear unimodular even.}.
Apart from the counterterms $N^{-3\chi(X^+)/2} e^{\im \pi \sigma(X^+)}$, this constant consists of three factors:
\begin{align}
    &\left( \frac{|H^0 (X^+, \partial X^+; \mathbb{Z}_N)|}{|H^1 (X^+, \partial X^+; \mathbb{Z}_N)|}  \right)
    \left( \frac{|H^0 (X^+, \partial X^+; \mathbb{Z}_N)|  |H^2 (X^+, \partial X^+; \mathbb{Z}_N)|}{|H^1 (X^+, \partial X^+; \mathbb{Z}_N)|} \right) \notag \\
    & \left(\frac{|H^0 (X^+, \partial X^+; \mathbb{Z}_N)|}{|H^1 (X^+, \partial X^+; \mathbb{Z}_N)|} \sum_{b \in H^2 (X^+, \partial X^+; \mathbb{Z}_N)} \rme^{-\frac{\im N}{4 \pi} \int b \wedge b}\right). \label{eq:fusion-rule-normalization-factorizing}
\end{align}
Each factor can be seen from the third line of (\ref{eq:charge-conjugation-B}):
\begin{align}
     \int \Diff b_1\, \calZ_{\mathcal{T}}[b_1]\rme^{-\frac{\im N}{4\pi}\int(b_1^2-B^2)} \int \Diff b_2\, \rme^{\frac{\im N}{2\pi}\int b_2\wedge (b_1+B)} \int \Diff b_3\,\rme^{-\frac{\im N}{4\pi}\int (b_3-b_2-B)^2}.
\end{align}
The $b_3$ integral corresponds to the last factor of (\ref{eq:fusion-rule-normalization-factorizing}).
If $b_1 = -B$ is satisfied, the $b_2$ integral becomes just $\int \Diff b_2$, which corresponds to the second factor of (\ref{eq:fusion-rule-normalization-factorizing}).
The $b_1$ integral is constrained by $b_1 = -B$ from the $b_2$ integral. Therefore, only the prefactor of the $\mathbb{Z}_N$ gauging contributes, which is the first factor of of (\ref{eq:fusion-rule-normalization-factorizing}).
This roughly explains why the normalization factor (\ref{eq:fusion-rule-normalization}) appears.
We have given a rough explanation on the fusion rules. For the full derivations, see Appendix \ref{sec:proof_claim_fusion_rule}.
As noted above, this normalization $\mathcal{N}_{ \mathscr{D}^3 }(M^{(3)})$ includes the partition function of the $U(1)$ level-$N$ Chern-Simons theory, which guarantees that the fusion rules are consistent with the associativity \cite{Choi:2022zal}.
In addition to the $U(1)$ level-$N$ Chern-Simons theory, we remark that the factor $\mathcal{N}_{ \mathscr{D}^3 }(M^{(3)})$ also contains the gravitational Chern-Simons theory $e^{\im \pi \sigma(X^+)}$ as a consequence of the gravitational mixed anomaly (Claim \ref{claim:self-dual-stinv}).

Note that the fusion rules, e.g., $\mathscr{D} (M^{(3)}) \times \eta(\Sigma) = (-1)^{Q(\Sigma)}\mathscr{D} (M^{(3)})$, imply the non-invertibility of $\mathscr{D} (M^{(3)})$.
Indeed, $\mathscr{D} (M^{(3)}) \times (1 - (-1)^{Q(\Sigma)} \eta(\Sigma)) = 0$ cannot be satisfied by an invertible operation.
Therefore, the topological defect constructed from the self-duality is a noninvertible symmetry.

\section{Anomaly matching constraints on dynamics}
\label{sec:constraints_dynamics}

In this section, we see how the self-duality or the non-invertible symmetry constrains the dynamics of the Cardy-Rabinovici model. 
We pay attention to the $ST^{-1}$ transformation, which has the fixed point $\tau_* = \rme^{\frac{\pi \im}{3}}$.
Similar discussions on other self-dualities are presented in Appendix \ref{App:other_parameters}.

We first show that Claim~\ref{claim:self-dual-stinv} can rule out the trivially gapped phase by evaluating both sides of the equality on a $K3$ surface, so it can be used as the anomaly matching constraint. 
Next, we discuss how the conjectured phase diagram of Cardy-Rabinovici model (Fig.~\ref{fig:phase_diagram}) satisfies the constraint.

\subsection{Mixed gravitational anomaly of the \texorpdfstring{$ST^{-1}$}{ST[-1]} duality on the \texorpdfstring{$K3$}{K3} surface}
\label{sec:no_triviallygapped}

We show the main claim that the self-duality constructed in the previous section leads to a constraint on the infrared dynamics.

If the theory is trivially gapped, its low-energy theory with a background gauge field can be described by an SPT phase.
Below, we will see that any partition function of $\mathbb{Z}_N^{[1]}$ SPT phases $\calZ_{\CR}^{\tau_*} [B]$ cannot satisfy the self-duality relation of Claim \ref{claim:self-dual-stinv} on a K3 surface, which excludes the possibility of the trivially gapped phase:

\begin{claim}[Mixed gravitational anomaly matching]
\label{claim:gravitational_anomaly}
Let $X$ be a $K3$ surface. The self-duality equation of Claim \ref{claim:self-dual-stinv} with $B= 0$ cannot be saturated by the SPT phases with the $\mathbb{Z}_N^{[1]}$ symmetry. 
In particular, the Cardy-Rabinovici model at $\tau_*=\rme^{\pi\im/3}$ cannot be trivially gapped. 
\end{claim}

Here, we would like to emphasize that the gravitational factor, especially the signature dependence, in \eqref{eq:self-duality-fixed-param} is essential to find this constraint.  
On the spin $4$-manifolds $X$, the partition function of SPT states with $\mathbb{Z}_N^{[1]}$ symmetry can be characterized as~\cite{Gukov:2013zka, Kapustin:2013uxa, Kapustin:2013qsa, Kapustin:2014gua} 
\begin{equation}
    \calZ_k[B]=\exp\left(\im \frac{N k}{4\pi}\int_X B\wedge B\right),  \label{eq:sec-4-classification-SPTs}
\end{equation}
where $B$ is the $\mathbb{Z}_N$ two-form gauge field, and $k\sim k+N$ is a discrete parameter.
For completeness,                            we classify SPT states with $\mathbb{Z}_N^{[1]}$ symmetry by computing the bordism group $\Omega^{\Spin}_4(B^2 \mathbb{Z}_N)$ in Appendix \ref{App:computation_bordism}.

Let us perform the $(\mathbb{Z}_N^{[1]})_p$ gauging to this SPT partition function, 
\begin{equation}
    \int\Diff b\, \calZ_k[b]\rme^{\im \frac{Np}{4\pi}\int b\wedge b+\im\frac{N}{2\pi}\int b\wedge B}
    =\int\Diff b\, \rme^{\im \frac{N(k+p)}{4\pi}\int b\wedge b+\im\frac{N}{2\pi}\int b\wedge B}. 
\end{equation}
In order to satisfy the relation~\eqref{eq:self-duality-fixed-param} with an SPT state, this path integral with $p=-1$ over the dynamical $\mathbb{Z}_N$ two-form gauge fields $b$ must come back to the SPT state. 
This requires that 
\begin{equation}
    \gcd(N,k+p)= 1. 
\end{equation}
Otherwise, this path integral gives an intrinsic topological order with spontaneously broken $\mathbb{Z}_N^{[1]}$ symmetry. 
When $\gcd(N,k+p)=1$, we can choose $\ell\in \mathbb{Z}$ that satisfies $\ell(k+p)=1 \bmod N$. 
We then get 
\begin{align}
    \int\Diff b\, \calZ_k[b]\rme^{\im \frac{Np}{4\pi}\int b\wedge b+\im\frac{N}{2\pi}\int b\wedge B}
    &=\int\Diff b\, \rme^{\im \frac{N(k+p)}{4\pi}\int (b+\ell B)^2-\im \frac{N\ell}{4\pi}\int B\wedge B}\notag\\
    &=\Bigl(\int \Diff b\, \rme^{\im\frac{N(k+p)}{4\pi}\int b\wedge b}\Bigr)\calZ_{-\ell}[B]. 
\end{align}
We note that $B$ dependence of \eqref{eq:self-duality-fixed-param} may be satisfied if we can choose $k,\ell$ so that $\ell(k-1)=1$ and $k=-\ell \bmod N$. 
For example, we may choose $N=3$, $k=2$, and $\ell=1$. 
Therefore, the $B$ dependence of \eqref{eq:self-duality-fixed-param} is not strong enough to rule out the trivially gapped vacuum for arbitrary $N$. 
What is rather important is the overall coefficient, 
\begin{equation}
    \int \Diff b\, \exp\left(\im \frac{N(k+p)}{4\pi}\int_X b\wedge b\right)=N^{\beta_0-\beta_1}\hspace{-1.2em}\sum_{b\in H^2(X;\mathbb{Z}_N)}\hspace{-1.2em}\exp\left(\im \frac{N(k+p)}{4\pi}\int_X b\wedge b\right).
\end{equation}
If this is not identical to $N^{\frac{\chi(X)}{2}}\rme^{-\frac{\pi \im}{3}\sigma(X)}$, then the trivially gapped phase is ruled out. 
We can use the $K3$ surface to achieve this.

We can indeed show that (see  Appendix~\ref{eq:proof_claim_gravitational} for the derivation)
\begin{align}
    \int\Diff b\,\rme^{\im\frac{N (k+p)}{4\pi}\int_{K3}b\wedge b}&=\left( \operatorname{gcd}(N,k+p) \right)^{11} N^{12}.
    \label{eq:Z_K3}
\end{align}
On the other hand, 
\begin{equation}
    N^{\frac{\chi(K3)}{2}}\rme^{-\frac{\pi \im}{3}\sigma(K3)}=N^{12}\rme^{\frac{16\pi\im}{3}}=N^{12}\rme^{-\frac{2\pi\im}{3}}. 
\end{equation}
Therefore, the magnitudes are the same when $\gcd(N,k+p)= 1$, but the phase factors are different. 
This proves Claim~\ref{claim:gravitational_anomaly}.

\subsection{Dynamics of Cardy-Rabinovici model at the \texorpdfstring{$ST^{-1}$}{ST[-1]} self-dual point}

We have shown that \eqref{eq:self-duality-fixed-param} excludes the trivially gapped phase, and it can be regarded as an anomaly matching condition that involves non-invertible $ST^{-1}$ duality symmetry. 
Let us discuss how the phase diagram of Cardy-Rabinovici model satisfies this constraint. 

As shown in Fig.~\ref{fig:phase_diagram}, Cardy-Rabinovici model at $\tau_*=\rme^{\pi\im/3}$ is the intersection point of three first-order phase transition lines. 
When $\tau$ is slightly away from $\tau_*$, $\tau=\tau_*+\delta \tau$, the $ST^{-1}$ transformation cyclically exchanges three gapped phases; Higgs phase, monopole-induced confinement phase, and dyon-induced confinement phase. 
Using these building blocks, we examine how the self-duality relation \eqref{eq:self-duality-fixed-param} can be obtained. 

For this purpose, we first need to compute the partition functions of these phases, and we denote them as $\calZ_{\mathrm{Higgs}}[B]$, $\calZ_{\mathrm{mon}}[B]$, and $\calZ_{\mathrm{dyon}}[B]$, respectively. 
The Higgs phase is the $\mathbb{Z}_N$ topological order described by the level-$N$ $BF$ theory, and confinement phases are SPT states. 
Then, the natural guess for these partition functions would be
\begin{align}
    \calZ_{\mathrm{Higgs}}[B]&=\int \Diff a \Diff b\exp\left(\frac{\im N}{2\pi}\int b\wedge (\diff a + B)\right), 
    \label{eq:Z_Higgs}\\
    \calZ_{\mathrm{mon}}[B] &= 1, 
    \label{eq:Z_monopole}\\
    \calZ_{\mathrm{dyon}}[B] &= \exp\left(\frac{\im N}{4\pi}\int B\wedge B\right).
    \label{eq:Z_dyon}
\end{align}
We note that these are natural candidates to satisfy the mixed anomaly (or global inconsistency~\cite{Gaiotto:2017yup, Kikuchi:2017pcp, Tanizaki:2017bam, Karasik:2019bxn, Tanizaki:2018xto, Cordova:2019jnf,Cordova:2019uob}) between $\mathbb{Z}_{N}^{[1]}$ and $\mathsf{CP}$ at $\theta=\pi$~\cite{Honda:2020txe}. 
We show that if we set 
\begin{equation}
    \calZ^{\tau_*}_{\CR}[B]= \calZ_{\mathrm{mon}}[B] + \rme^{\frac{\pi \im}{3}\sigma(X)}\calZ_{\mathrm{dyon}}[B] + N^{-\frac{\chi(X)}{2}}\rme^{\frac{2\pi\im}{3}\sigma(X)}\calZ_{\mathrm{Higgs}}[B], 
    \label{eq:CR_Z_IR}
\end{equation}
then \eqref{eq:self-duality-fixed-param} is indeed satisfied. 

First, let us compute the partition function of the Higgs phase explicitly. 
Performing the $a$ integration, $b$ is restricted to $\mathbb{Z}_N$ two-form gauge fields, and we obtain\footnote{We determine the normalization of $\calZ_{\mathrm{Higgs}}[B]$ by the first line of this equation.}
\begin{align}
    \calZ_{\mathrm{Higgs}}[B]&=\frac{|H^0(X;\mathbb{Z}_N)|}{|H^1(X;\mathbb{Z}_N)|}\sum_{b\in H^2(X;\mathbb{Z}_N)}\rme^{\im\frac{N}{2\pi}\int_X b\wedge B} \nonumber\\
    &=N^{\beta_0-\beta_1+\beta_2}\,\delta(B) \nonumber\\
    &=N^{\chi(X)+\beta_1-\beta_0}\, \delta(B),
\end{align}
where $\delta(B)$ is the Kronecker delta in $H^2(X;\mathbb{Z}_N)$, namely, $\delta(B=0) = 1$ and $\delta(B) = 0$ for a nontrivial $B$.
Here, we again assume that $H^*(X;\mathbb{Z})$ is torsion free. 
We note that the partition function of the Higgs phase vanishes unless $B=0\in H^2(X;\mathbb{Z}_N)$. 
Physically, nonzero $B$ requires the existence of the vortex for the gauge field $a$ in the Higgs phase. Since the $BF$ theory is an effective theory regarding that the vacuum expectation value of the Higgs field is sufficiently large, the partition function with the vortex configuration is exponentially small as it costs large action density. 

Let us then perform the $(\mathbb{Z}_N^{[1]})_{-1}$ gauging to these partition functions: 
\begin{align}
    \int\Diff b\, \calZ_{\mathrm{Higgs}}[b]\, \rme^{-\im\frac{N}{4\pi}\int_X b^2+\im\frac{N}{2\pi}\int_X b\wedge B}&=N^{\chi(X)}\calZ_{\mathrm{mon}}[B], \\
    \int \Diff b\, \calZ_{\mathrm{mon}}[b]\, \rme^{-\im\frac{N}{4\pi}\int_X b^2+\im\frac{N}{2\pi}\int_X b\wedge B}&=N^{\frac{\chi(X)}{2}}\calZ_{\mathrm{dyon}}[B],\\ \int \Diff b\, \calZ_{\mathrm{dyon}}[b]\, \rme^{-\im\frac{N}{4\pi}\int_X b^2+\im\frac{N}{2\pi}\int_X b\wedge B}&=\calZ_{\mathrm{Higgs}}[B]. 
\end{align}
This is consistent with the observation that the $ST^{-1}$ transformation cyclically permutes these phases. 
Assuming \eqref{eq:CR_Z_IR}, we obtain
\begin{align}
    \calZ^{\tau_*}_{\CR/(\mathbb{Z}_N)_{-1}}[B]
    &=N^{\frac{\chi(X)}{2}}\calZ_{\mathrm{dyon}}[B]
    +\rme^{\frac{\pi \im}{3}\sigma(X)}\cdot \calZ_{\mathrm{Higgs}}[B]
    +N^{-\frac{\chi(X)}{2}}\rme^{\frac{2\pi\im}{3}\sigma(X)}\cdot N^{\chi(X)}\calZ_{\mathrm{mon}}[B]\nonumber\\
    &=N^{\frac{\chi(X)}{2}}\rme^{-\frac{\pi\im}{3}\sigma(X)}\left(\calZ_{\mathrm{mon}}[B] + \rme^{\frac{\pi \im}{3}\sigma(X)}\calZ_{\mathrm{dyon}}[B] + N^{-\frac{\chi(X)}{2}}\rme^{\frac{2\pi\im}{3}\sigma(X)}\calZ_{\mathrm{Higgs}}[B]\right)\nonumber\\
    &=N^{\frac{\chi(X)}{2}}\rme^{-\frac{\pi\im}{3}\sigma(X)} \calZ^{\tau_*}_{\CR}[B]. 
\end{align}
Here, we used the fact that $\sigma(X)\in 16\mathbb{Z}$ on the spin $4$-manifolds. 
This is nothing but the self-duality relation~\eqref{eq:self-duality-fixed-param}, and thus the conjectured phase diagram in Fig.~\ref{fig:phase_diagram} is consistent with the new anomaly matching constraint. 

Since $\tau_*=\rme^{\pi \im/3}$ has $\theta=\pi$, we can also check if the above partition function satisfies the mixed 't~Hooft anomaly between $\mathbb{Z}_{N}^{[1]}$ and $\mathsf{CP}$. 
Since $\mathsf{CP}$ flips the orientation of the spacetime, it gives 
\begin{equation}
    \mathsf{CP}: \rme^{\im \frac{Nk}{4\pi}\int B^2}\mapsto \rme^{-\im \frac{Nk}{4\pi}\int B^2},\quad \sigma(X)\mapsto -\sigma(X). 
\end{equation}
We note that $\chi(X)$ is invariant under $\mathsf{CP}$. 
Then, we can readily find that the partition function~\eqref{eq:CR_Z_IR} satisfies 
\begin{equation}
    \mathsf{CP}:\calZ_{\CR}^{\tau_*}[B]\mapsto \rme^{-\frac{\pi\im}{3}\sigma(X)}\rme^{-\im \frac{N}{4\pi}\int_{X} B\wedge B}\calZ_{\CR}^{\tau_*}[B]. 
\end{equation}
By adjusting the continuous gravitational $\theta$ term, the first factor on the right-hand side can be eliminated, and thus we may neglect it. 
The second factor reproduces the mixed anomaly between $\mathbb{Z}_{N}^{[1]}$ and $\mathsf{CP}$ at $\theta=\pi$ correctly~\cite{Honda:2020txe}, so the partition function~\eqref{eq:CR_Z_IR} also satisfies the anomaly matching of invertible symmetries.

\section{Conclusion and discussion}
\label{sec:conclusion}

In this paper, we have studied the properties of self-duality of the Cardy-Rabinovici model. 
This model does not naively realize the self-duality because of the imbalance between the electric and magnetic $1$-form symmetries. 
Since the $S$ transformation exchanges electric and magnetic charges, the dual theory acquires the magnetic $\mathbb{Z}_{N}^{[1]}$ symmetry while the original one has the electric $\mathbb{Z}_{N}^{[1]}$ symmetry. 

We consider the $\mathbb{Z}_{N}^{[1]}$-gauging with the level-$p$ discrete topological term, and then it turns out that the $(\mathbb{Z}_N^{[1]})_p$-gauged Cardy-Rabinovici model at the complex gauge coupling $\tau$ becomes dual to the original Cardy-Rabinovici model at $ST^p(\tau)$. 
We can repeat this $(\mathbb{Z}_{N}^{[1]})_p$ gauging $n$ times with $p=p_1,\ldots, p_n$, we can find the duality relation between the theories at $\tau$ and $ST^{p_n}\cdots ST^{p_1}(\tau)$. 
When the theory is on the fixed point under this duality transformation, we can obtain the codimension-$1$ topological defect by considering the half-space gauging procedure, which gives the topological self-duality defect. 
Such self-duality defects generically form the non-group-like fusion rule, and thus they can be regarded as generators of a non-invertible symmetry. 

As the simplest but nontrivial example, we have discussed the Cardy-Rabinovici model at $\tau=\tau_{*}=\rme^{\pi\im/3}$ in detail, which is the fixed point of $ST^{-1}$ transformation. 
Since we have $(ST^{-1})^3=\mathsf{C}$ at the naive level, it is natural to expect that the fusion of three $ST^{-1}$ defects can be written using the charge conjugation. 
We confirm that this is partly true, but the fusion of three $ST^{-1}$ defects does not become the single charge conjugation $\mathsf{C}$. 
Instead, it involves the summation of the $\mathbb{Z}_{N}^{[1]}$ topological defects over nontrivial $2$-cycles of the $3$-manifold that supports $ST^{-1}$ duality defects. 

What would be most interesting in our findings is the mixed gravitational anomaly. 
We find that the partition functions of dual theories do not coincide completely, and, instead, they are identical with including the specific gravitational counterterms. 
We may regard this property as the mixed gravitational anomaly of the non-invertible self-duality symmetry. 
Indeed, we have shown that the SPT state with $\mathbb{Z}_N^{[1]}$ symmetry cannot reproduce the signature dependence of the gravitational counterterm on a $K3$ surface, and thus the trivially gapped phase is ruled out from the possible ground states. 
We also show that the conjectured phase diagram of the Cardy-Rabinovici model satisfies this new anomaly-matching condition. 

It would be an interesting future study if our results can be generalized to the duality symmetry of other $4$d gauge theories. 
For example, the $S$ and $T$ transformations do not act as the self-duality for $\mathcal{N}=4$ $\mathfrak{su}(N)$ super Yang-Mills theory~\cite{Aharony:2013hda}, but they relate the theories with the appropriate gauging of $1$-form symmetries. 
Indeed, the idea of the half-space gauging has been already used to obtain the $S$-duality defect for $\mathcal{N}=4$ $SU(2)$ super Yang-Mills theory at $\tau=\im$ in Ref.~\cite{Kaidi:2021xfk}, so it would be natural to expect that we can obtain various $ST^p$ dualities for $\mathcal{N}=4$ theories from our results. 

Another interesting question is whether there exists an 't~Hooft anomaly that depends only on the non-invertible self-duality symmetry. 
In the case of pure Maxwell theory, when we consider it as the all-fermion electrodynamics, such an anomaly has been found in Refs.~\cite{Hsieh:2019iba, Hsieh:2020jpj}. 
It seems to be quite nontrivial if such anomaly is still present when the matter fields are included and the self-duality becomes a non-invertible operation.

\acknowledgments
The work of Y. T. was supported by Japan Society for the Promotion of Science (JSPS) KAKENHI Grant numbers, 22H01218 and 20K22350, and by Center for Gravitational Physics and Quantum Information (CGPQI) at Yukawa Institute for Theoretical Physics.
Y.~H. was supported by JSPS Research Fellowship for Young Scientists Grant No.~20J20215

\appendix

\section{Proofs of main claims}\label{app:proofs}

Here, we give proofs of the main claims in this paper, which are omitted to streamline discussions. 

\subsection{Proof of  Claim~\ref{claim:self-duality-1}}
\label{sec:proof_claim1}

In this section, let us prove Claim~\ref{claim:self-duality-1}. 
We shall derive (\ref{eq:self-duality-1}) with the following steps.
\begin{description}
    \item[Step 1.]  Rescaling of the gauge field $a \rightarrow a/N$.
    \item[Step 2.]  ($T^p$- and) $S$-dual transformation.
\end{description}
At the same time, we derive the transformation law of the line operators [statement \textsc{(ii)}] in these steps: (\ref{eq:line_opers_step1}) and (\ref{eq:line_opers_step2}).
In Step 2, we follow the discussion given in Ref.~\cite{Witten:1995gf} especially when we apply the $S$ transformation (see also Ref.~\cite{Verlinde:1995mz}). 

In the following discussion, we fix a UV regularization scheme, and we denote the dimension of the space of $k$-forms on $X$ as $\mathcal{B}_k$. 
For $2$-forms, we decompose $\calB_2=\calB_2^+ + \calB_2^-$, where $\calB_2^{\pm}$ describe the dimension of (anti-)self-dual $2$-forms. 
These quantities enjoy the relation
\begin{equation}
    \calB_2^\pm = \beta_2^{\pm} + (\calB_1 - \calB_0 -\beta_1 +\beta_0) = \calB_1 - \calB_0 + \frac{\chi(X) \pm \sigma(X)}{2}, 
    \label{eq:uv_euler_signature}
\end{equation}
where $\beta_k$ are Betti numbers, $\chi(X)$ is the Euler number and $\sigma(X)$ is the signature.

\subsubsection*{Step 1. Rescaling of the gauge field $a \rightarrow a/N$}

Let us begin with the description of the path integral.
The path integral over the $U(1)$ gauge field consists of 
\begin{itemize}
    \item[(1)] the sum over the $U(1)$ gauge bundles classified by the Chern class $H^2(X;\mathbb{Z})$, 
    \item[(2)] the integration over constant holonomies, and
    \item[(3)] the integration over local fluctuations for each bundle.
\end{itemize}
Accordingly, we decompose the field strength of the $U(1)$ gauge field $a$ as 
\begin{equation}
    \diff a=\diff(\delta a)+m, 
\end{equation}
where $\delta a$ is a globally-defined $1$-form and $m\in H^2(X;\mathbb{Z})$ characterizes the topological sector. 
The holonomy integration gives the restriction that the Wilson loop configurations with nontrivial net winding does not contribute to the partition function, and thus we may assume that the closed loops $C$ in the summation $\sum_{C:\,\mathrm{loops}} W^N(C)$ always have a surface $\Sigma$ with $\p \Sigma=C$. 
We can neglect the integration over constant holonomies after this restriction as it only gives an overall constant independent of the parameters such as $\tau$ and $N$. 
Then, the space of $\delta a$ has the dimension $(\calB_1-\beta_1)-(\calB_0-\beta_0)$, where $\calB_0-\beta_0$ denotes the dimension of the local gauge transformations and $\beta_1$ denotes the dimension of the space of constant holonomies.
We then write the $U(1)$-gauge path integral as
\begin{align}
    \int \mathcal{D}a~ (\cdots) = \sum_{m \in H^2 (X; \mathbb{Z})} \int_{\Omega_1^\perp (X)} \hspace{-1.5em}\mathcal{D}(\delta a) ~  (\cdots)\Bigr|_{\diff a = \diff(\delta a)+m}, \label{eq:u1-path-int}
\end{align}
where $\Omega_1^\perp (X)$ denotes a complement linear space to $\operatorname{Ker} (\diff:\{1\mbox{-}\mathrm{forms}\}\to\{2\mbox{-}\mathrm{forms}\})$. 

Next, we proceed to a computation of $\calZ_{\CR/(\mathbb{Z}_N^{[1]})_p}^{\tau} [B]$ defined in \eqref{eq:Z_N-gauging}.
From the above observation, this reads,
\begin{align}
    \calZ_{\CR/(\mathbb{Z}_N^{[1]})_p}^{\tau} [B] &= \frac{|H^0 (X; \mathbb{Z}_N)|}{|H^1 (X; \mathbb{Z}_N)|} \sum_{b \in H^2 (X; \mathbb{Z}_N)} \sum_{m \in H^2 (X; \mathbb{Z})} \rme^{\frac{\im pN}{4 \pi} \int b \wedge b+ \frac{\im N}{2 \pi} \int b \wedge B}  \notag \\
    &\,\, \times  \int_{\Omega_1^\perp (X)} \hspace{-1.5em}\mathcal{D}(\delta a)~ \rme^{- S^\tau_{U(1)}[\diff a+ b]}\sum_{C, C':\,\mathrm{loops}} \hspace{-0.8em} W^N_{\diff a +b}(C) H_{\diff a + b }(C')\Bigr|_{\diff a = \diff(\delta a)+m}.
\end{align}
We note that $b$ and $m$ always appear in the combination of $b+m$. In the integrand of $\int \Diff(\delta a)$, this is almost trivial since $b$ appears in the form of $\diff a+b$. For the rest, we have to notice that 
\begin{equation}
    \rme^{\frac{\im pN}{4 \pi} \int b \wedge b+ \frac{\im N}{2 \pi} \int b \wedge B}
    =\rme^{\frac{\im pN}{4 \pi} \int (m+b) \wedge (m+b)+ \frac{\im N}{2 \pi} \int (m+b) \wedge B}. 
\end{equation}
Since we are assuming that $H^*(X;\mathbb{Z})$ is torsion free, the sum over $b$ and $m$ can be combined into
\begin{align}
    \frac{|H^0 (X; \mathbb{Z}_N)|}{|H^1 (X; \mathbb{Z}_N)|}\sum_{b \in H^2 (X; \mathbb{Z}_N)} \sum_{m \in H^2 (X; \mathbb{Z})} (\cdots) = N^{\beta_0-\beta_1}\sum_{m \in H^2 (X; \mathbb{Z})} (\cdots) \Bigr|_{(m + b) \rightarrow \frac{1}{N} m}, 
\end{align}
and we obtain
\begin{align}
    \calZ_{\CR/(\mathbb{Z}_N^{[1]})_p}^{\tau} [B] &= N^{\beta_0-\beta_1} \sum_{m \in H^2 (X; \mathbb{Z})} \rme^{\frac{\im p}{4 \pi N} \int m\wedge  m+\frac{\im}{2 \pi} \int  m \wedge B}  \notag \\
    &\,\, \times  \int_{\Omega_1^\perp (X)} \hspace{-1.5em}\mathcal{D}(\delta a)~ \rme^{- S^\tau_{U(1)}[\diff(\delta a)+\frac{1}{N}m]}\sum_{C, C':\,\mathrm{loops}} \hspace{-0.8em} W^N_{\diff(\delta a)+\frac{1}{N}m}(C) H_{\diff(\delta a)+\frac{1}{N}m}(C').
\end{align}
Performing the rescaling of the local fluctuation as $\delta a \rightarrow \delta a /N$, the path-integral measure $\int \Diff(\delta a)$ gives the overall factor $1/{N^{\calB_1-\calB_0-\beta_1+\beta_0}}$, and we have
\begin{align}
    \calZ_{\CR/(\mathbb{Z}_N^{[1]})_p}^{\tau} [B] 
    &= \frac{1}{N^{\calB_1-\calB_0}} 
    \sum_{m \in H^2 (X; \mathbb{Z})} \rme^{\frac{\im p}{4 \pi N} \int m\wedge  m+\frac{\im}{2 \pi} \int  m \wedge B}  \notag \\
    &\qquad\times  \int_{\Omega_1^\perp (X)} \hspace{-1.5em}\mathcal{D}(\delta a)~ \rme^{- S^{\tau/N^2}_{U(1)}[\diff (\delta a)+ m]} 
    \sum_{C, C':\,\mathrm{loops}} \hspace{-0.8em} W_{(\diff (\delta a)+m)}(C) H^N_{\left( \diff(\delta a) + m\right)}(C') \notag \\
    &= \frac{1}{N^{\calB_1-\calB_0}} \int \mathcal{D}a ~ \rme^{- S^{(\tau+p)/N^2}_{U(1)}[\diff a]+\frac{\im}{2 \pi} \int  \diff a\wedge B}  
    \sum_{C, C':\,\mathrm{loops}} \hspace{-0.8em} W_{\diff a}(C) (H_{\diff a}^N (C')W_{\diff a}^{-p}(C'))\notag\\
    &= \frac{1}{N^{\calB_1-\calB_0}} \int \mathcal{D}a ~ \rme^{- S^{(\tau+p)/N^2}_{U(1)}[\diff a]+\frac{\im}{2 \pi} \int  \diff a\wedge B}  
    \sum_{C, C':\,\mathrm{loops}} \hspace{-0.8em} W_{\diff a}(C) H_{\diff a}^N (C'). 
    \label{eq:step1}
\end{align}
Here, we have used (\ref{eq:u1-path-int}) and $H_{f/N}(C') = H_{f}^N (C')$.\footnote{Since $H_{f/N}(C')$ represents a defect imposing
\begin{align}
    \int_{M^{(2)}} f/N = 2 \pi \operatorname{Link}(C, M^{(2)}) \Leftrightarrow \int_{M^{(2)}} f = 2 \pi N \operatorname{Link}(C, M^{(2)}),
\end{align}
it is equivalent to $H_{f}^N (C')$.} We have also used the fact that the local part $\diff (\delta a)$ does not contribute to $\rme^{\frac{\im p}{4 \pi N} \int \diff a \wedge  \diff a +\frac{\im}{2 \pi} \int  \diff a \wedge B}$, and its first term in the exponent can be combined with the continuous $\theta$ angle as $\tau+p$. 
Along with this change, we note that $H^N(C')$ should be replaced by $H^N(C')W^{-p}(C')$ due to the Witten effect:
\begin{align}
    H^N(C') \mapsto H^N(C')W^{-p}(C') \label{eq:App-W-effect-HN}
\end{align}
Indeed, since the 't Hooft operator inserts a magnetic defect [$da \rightarrow da + 2 \pi \delta (\Sigma)$ with $\partial \Sigma = C$], the integral $\int \diff a \wedge \diff a$ is not simply $\int m \wedge m$ in the presence of the 't Hooft operator. More explicitly, we get\footnote{Strictly speaking, there is an extra factor because $C$ can have some disconnected loops (Recall that ``the sum over all possible loops'' arises from the sum over worldline configurations $\{ n_\mu \}$):
\begin{align}
    \rme^{ \frac{\im p}{4 \pi N}  \int \diff a \wedge \diff a}H_{\diff a}(C) = \rme^{ \frac{\im p}{4 \pi N}  \int m \wedge m + \frac{\im p }{N} \int_C a + \frac{\im \pi p}{N} \operatorname{Int}(\Sigma, \Sigma)} H_{\diff a}(C),
\end{align}
where $\operatorname{Int}(\Sigma, \Sigma) = \int \delta(\Sigma) \wedge \delta(\Sigma)$ denotes the intersection number of $\Sigma$, which can be non-zero when $C$ contains multiple loops.
If $C$ contains multiple loops, say $C = C_1 + C_2$ with $\partial \Sigma_1 = C_1$ and $\partial \Sigma_2 = C_2$, then the intersection number can be a nonvanishing even integer $\operatorname{Int}(\Sigma, \Sigma) = 2 \operatorname{Int}(\Sigma_1, \Sigma_2)$.
Note that, for $H^N(C')$, the extra factor becomes $\rme^{\im \pi p \operatorname{Int}(\Sigma, \Sigma)} = 1$, and the conclusion (\ref{eq:App-W-effect-HN}) is still valid.
Note also that the transformation law (\ref{eq:line_opers_step1}) is understood for a line operator on a single loop.
}
\begin{align}
    \rme^{ \frac{\im p}{4 \pi N}  \int \diff a \wedge \diff a}H_{\diff a}(C) = \rme^{ \frac{\im p}{4 \pi N}  \int m \wedge m + \frac{\im p }{N} \int_C a} H_{\diff a}(C).
\end{align}
Therefore, the 't Hooft loop $H^N(C')$ acquires the electric line $W^{-p}(C')$ in the second equality of (\ref{eq:step1}).
This completes the first step.

For the later use, we summarize how the line operators are transformed in this step:
    \begin{align}
        \begin{pmatrix}
        H_{N(\diff a+ b)}(C')\\
        W_{N(\diff a+ b)}(C)
        \end{pmatrix}\mapsto 
        \begin{pmatrix}
        H_{\diff a}(C') W^{-p/N}_{\diff a}(C')\\
        W_{\diff a}(C)
        \end{pmatrix}. \label{eq:line_opers_step1}
    \end{align}
Incidentally, this relation is consistent with the fact that the genuine lines of the level-$p$ $\mathbb{Z}_N^{[1]}$-gauged model are $W_{N(\diff a+ b)}(C)$ and $H_{N(\diff a+ b)}(C') W^{p/N}_{N(\diff a+ b)}(C')$.

\subsubsection*{Step 2.  ($T^p$- and) $S$-dual transformation.}

When we obtain \eqref{eq:step1}, we include the discrete $\theta$ parameter $p$ into the continuous $\theta$ parameter. This means the discrete $\theta$-term $\rme^{\frac{\im pN}{4 \pi} \int b \wedge b}$ can be identified with the $T^p$ transformation: $\theta \rightarrow \theta + 2 \pi p$, or $\tau\to \tau+p$.
We then implement the $S$-dual transformation following Ref.~\cite{Witten:1995gf}.
We first describe the $S$-dual transformation by the $BF$ coupling up to a constant, and then determine its overall constant by the computation in the free Maxwell theory.

We consider the following path integral that includes another $U(1)$ gauge field $\widetilde{a}$ and $\mathbb{R}$-valued $2$-form field $\widetilde{h}$: 
\begin{align}
    \int \mathcal{D}\widetilde{a} \mathcal{D} \widetilde{h} \mathcal{D}a ~\rme^{\frac{\im}{2 \pi} \int \diff \widetilde{a}\wedge \widetilde{h}+\frac{\im}{2 \pi} \int  (\diff a+\widetilde{h}) \wedge B}\rme^{- S^{(\tau + p)/N^2}_{U(1)}[\diff a+\widetilde{h}]}  \sum_{C, C':\,\mathrm{loops}} \hspace{-0.8em} W_{\diff a+\widetilde{h}}(C) H_{\diff a+\widetilde{h}}^N (C'). 
    \label{eq:S-duality-BF}
\end{align}
Here, we require that the $2$-form field $\widetilde{h}$ includes the magnetic defects caused by the 't Hooft loop $H_{\diff a+\widetilde{h}}^N (C')$ so that $a$ is the $U(1)$ gauge field without defects.

We can see that this quantity is proportional to \eqref{eq:step1} by performing the path integral of $\widetilde{a}$ first. 
The integration over the additional gauge field $\widetilde{a}$ imposes $\diff \widetilde{h} = 0$ and also the quantization condition $\int_{M^{(2)}} \frac{\widetilde{h}}{2 \pi} \in \mathbb{Z}$, which guarantees that $W_{\diff a+\widetilde{h}}(C) $ is a genuine line.
Therefore, it is reduced to 
\begin{align}
    \int \mathcal{D}\widetilde{a}\mathcal{D} \widetilde{h} \mathcal{D}a&~\rme^{\frac{\im}{2 \pi} \int \diff\widetilde{a}\wedge \widetilde{h} + \frac{\im}{2 \pi} \int  (\diff a+\widetilde{h}) \wedge B}\rme^{- S^{(\tau + p)/N^2}_{U(1)}[\diff a+\widetilde{h}]}  \sum_{C, C':\,\mathrm{loops}} \hspace{-0.8em} W_{\diff a+\widetilde{h}}(C) H_{\diff a+\widetilde{h}}^N (C') \notag \\
    &= \mathcal{N}_1 \int \mathcal{D}a ~ \rme^{\frac{\im}{2 \pi} \int  \diff a \wedge B} \rme^{- S^{(\tau + p)/N^2}_{U(1)}[\diff a]}  \sum_{C, C':\,\mathrm{loops}} \hspace{-0.8em} W_{\diff a}(C) H_{\diff a}^N (C') , 
    \label{eq:Sdual_derivation_1}
\end{align}
and this is identical to the path integral that appears in \eqref{eq:step1} up to the overall constant. 
Note that the constant $\mathcal{N}_1$ appearing here has nothing to do with the matter part\footnote{Roughly, $\mathcal{N}_1$ can be expressed as
$\int \mathcal{D}\widetilde{a} \mathcal{D} \widetilde{h}~\rme^{\frac{\im}{2 \pi} \int \diff \widetilde{a}\wedge \widetilde{h}} (\cdots) = \mathcal{N}_1 \frac{1}{|H^2(X;\mathbb{Z})|} \sum_{\widetilde{h} \in H^2(X;\mathbb{Z})}(\cdots).
$}.

On the other hand, we can use the $1$-form gauge redundancy to gauge away $\diff a$ from (\ref{eq:S-duality-BF}):
\begin{align}
    \int& \mathcal{D}\widetilde{a} \mathcal{D} \widetilde{h} \mathcal{D}a ~\rme^{\frac{\im}{2 \pi} \int \diff \widetilde{a}\wedge \widetilde{h} + \frac{\im}{2 \pi} \int  (\diff a+\widetilde{h}) \wedge B} \rme^{- S^{(\tau + p)/N^2}_{U(1)}[\diff a+\widetilde{h}]}  \sum_{C, C':\,\mathrm{loops}} \hspace{-0.8em} W_{\diff a+\widetilde{h}}(C) H_{\diff a+\widetilde{h}}^N (C') \notag \\
    &= \left( \int \mathcal{D}a \right) \int \mathcal{D}\Tilde{a}\mathcal{D} \widetilde{h} ~ \rme^{\frac{\im}{2 \pi} \int (\diff \widetilde{a}+B)\wedge \widetilde{h}}\rme^{- S^{(\tau + p)/N^2}_{U(1)}[\widetilde{h}]}  \sum_{C, C':\,\mathrm{loops}} \hspace{-0.8em} W_{\widetilde{h}}(C) H_{\widetilde{h}}^N (C').
    \label{eq:before_b_integral}
\end{align}
In the derivation, we used\footnote{The $U(1)$ gauge field $a$ has no magnetic defects in this setup, where the 2-form field $\widetilde{h}$ is in charge of the magnetic defects from $ H_{\diff a+\widetilde{h}}^N (C')$. This guarantees $\int \diff \widetilde{a} \wedge \diff a \in 4 \pi^2 \mathbb{Z}$.} $\rme^{\frac{\im}{2 \pi} \int \diff \widetilde{a} \wedge \diff a}=1$. 
We note that $\diff \widetilde{a}$ and $B$ always appears in the combination of $\diff \widetilde{a}+B$ in this expression. 
Next, we would like to perform the path integral of $\widetilde{h}$ explicitly, and we need to rewrite the loop operators for this purpose. 

We show that the Wilson and 't Hooft loops for $\widetilde{h}$, $W_{\widetilde{h}}(C)$ and $H_{\widetilde{h}}(C',\Sigma')$, can be represented as the 't~Hooft and Wilson loops of $\widetilde{a}$, respectively: 
\begin{align}
    & W_{\widetilde{h}}(C) = H_{\diff \widetilde{a}+B} (C),
    \label{eq:claim-1-wilson-loop}\\
    & H_{\widetilde{h}}(C',\Sigma') = W_{\diff \widetilde{a}+B}^{-1}(C',\Sigma') 
    = \rme^{\im \int_{\Sigma':\partial \Sigma' = C'} (\diff\widetilde{a} + B)}. 
    \label{eq:claim-1-thooft-loop}
\end{align}
In order to see \eqref{eq:claim-1-wilson-loop}, we express the left-hand side as $W_{\widetilde{h}}(C)=\exp(\im \int_\Sigma \widetilde{h})=\exp(\im\int \delta(\Sigma)\wedge \widetilde{h})$ with $\p\Sigma=C$, and thus 
\begin{equation}
    \rme^{\frac{\im}{2\pi}\int (\diff \widetilde{a}+B)\wedge \widetilde{h}} W_{\widetilde{h}}(C)
    =\rme^{\frac{\im}{2\pi}\int (\diff \widetilde{a}+B+2\pi \delta(\Sigma))\wedge \widetilde{h}}. 
\end{equation}
We may absorb $2\pi\delta(\Sigma)$ into the $U(1)$ gauge field $\widetilde{a}$ so that $\diff \widetilde{a}'=\diff \widetilde{a}+2\pi \delta(\Sigma)$, but then the $U(1)$ gauge field $\widetilde{a}'$ acquires the monopole singularity, 
\begin{equation}
    \int_{M_2} (\diff \widetilde{a}'+B)=  2\pi\, \mathrm{Link}(C, M_2). 
\end{equation}
This proves \eqref{eq:claim-1-wilson-loop}. 
To see \eqref{eq:claim-1-thooft-loop}, we consider the equation of motion of $\widetilde{a}$ under the presence of $W_{\diff \widetilde{a}+B}(C',\Sigma')$, then 
\begin{equation}
    \diff \widetilde{h}= - 2\pi \delta(C'). 
\end{equation}
This means that $\widetilde{h}$ can be understood as the $U(1)$ gauge field with the monopole singularity on $C'$ of magnetic charge $(-1)$, and we find \eqref{eq:claim-1-thooft-loop}. 
Using this correspondence, we can rewrite the loop operators in \eqref{eq:before_b_integral} as 
\begin{align}
    \sum_{C,C':\,\mathrm{loops}} \hspace{-0.8em} W_{\widetilde{h}}(C) H_{\widetilde{h}}^N (C') 
    &= \sum_{C,C':\,\mathrm{loops}} \hspace{-0.8em} H_{\diff \widetilde{a}+B}(C) W^{-N}_{\diff \widetilde{a}+B} (C')
     \notag \\
    &= \sum_{C,C':\,\mathrm{loops}} \hspace{-0.8em} H_{\diff \widetilde{a}+B}(C) W^N_{\diff \widetilde{a}+B}(C'),\label{eq:b_integration_loops_Sdual}
\end{align}
where we have used $\sum_{C'} W^{-N} (C') = \sum_{C'} W^{N} ({C'}^{-1}) = \sum_{C} W^{N} ({C'}) $ in the last line.
Then, the path integral of $\widetilde{h}$ becomes a Gaussian integral.

We can now integrate out $\widetilde{h}$:
\begin{align}
    \int \mathcal{D} \widetilde{h} &~ \rme^{\frac{\im}{2 \pi} \int (\diff \widetilde{a} + B) \wedge \widetilde{h}} \rme^{- S^{(\tau + p)/N^2}_{U(1)}[\widetilde{h}]} \notag \\
    &= \left( \int \mathcal{D}\widetilde{h}^+ e^{- \frac{1}{4 \pi} \int \widetilde{h}^+ \wedge \widetilde{h}^+} \right) 
    \left( \int \mathcal{D}\widetilde{h}^- e^{- \frac{1}{4 \pi} \int \widetilde{h}^- \wedge \widetilde{h}^-} \right) \notag \\
    &\times \left( \frac{-\im(\tau + p)}{N} \right)^{- \frac{\calB_2^+}{2}} \left( \frac{\im(\Bar{\tau} + p)}{N} \right)^{- \frac{\calB_2^-}{2}} e^{- S^{-(\tau + p)^{-1}}_{U(1)}[\diff \widetilde{a} + B]},
    \label{eq:b_integration_Sdual}
\end{align}
where $\widetilde{h}$ is decomposed into the self-dual part $\widetilde{h}^+$ and the anti-self-dual part $b^-$ with the rescaling.
Combining \eqref{eq:before_b_integral}, \eqref{eq:b_integration_loops_Sdual}, and \eqref{eq:b_integration_Sdual}, we obtain 
\begin{align}
    &\quad \int \mathcal{D}\widetilde{a} \mathcal{D} \widetilde{h} \mathcal{D}a ~\rme^{\frac{\im}{2 \pi} \int \diff \widetilde{a}\wedge \widetilde{h} + \frac{\im}{2 \pi} \int  (\diff a+\widetilde{h}) \wedge B} \rme^{- S^{(\tau + p)/N^2}_{U(1)}[\diff a+\widetilde{h}]}  \sum_{C, C':\,\mathrm{loops}} \hspace{-0.8em} W_{\diff a+\widetilde{h}}(C) H_{\diff a+\widetilde{h}}^N (C') \notag \\
    &= \mathcal{N}_2 \left( \frac{\tau + p}{N} \right)^{- \frac{\calB_2^+}{2}} \left( \frac{\Bar{\tau} + p}{N} \right)^{- \frac{\calB_2^-}{2}} \int \mathcal{D}\widetilde{a}~e^{- S^{-(\tau + p)^{-1}}_{U(1)}[\diff \widetilde{a} + B]} \sum_{C, C':\,\mathrm{loops}} \hspace{-0.8em} W^N_{\diff \widetilde{a} + B}(C') H_{\diff \widetilde{a} + B}(C),
    \label{eq:Sdual_derivation_2}
\end{align}
where $\mathcal{N}_2$ is a constant that is independent of $N$, the coupling $\tau$, and the matter sector.
Using \eqref{eq:Sdual_derivation_1} and \eqref{eq:Sdual_derivation_2}, we find the S-dual relation, 
\begin{align}
    & \quad \int \mathcal{D}a ~ \rme^{\frac{\im}{2 \pi} \int  \diff a \wedge B} \rme^{- S^{(\tau + p)/N^2}_{U(1)}[\diff a]}  \sum_{C, C':\,\mathrm{loops}} \hspace{-0.8em} W_{\diff a}(C) H_{\diff a}^N (C')  \notag \\
    &= \frac{\mathcal{N}_2}{\mathcal{N}_1} \left( \frac{\tau + p}{N} \right)^{- \frac{\calB_2^+}{2}} \left( \frac{\Bar{\tau} + p}{N} \right)^{- \frac{\calB_2^-}{2}} \int \mathcal{D}\widetilde{a}~e^{- S^{-(\tau + p)^{-1}}_{U(1)}[\diff \widetilde{a} + B]} \sum_{C, C':\,\mathrm{loops}} \hspace{-0.8em} W^N_{\diff \widetilde{a} + B}(C') H_{\diff \widetilde{a} + B}(C). \label{eq:S-dual-procedure}
\end{align}
Lastly, we must determine the constant $\mathcal{N}_2/\mathcal{N}_1$. Since both $\mathcal{N}_1$ and $\mathcal{N}_2$ are independent of the matter sector, we can compute its ratio using the free Maxwell theory, 
\begin{align}
    \int \mathcal{D}a ~ \rme^{- S^{(\tau + p)/N^2}_{U(1)}[\diff a]}  &= \frac{\mathcal{N}_2}{\mathcal{N}_1} \left( \frac{\tau + p}{N} \right)^{- \frac{\calB_2^+}{2}}  \left( \frac{\Bar{\tau} + p}{N} \right)^{- \frac{\calB_2^-}{2}}  \int \mathcal{D}\widetilde{a}~e^{- S^{-(\tau + p)^{-1}}_{U(1)}[\diff \widetilde{a}]}.
\end{align}
This can be done by an explicit computation~\cite{Witten:1995gf}.
Indeed,
\begin{align}
    \int \mathcal{D}a ~ \rme^{- S^{(\tau + p)/N^2}_{U(1)}[\diff a]}  &= \sum_{m \in H^2 (X; \mathbb{Z})} \hspace{-1.1em}\rme^{-\frac{1}{2g^2 N^2} \int m \wedge * m + \frac{\im (\theta + 2 \pi p)}{8 \pi^2 N} \int m \wedge m}  
    \int_{\Omega_1^\perp (X)} \hspace{-1.5em}\mathcal{D}(\delta a) \rme^{-\frac{1}{2g^2 N^2} \int \diff (\delta a) \wedge * \diff (\delta a)} \notag \\
    &= \sum_{m \in H^2 (X; \mathbb{Z})} \left( \rme^{2 \pi \im \frac{\tau + p}{N}} \right)^{\frac{1}{8 \pi^2}\int m \wedge \left( \frac{1+*}{2} \right) m} \left( \rme^{2 \pi \im \frac{\Bar{\tau} + p}{N}} \right)^{\frac{1}{8 \pi^2}\int m \wedge \left( \frac{1-*}{2} \right) m}   \notag \\
    &\times \left(\operatorname{Im} \frac{\tau + p}{N} \right)^{-\frac{\calB_1 - \calB_0 - \beta_1 +\beta_0}{2}} 
    \int_{\Omega_1^\perp (X)} \hspace{-1.5em} \mathcal{D}(\delta a) \rme^{-\frac{1}{4 \pi} \int \diff (\delta a) \wedge * \diff (\delta a)}. \label{eq:computation_free_Maxwell}
\end{align}
The Jacobi identity for the theta function guarantees that the sum over topological sectors, $m\in H^2(X;\mathbb{Z})$, gives a modular form of weight $(\beta_2^+/2, \beta_2^-/2)$ in terms of $\frac{\tau + p}{N}$. Therefore, the partition function (\ref{eq:computation_free_Maxwell}) is a modular form of weight
\begin{align}
    \left(\frac{\beta_2^+ + \calB_1 - \calB_0 - \beta_1 + \beta_0}{2} , \frac{\beta_2^- + \calB_1 - \calB_0 - \beta_1 +\beta_0}{2} \right) = \left(\frac{\calB_2^+}{2} , \frac{\calB_2^-}{2}\right),
\end{align}
and we find $\mathcal{N}_2/\mathcal{N}_1 = 1$.

Using (\ref{eq:step1}) and (\ref{eq:S-dual-procedure}) with $\mathcal{N}_2/\mathcal{N}_1 = 1$, we obtain 
\begin{align}
    \calZ_{\CR/(\mathbb{Z}_N^{[1]})_p}^{\tau} [B] 
    &= \frac{1}{N^{\calB_1-\calB_0}} \int \mathcal{D}a ~ \rme^{- S^{(\tau+p)/N^2}_{U(1)}[\diff a]+\frac{\im}{2 \pi} \int  \diff a\wedge B}  
    \sum_{C, C':\,\mathrm{loops}} \hspace{-0.8em} W_{\diff a}(C) H_{\diff a}^N (C') \notag\\
    &= \frac{N^{\frac{\calB_2^+ + \calB_2^-}{2}-\calB_1+\calB_0} }{(\tau + p)^{\frac{\calB_2^+}{2}} (\Bar{\tau} + p)^{\frac{\calB_2^-}{2}}}
    \int \mathcal{D}\widetilde{a}~e^{- S^{-(\tau + p)^{-1}}_{U(1)}[\diff \widetilde{a} + B]} \sum_{C, C':\,\mathrm{loops}} \hspace{-0.8em} W^N_{\diff \widetilde{a} + B}(C') H_{\diff \widetilde{a} + B}(C)\notag\\
    &=\frac{N^{\chi(X)/2}}{\left( \tau+p\right)^{\calB_{2}^+/2} \left( \overline{\tau}+p \right)^{\calB_{2}^-/2}}\, \calZ_{\CR}^{-(\tau+p)^{-1}} [B]. 
\end{align}
This is almost identical with \eqref{eq:self-duality-1}, but the coefficient is UV divergent. 
To fix this, we introduce the following UV counterterm to the Maxwell action, 
\begin{equation}
    S^\tau_{U(1)}[\diff a]\to S^\tau_{U(1)}[\diff a]-\frac{(\calB_1-\calB_0)}{2} \ln(\mathrm{Im}(\tau)), 
\end{equation}
and then the above duality relation becomes 
\begin{align}
    \calZ_{\CR/(\mathbb{Z}_N^{[1]})_p}^{\tau} [B]
    &=\frac{\left(\mathrm{Im}(\tau)\right)^{(\calB_1-\calB_0)/2}}{\left(\mathrm{Im}(-(\tau+p)^{-1})\right)^{(\calB_1-\calB_0)/2}}\frac{N^{\chi(X)/2}}{\left( \tau+p\right)^{\calB_{2}^+/2} \left( \overline{\tau}+p \right)^{\calB_{2}^-/2}}\, \calZ_{\CR}^{-(\tau+p)^{-1}} [B] \notag\\
    &=\frac{N^{\chi(X)/2}}{\left( \tau+p\right)^{\frac{\chi(X)+\sigma(X)}{4} } \left( \overline{\tau}+p \right)^{\frac{\chi(X)-\sigma(X)}{4} } } \, \calZ_{\CR}^{-(\tau+p)^{-1}} [B].
\end{align}
Here, we have used \eqref{eq:uv_euler_signature} to obtain the last expression, and this gives \eqref{eq:self-duality-1} of Claim~\ref{claim:self-duality-1}. 
We note that the UV counterterms are identical on both sides at the self-dual point, $\tau=\tau_*$ and $p=-1$, and thus we can obtain Claim~\ref{claim:self-dual-stinv} whether or not including the above UV counterterm. 

In this step, the line operators are transformed as, from \eqref{eq:claim-1-wilson-loop} and \eqref{eq:claim-1-thooft-loop},
    \begin{align}
        \begin{pmatrix}
        H_{\diff a}(C') \\
        W_{\diff a}(C)
        \end{pmatrix}\mapsto 
        \begin{pmatrix}
        H_{\widetilde{h}}(C') \\
        W_{\widetilde{h}}(C)        
        \end{pmatrix}\mapsto 
        \begin{pmatrix}
        W^{-1}_{\diff \widetilde{a} + B}(C') \\
        H_{\diff \widetilde{a} + B}(C)        
        \end{pmatrix}\label{eq:line_opers_step2}
    \end{align}
By combining (\ref{eq:line_opers_step1}) and (\ref{eq:line_opers_step2}), we finally obtain the statement~\textsc{(ii)} of Claim~\ref{claim:self-duality-1}.
This completes the proof.

\subsection{Supplements for the proof of Claim~\ref{claim:gravitational_anomaly}}
\label{eq:proof_claim_gravitational}

In this appendix, let us show \eqref{eq:Z_K3} including the dependence on the background gauge field $B$: 
\begin{align}
&\quad \frac{|H^0 (K3; \mathbb{Z}_N)|}{|H^1 (K3; \mathbb{Z}_N)|} \sum_{b \in H^2 (K3; \mathbb{Z}_N)} \hspace{-1.5em} \rme^{\frac{\im N k}{4 \pi} \int_{K3} b \wedge b+\im\frac{N}{2\pi}\int_{K3}b\wedge B} \notag\\
&= \left( \operatorname{gcd}(N,k) \right)^{11} N^{12}\, \exp\left(-\im\frac{\gcd(N,k)\ell}{4\pi N}\int_{K3}\left(\frac{N}{\gcd(N,k)}B\right)^2\right)\notag\\
&\qquad \times \delta\left(\frac{N}{\gcd(N,k)}B\in 2\pi H^2(K3;\mathbb{Z})\right), 
\label{eq:SPT_K3}
\end{align}
where $\ell k=\gcd(N,k) \bmod N$. 
We note that such $\ell$ can be uniquely determined in modulo $\frac{N}{\gcd(N,k)}$, so the right-hand side is well-defined. 
As discussed in Sec.~\ref{sec:no_triviallygapped}, it is sufficient to show this equality for $B=0$ to prove Claim~\ref{claim:gravitational_anomaly}. 
However, the $B$ dependence becomes important for studying other self-dual parameters. 

Let us first show that the path integral in \eqref{eq:SPT_K3} vanishes unless 
\begin{equation}
    \exp\left(\im \frac{N}{\gcd(N,k)}\int_{M_2}B\right)=1
\end{equation} 
for any $M_2\subset X$. For simplicity of notation, let us set $d=\gcd(N,k)$. In order to see this, we consider the following trick of the path integral:
\begin{align}
    \int \Diff b\, \rme^{\frac{\im N k}{4\pi}\int_{X}b\wedge b+\frac{\im N}{2\pi}\int_{X}b\wedge B}
    &\propto \int \Diff b \Diff b'\,\rme^{\frac{\im N k}{4\pi}\int_{X}\left(b+\frac{N}{d}b'\right)^2+\frac{\im N}{2\pi}\int_{X}\left(b+\frac{N}{d}b'\right)\wedge B}.
\end{align}
This is the trivial identity since we can eliminate $b'$ on the right-hand-side by the redefinition of $b$. 
We then note that 
\begin{align}
    \rme^{\frac{\im N k}{4\pi}\int\left(b+\frac{N}{d}b'\right)^2}
    &=\rme^{\frac{\im N k}{4\pi}\int b^2+\frac{k}{d}\frac{\im N^2 }{2\pi}\int b\wedge b'+\frac{Nk}{d^2}\frac{\im N^2}{4\pi}\int {b'}^2}\notag\\
    &=\rme^{\frac{\im N k}{4\pi}\int b^2}. 
\end{align}
As a result, we find 
\begin{align}
    \int \Diff b\, \rme^{\frac{\im N k}{4\pi}\int_{X}b\wedge b+\frac{\im N}{2\pi}\int_{X}b\wedge B}
    &\propto \int \Diff b \Diff b'\,\rme^{\frac{\im N k}{4\pi}\int_{X} b^2+\frac{\im N}{2\pi}\int_{X}\left(b+\frac{N}{d}b'\right)\wedge B}.
\end{align}
Then, the $b'$ path integral gives the delta functional constraint on $\frac{N}{d}B$. 

Due to this constraint, we can set $B=d\tilde{B}$ with another $\mathbb{Z}_N$ two-form gauge field $\tilde{B}$. 
We then find that 
\begin{align}
    \int \Diff b\, \rme^{\frac{\im N k}{4\pi}\int_{X}b\wedge b+\frac{\im N}{2\pi}\int_{X}b\wedge B}
    &=\int \Diff b\, \rme^{\frac{\im N k}{4\pi}\int_{X}b\wedge b+\frac{\im N d}{2\pi}\int_{X}b\wedge \tilde{B}} \notag\\
    &=\int \Diff b\, \rme^{\frac{\im N k}{4\pi}\int_{X}(b+\ell \tilde{B})^2-\frac{\im N k \ell^2}{4\pi}\int_{X}\tilde{B}^2} \notag\\
    &=\int \Diff b\, \rme^{\frac{\im N k}{4\pi}\int_{X}b^2-\frac{\im N d \ell}{4\pi}\int_{X}\tilde{B}^2}. 
\end{align}
We then obtain the $B$ dependence of \eqref{eq:SPT_K3}. Therefore, we may set $B=0$ in the following to determine the overall coefficient. 

As a preparation, we note the following properties (see Sec.~3 of Ref.\cite{scorpan2005wild}) of a K3 surface.
The cohomology group $H^*(K3;\mathbb{Z})$ is torsion free, and it has $\beta_0=\beta_4=1$, $\beta_1=\beta_3=0$, $\beta_2=22=\beta_2^+ + \beta_2^-$ with $\beta_2^{+}=3$, $\beta_2^{-}=19$. Thus, $\chi(K3)=24$ and $\sigma(K3)=-16$. 
By writing $\vec{n}_1,\vec{n}_2\in H^2(X;\mathbb{Z})\simeq \mathbb{Z}^{22}$ as integer-valued vectors, the intersection form can be expressed as
\begin{align}
    &\langle\vec{n}_1, \vec{n}_2\rangle = \vec{n}_1 \cdot Q_{K3}\cdot \vec{n}_2, \notag \\
    &Q_{K3} = U \oplus U \oplus U \oplus (- E_8) \oplus (- E_8),
\end{align}
    where $U$ is the hyperbolic matrix and $E_8$ is the $E_8$ matrix:
\begin{align}
    U = \begin{pmatrix}
0 & 1 \\
1 & 0 \\
\end{pmatrix},
~~~
E_8 = 
\begin{pmatrix}
2 & 1 & 0 & 0 & 0 & 0 & 0 & 0 \\
1 & 2 & 1 & 0 & 0 & 0 & 0 & 0\\
0 & 1 & 2 & 1 & 0 & 0 & 0 & 0\\
0 & 0 & 1 & 2 & 1 & 0 & 0 & 0\\
0 & 0 & 0 & 1 & 2 & 1 & 0 & 1\\
0 & 0 & 0 & 0 & 1 & 2 & 1 & 0\\
0 & 0 & 0 & 0 & 0 & 1 & 2 & 0\\
0 & 0 & 0 & 0 & 1 & 0 & 0 & 2\\
\end{pmatrix}. 
\end{align}
We later use $E_8 \oplus (-1) \simeq (-1) \oplus 8 (+1)$, where $8(+1)$ denotes the $8 \times 8$ identity matrix.\footnote{This stems from the following construction of $E_8$ matrix.
Let us start with a 9-dimensional space with a form $Q = (-1) \oplus 8 (+1)$. We write its basis as $\{ e_0, e_1, \cdots, e_8\}$. A vector $\kappa = 3 e_0 + e_1 + \cdots + e_8 $ has a negative intersection $\kappa^T Q \kappa = -1$. Its $Q$-orthogonal complement space has a basis $\{ e_1', \cdots, e_8'\}$: $e_i' = e_i - e_{i+1}$ for $i = 1, \cdots, 7$ and $e_8' = e_0+e_6+e_7+ e_8$. 
In terms of this basis, the intersection form reduces to the $E_8$ matrix: $E_8 \oplus (-1) \simeq (-1) \oplus 8 (+1)$.
Notice that the change-of-basis matrix from $\{ e_0, e_1, \cdots, e_8\}$ to $\{ \kappa, e_1', \cdots, e_8'\}$ is in $SL(9,\mathbb{Z})$ and is invertible with $\mathbb{Z}$ coefficients.}

Let us now compute the left-hand side of \eqref{eq:SPT_K3},
\begin{align}
&N \hspace{-1.2em} \sum_{b \in H^2 (K3; \mathbb{Z}_N)} \hspace{-1.5em}\rme^{\frac{\im Nk}{4 \pi} \int b \wedge b} = N \left( \sum_{(n_1,n_2) \in \mathbb{Z}_N^{2}}  \rme^{ \frac{2 \pi \im k}{N} n_1 n_2 } \right)^3 \left( \sum_{\vec{n} \in \mathbb{Z}_N^{8}}  \rme^{-\frac{2 \pi \im k}{N} \left(\frac{1}{2}\vec{n}\cdot E_8\cdot \vec{n}\right) }  \right)^2.
\end{align}
The hyperbolic factor can be easily computed, 
\begin{equation}
    \sum_{n_1,n_2\in\mathbb{Z}_N}\rme^{\frac{2\pi \im k}{N}n_1n_2}=\gcd(N,k)N, 
\end{equation}
and then proving \eqref{eq:SPT_K3} can be reduced to the problem to show that the $E_8$ factor becomes
\begin{equation}
    \sum_{\vec{n} \in \mathbb{Z}_N^{8}}  \rme^{-\frac{2 \pi \im k}{N} \left(\frac{1}{2}\vec{n}\cdot E_8\cdot \vec{n}\right)}=\gcd(N,k)^4 N^4. 
    \label{eq:lattice-sum-e8}
\end{equation}

It is sufficient to show \eqref{eq:lattice-sum-e8} for the case $\operatorname{gcd}(N,k) = 1$.
Indeed, if $\operatorname{gcd}(N,k)=d > 1$, we can decompose $\vec{n}=\vec{n}'+d\,\vec{m}$ with $\vec{n}'\in \mathbb{Z}_{N/d}^8$ and $\vec{m}\in\mathbb{Z}_d^8$ so that 
\begin{align}
    \sum_{\vec{n} \in \mathbb{Z}_N^{8}}  \rme^{-\frac{2 \pi \im k}{N} \left(\frac{1}{2}\vec{n}\cdot E_8\cdot \vec{n}\right)} 
    &=\sum_{\vec{m}\in \mathbb{Z}_d^8}\sum_{\vec{n}'\in \mathbb{Z}_{N/d}^8}  \rme^{-\frac{2 \pi \im (k/d)}{(N/d)} \left(\frac{1}{2}\vec{n}'\cdot E_8\cdot \vec{n}'\right)} \notag\\
    &= \gcd(N,k)^8 \sum_{\vec{n}'\in \mathbb{Z}_{N/d}^8}  \rme^{-\frac{2 \pi \im (k/d)}{(N/d)} \left(\frac{1}{2}\vec{n}'\cdot E_8\cdot \vec{n}'\right)}. 
\end{align}
If \eqref{eq:lattice-sum-e8} is true for $\gcd(N,k)=1$, then the sum over $\vec{n}'$ gives $(N/d)^4=(N/\gcd(N,k))^4$, and thus \eqref{eq:lattice-sum-e8} for the general case is obtained. 
In the following, we assume $\gcd(N,k)=1$.

Let us factorize $N = 2^r N'$ with odd $N'$, and then Chinese remainder theorem gives $\mathbb{Z}_N\simeq \mathbb{Z}_{2^r}\times \mathbb{Z}_{N'}$. 
Accordingly, we can decompose $\vec{n}\in \mathbb{Z}_N^8$ as 
\begin{align}
    \vec{n}=N' \vec{\ell}+2^r \vec{m},
\end{align}
where $\vec{\ell}\in \mathbb{Z}_{2^r}^8$ and $\vec{m}\in \mathbb{Z}_{N'}^8$. We therefore have,
\begin{align}
    \sum_{\vec{n} \in \mathbb{Z}_N^{8}}  \rme^{-\frac{2 \pi \im k}{N} \left(\frac{1}{2}\vec{n}\cdot E_8\cdot \vec{n}\right)}
    =\left(\sum_{\vec{\ell} \in \mathbb{Z}_{2^r}^{8}}  \rme^{-\frac{2 \pi \im (N'k)}{2^r} \left(\frac{1}{2}\vec{\ell}\cdot E_8\cdot \vec{\ell}\right)}\right)
    \left(\sum_{\vec{m} \in \mathbb{Z}_{N'}^{8}}  \rme^{-\frac{2 \pi \im (2^r k)}{N'} \left(\frac{1}{2}\vec{m}\cdot E_8\cdot \vec{m}\right)}\right). 
\end{align}
By assumptions, $\gcd(N' k,2^r)=1$ and $\gcd(2^r k,N')=1$, and thus each factor on the right hand side has the form of \eqref{eq:lattice-sum-e8} with $\gcd(N,k)$=1. 
Therefore, it is sufficient to show \eqref{eq:lattice-sum-e8} for the cases $N=2^r$ and $N=N'$ with odd $N'$ assuming $\gcd(N,k)=1$. 
\begin{itemize}
    \item We shall verify, for $\gcd(2^r,k)=1$,
\begin{align}
    \sum_{\vec{n} \in \mathbb{Z}_{2^r}^{8}}  \rme^{-\frac{2 \pi \im k}{2^r} \left(\frac{1}{2}\vec{n}\cdot E_8\cdot \vec{n}\right)}
    = (2^r)^4.
    \label{eq:e8-sum-even}
\end{align}
For $r=0$, this is evident, and we can check this for $r=1$ by an explicit calculation. 
Let us focus on the cases of $r \geq 2$. In this case, we can decompose $\vec{n}$ into
\begin{align}
    \vec{n} = 2^{r-1} \vec{n}' + \vec{n}''~~~(\vec{n}' \in \mathbb{Z}_2^8,~\vec{n}'' \in \mathbb{Z}_{2^{r-1}}^8).
\end{align}
Substituting this decomposition, we have 
\begin{align}
    \sum_{\vec{n} \in \mathbb{Z}_{2^r}^{8}}  \rme^{-\frac{2 \pi \im k}{2^r} \left(\frac{1}{2}\vec{n}\cdot E_8\cdot \vec{n}\right)}
    &= \sum_{\vec{n}'' \in \mathbb{Z}_{2^{r-1}}^{8}}  \rme^{-\frac{2 \pi \im k}{2^r} \left(\frac{1}{2}\vec{n}''\cdot E_8\cdot \vec{n}''\right)}
    \sum_{\vec{n}' \in \mathbb{Z}_{2}^{8}} \rme^{-\frac{2 \pi \im k}{2} \left(\vec{n}'\cdot E_8\cdot \vec{n}''\right)}, 
\end{align}
and the summation over $\vec{n}'\in\mathbb{Z}_2^8$ gives the restriction that $\vec{n}''$ has to be even. 
Therefore, we may set $\vec{n}''=2\vec{\ell}$ with $\ell\in \mathbb{Z}_{2^{r-2}}^8$, and we get 
\begin{align}
    \sum_{\vec{n} \in \mathbb{Z}_{2^r}^{8}}  \rme^{-\frac{2 \pi \im k}{2^r} \left(\frac{1}{2}\vec{n}\cdot E_8\cdot \vec{n}\right)}
    &= 2^8\sum_{\vec{\ell} \in \mathbb{Z}_{2^{r-2}}^{8}}  \rme^{-\frac{2 \pi \im k}{2^{r-2}} \left(\frac{1}{2}\vec{\ell}\cdot E_8\cdot \vec{\ell}\right)}. 
\end{align}
By repeating this procedure, we obtain (\ref{eq:e8-sum-even}) by induction.

\item For the latter factor, we shall derive the result for odd $N'$ and $\gcd(N',k)=1$. 
We note that $\frac{N'+1}{2}$ plays the role of $2^{-1}$ for the multiplication in $\mathbb{Z}_{N'}$, and thus we may change $\vec{n}=2\vec{n}'$ in the summation,
\begin{align}
    \sum_{\vec{n} \in \mathbb{Z}_{N'}^{8}}  \rme^{-\frac{2 \pi \im k}{N'} \left(\frac{1}{2}\vec{n}\cdot E_8\cdot \vec{n}\right)}
    = \sum_{\vec{n}' \in \mathbb{Z}_{N'}^{8}}  \rme^{-\frac{2 \pi \im (2k)}{N'} \left(\vec{n}'\cdot E_8\cdot \vec{n}'\right)}
\end{align}

The relation $E_8 \oplus (-1) \simeq (-1) \oplus 8 (+1)$ implies
\begin{align}
&\left( \sum_{\vec{n} \in \mathbb{Z}_{N'}^{8}}  \rme^{-\frac{2 \pi \im (2k)}{N'} \left(\vec{n}\cdot E_8\cdot \vec{n}\right)} \right) 
\left( \sum_{n \in \mathbb{Z}_{N'}}  \rme^{\frac{2\pi\im(2k)}{N'} n^2 } \right) \nonumber\\
&= \left( \sum_{n \in \mathbb{Z}_{N'}}  \rme^{\frac{2\pi\im(2k)}{N'} n^2 } \right) \left( \sum_{n \in \mathbb{Z}_{N'}}  \rme^{-\frac{2\pi\im(2k)}{N'} n^2 } \right) ^8.
\end{align}
The sums on the right hand side are known as the quadratic Gauss sum. For odd $N'$ and $\operatorname{gcd}(2k,N') = 1$,
\begin{align}
\left( \sum_{n \in \mathbb{Z}_{N'}}  \rme^{\frac{2 \pi \im(2k)}{N'} n^2 } \right) = 
\begin{cases}
\sqrt{N'} ~~&(N' = 1~\mathrm{mod}~4) \\
\im\sqrt{N'} ~~&(N' = 3~\mathrm{mod}~4).
\end{cases}
\end{align}
This also ensures that the sum related to the extra factor $(-1)$ does not vanish, and thus we get 
\begin{align}
    \left( \sum_{\vec{n} \in \mathbb{Z}_{N'}^{8}}  \rme^{-\frac{2 \pi \im (2k)}{N'} \left(\vec{n}\cdot E_8\cdot \vec{n}\right)} \right)
    &=\left( \sum_{n \in \mathbb{Z}_{N'}}  \rme^{-\frac{2\pi\im(2k)}{N'} n^2 } \right) ^8=(N')^4.
    \label{eq:e8-sum-odd}
\end{align}
\end{itemize}
This completes the proof of \eqref{eq:SPT_K3}, and then we obtain Claim~\ref{claim:gravitational_anomaly}. 

\section{Detailed discussion on the fusion rule given in Claim~\ref{claim:fusion_rule}}
\label{sec:proof_claim_fusion_rule}

\subsection{Derivation of (\ref{eq:Deta-fusion-rule})}

Beforehand, it is useful to note that the $\mathbb{Z}_N^{[1]}$ background $B$ dependence can be found in (\ref{eq:CR-partition-func-w-background}) for the ungauged side and in (\ref{eq:Z_N-gauging}) for the gauged side.

First, let us derive
\begin{align}
    \mathscr{D} (M^{(3)}) \times \eta(\Sigma) = \mathscr{D} (M^{(3)})
\end{align}
with $\Sigma$ is a two-cycle embedded in $M^{(3)}$.
In this product, the symmetry defect $\eta(\Sigma)$ is in the gauged side, so it can be expressed by $\eta(\Sigma) = e^{\frac{\im N}{2 \pi} \int_{\Sigma} b}$, see (\ref{eq:Z_N-gauging}).
Therefore, from the Dirichlet boundary condition $\left. b \right|_{M^{(3)}} = 0$, $\eta(\Sigma)$ can be completely absorbed by the defect $\mathscr{D} (M^{(3)})$.

Next, we consider the product with $\eta(\Sigma)$ in the ungauged side:
\begin{align}
    \eta(\Sigma) \times \mathscr{D} (M^{(3)}) 
\end{align}
Since these defects are topological, we can slightly deform the configuration so that $\Sigma$ is inside $I \times M^{(3)} \subset X^+$.
We can regard $\Sigma \in H_2(I \times M^{(3)}; \mathbb{Z}_N) \simeq H_2(M^{(3)}; \mathbb{Z}_N)$, which corresponds to $\delta_\Sigma \simeq H^2(I \times M^{(3)}, \partial (I \times M^{(3)}); \mathbb{Z}_N)$.
Here, note that the Poincar\'e-Lefschetz duality and shrinking the interval imply
\begin{align}
    H^k (M^{(3)} \times I, \partial (M^{(3)} \times I); \mathbb{Z}_N) \simeq H_{4-k} (M^{(3)} \times I; \mathbb{Z}_N) \simeq H_{4-k} (M^{(3)}; \mathbb{Z}_N) \simeq H^{k-1} (M^{(3)}; \mathbb{Z}_N), \label{eq:isomorphism-shrinking-interval-2}
\end{align}

Since $\Sigma$ is also a two-cycle in $X^+$, $\eta(\Sigma)$ can be absorbed by the replacement $b \rightarrow b - \delta_\Sigma$.
Then, the discrete $\theta$-term becomes,
\begin{align}
    \rme^{-\frac{\im N}{4 \pi} \int b \wedge b} &\rightarrow   \rme^{-\frac{\im N}{4 \pi} \int b \wedge b + \frac{\im N}{2 \pi} \int b \wedge \delta_\Sigma -  \frac{\im N}{4 \pi}  \int \delta_\Sigma \wedge \delta_\Sigma} 
\end{align}
The mixing term $\int b \wedge \delta_\Sigma$ expresses the intersection number between $\Sigma$ and dual of $b$.
Since we can separate them by a continuous deformation, we have $\int b \wedge \delta_\Sigma = 0$.
This can also be obtained by $\int b \wedge \delta_\Sigma \propto \int_\Sigma b$ and the Dirichlet boundary condition $\left. b \right|_{M^{(3)}} = 0$.

Therefore, we have
\begin{align}
    \eta(\Sigma) \times \mathscr{D} (M^{(3)}) = \rme^{-  \frac{\im N}{4 \pi}  \int \delta_\Sigma \wedge \delta_\Sigma} \mathscr{D} (M^{(3)}).
\end{align}
The extra factor can be expressed as
\begin{align}
     \rme^{-  \frac{\im N}{4 \pi}  \int \delta_\Sigma \wedge \delta_\Sigma} = (-1)^{Q(\Sigma)}.
\end{align}
We derive this equation below at (\ref{eq:M3_discrete-theta}).
This completes the proof of (\ref{eq:Deta-fusion-rule}).

\subsection{Derivation of (\ref{eq:orientation_inv_fusion_rule})}

Here, we consider $\mathscr{D} (M^{(3)}) \times \Bar{\mathscr{D}} (M^{(3)}) $ and $\Bar{\mathscr{D}} (M^{(3)}) \times \mathscr{D} (M^{(3)})$.
We first note that the self-duality (Claim \ref{claim:self-dual-stinv}) yields,
\begin{align}
\calZ_{\CR}^{\tau_*} [B] = N^{\frac{-\chi(X)}{2}} \rme^{-\frac{\pi \im}{3} \sigma(X)} \int \mathcal{D}b
     \,\calZ_{\CR}^{\tau_*} [b] \exp\left(- \frac{\im N}{2\pi}\int_X b\wedge B + \frac{\im N }{4\pi}\int_X B \wedge B \right), \label{eq:duality-sinvt}
\end{align}
where we have used $\frac{|H^0(X;\mathbb{Z}_N)|^2 |H^2(X;\mathbb{Z}_N)|}{|H^1(X;\mathbb{Z}_N)|^2} = N^{\chi(X)}$.

\begin{figure}[t]
    \centering
    \includegraphics[width =  0.7 \linewidth]{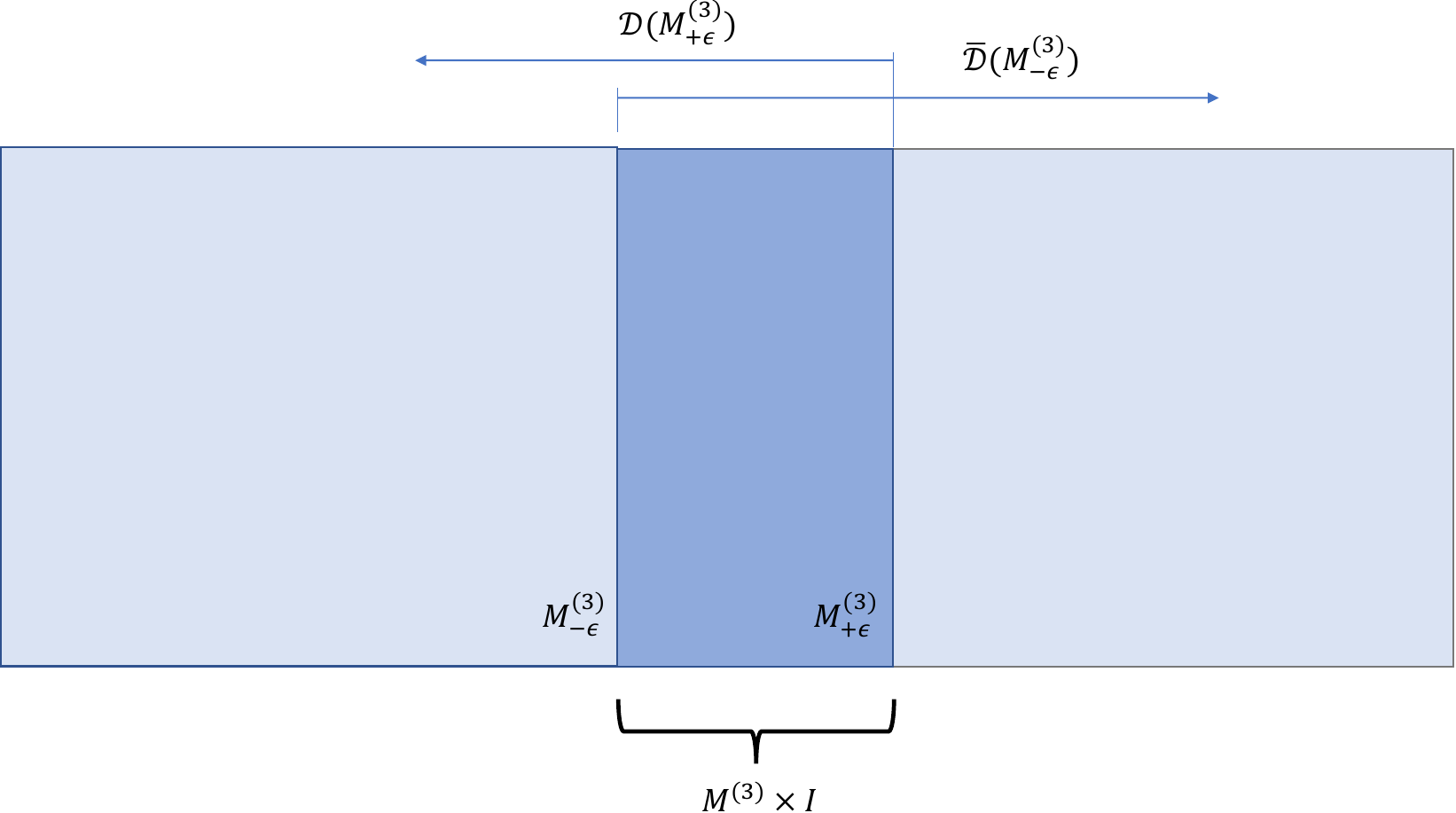}
    \caption{Fusion rule of the duality defect $\mathscr{D}(M^{(3)})$ and its orientation reverse $\Bar{\mathscr{D}}(M^{(3)})$. We infinitesimally displace those hypersurfaces as $\mathscr{D}(M^{(3)}_{+\ve})$ and $\Bar{\mathscr{D}}(M^{(3)}_{-\ve})$. }
    \label{fig:DDbar_fusion_rule}
\end{figure}

To evaluate the product $\mathscr{D} (M^{(3)}) \times \Bar{\mathscr{D}} (M^{(3)}) $, we regularize this product by slightly separating the hypersurface $M^{(3)}$:
$\mathscr{D} (M^{(3)}_{+\varepsilon}) \times \Bar{\mathscr{D}} (M^{(3)}_{-\varepsilon})$, see Fig.~\ref{fig:DDbar_fusion_rule}.
Correspondingly, we divide $X$ into three parts: (A) $X^+_{-\ve}$ satisfying $\partial X^+_{-\ve} = M^{(3)}_{-\ve}$ (B) the overlapping interval $M^{(3)} \times I$ (C) $X^-_{+\ve}$ satisfying $\partial X^-_{+\ve} = -M^{(3)}_{+\ve}$.

For each region, we have:
\begin{align}
    &\mathrm{(A):}~ \calZ_{A} [b_1] \rme^{ \frac{\im N}{2\pi} \left(\int b_1 \wedge B - \frac{1}{2} \int b_1^2 \right)},\notag \\
    &\mathrm{(B):}~ \calZ_{B} [b_1] \rme^{ \frac{\im N}{2\pi} \left(\int b_1 \wedge b_2 - \frac{1}{2} \int b_1^2 + \int b_2 \wedge B - \frac{1}{2} \int b_2^2 \right)}, \notag  \\
    &\mathrm{(C):}~ \calZ_{C} [b_2] \rme^{ \frac{\im N}{2\pi} \left(\int b_2 \wedge B - \frac{1}{2} \int b_2^2 \right)},
\end{align}
where $b_1$ (resp.~$b_2$) is the $\mathbb{Z}_N^{[1]}$ gauge field introduced by $\mathscr{D}(M^{(3)}_{+\ve})$ (resp.~$\Bar{\mathscr{D}}(M^{(3)}_{-\ve})$), and $\calZ_{A},~\calZ_{B}$ and $\calZ_{C}$ are the partition functions of the respective parts.
Here, the background field $B$ vanishes around $M^{(3)}$, since it should be evaluated in more complicated fusion rules if $B$ does not vanish.

From (\ref{eq:duality-sinvt}), the expressions become
\begin{align}
    &\mathrm{(A):}~ \calZ_{A} [b] \rme^{ \frac{\im N}{2\pi} \left(\int b_1 \wedge (B - b)\right)},\notag  \\
    &\mathrm{(B):}~ \calZ_{B} [b] \rme^{ \frac{\im N}{2\pi} \left(\int b_1 \wedge (b_2 - b) + \int b_2 \wedge B - \frac{1}{2} \int b_2^2 \right)}, \notag  \\
    &\mathrm{(C):}~ \calZ_{C} [b] \rme^{ \frac{\im N}{2\pi} \left(\int b_2 \wedge (B-b) \right)},
\end{align}
with the $\mathbb{Z}_N^{[1]}$ gauge field $b$ on $X$.
By integrating out $b_1$, $b$ is fix to $B$ on the (A) region and $b_2$ on the (B) region.
Since $b$ is fix to $B$ on the (A) region, we can regard $b \in H^2(X^-_{+\ve}, \partial X^-_{+\ve}; \mathbb{Z}_N)$
In addition, it is imposed that $b=b_2$ on the (B) region.
We can fix $b_2$ on (B) by representing $b_2 = b + b'$ with $b' \in H^2(X^-_{+\ve}, \partial X^-_{+\ve}; \mathbb{Z}_N)$.
Therefore, we have
\begin{align}
    &\mathrm{(A):}~ \calZ_{A} [B] ,\notag  \\
    &\mathrm{(B):}~ \calZ_{B} [b] \rme^{ \frac{\im N}{2\pi} \left(\int b \wedge B - \frac{1}{2} \int b^2 \right)}, \notag  \\
    &\mathrm{(C):}~ \calZ_{C} [b] \rme^{ \frac{\im N}{2\pi} \left(\int (b + b') \wedge (B-b) \right)},
\end{align}
We further integrate out $b'$, which imposes $b = B$ on the (C) region.
Since $b$ is fix to $B$ on both (A) and (C) regions, we can express $b \in H^2 (M^{(3)} \times I, \partial (M^{(3)} \times I); \mathbb{Z}_N)$.
To sum up, we obtain, noting the locality of the counterterms $\chi$ and $\sigma$,
\begin{align}
    \mathscr{D} (M^{(3)}) \times \Bar{\mathscr{D}} (M^{(3)}) &= \left( N^{-\frac{\chi(X^+)}{2}}\rme^{\frac{\pi\im}{3}\sigma(X^+)}\right) \left( N^{-\frac{\chi(X^-)}{2}}\rme^{\frac{\pi\im}{3}\sigma(X^-)}\right) \left( N^{-\frac{\chi(X)}{2}}\rme^{-\frac{\pi\im}{3}\sigma(X)}\right) \notag \\
    &\times \left( \int_{X^+} \mathcal{D}b_1  \int_{X^-} \mathcal{D}b_2 \right) \frac{|H^0(X;\mathbb{Z}_N)|}{|H^1(X;\mathbb{Z}_N)|} \sum_{\Sigma \in H_2 (M^{(3)} ;\mathbb{Z}_N)} \eta(\Sigma) \rme^{-\frac{\im N}{4 \pi} \int \delta_\Sigma \wedge \delta_\Sigma} \notag \\
    &= \left( \frac{\int_{X^+} \mathcal{D}b_1  \int_{X^-} \mathcal{D}b_2}{\int \mathcal{D}b} \right) \sum_{\Sigma \in H_2 (M^{(3)} ;\mathbb{Z}_N)} \eta(\Sigma) \rme^{-\frac{\im N}{4 \pi} \int \delta_\Sigma \wedge \delta_\Sigma}
\end{align}
where $\int_{X^\pm} \mathcal{D}b = \frac{|H^0 (X^\pm, \partial X^\pm; \mathbb{Z}_N)| |H^2 (X^\pm, \partial X^\pm; \mathbb{Z}_N)|}{|H^1 (X^\pm, \partial X^\pm; \mathbb{Z}_N)|}$, and $\delta_\Sigma \in H^2 (M^{(3)} \times I, \partial (M^{(3)} \times I); \mathbb{Z}_N)$ is the Poincar\'e-Lefschetz dual of $\Sigma \in H_2 (M^{(3)} \times I;\mathbb{Z}_N) \simeq H_2 (M^{(3)};\mathbb{Z}_N)$.

The remaining work is to determine the prefactor $\left( \frac{\int_{X^+} \mathcal{D}b_1  \int_{X^-} \mathcal{D}b_2}{\int \mathcal{D}b} \right)$. 
We note that this prefactor turns out to be exactly the same with the one for the fusion rule of $S$-duality defect discussed in Refs.~\cite{Choi:2021kmx, Kaidi:2021xfk}, which suggests 
\begin{equation}
    \left( \frac{\int_{X^+} \mathcal{D}b_1  \int_{X^-} \mathcal{D}b_2}{\int \mathcal{D}b} \right)=\frac{|H^0 (M^{(3)} \times I, \partial (M^{(3)} \times I); \mathbb{Z}_N)|}{|H^1 (M^{(3)} \times I, \partial (M^{(3)} \times I); \mathbb{Z}_N)|} = \frac{1}{N}.
\end{equation}
As a consistency check, we can confirm the bulk independence of this factor, so it depends only on the topology of $M^{(3)}$.
When $M^{(3)}$ has no nontrivial cycle, i.e., $H_1(M^{(3)}; \mathbb{Z}_N) = \{ 0 \}$ and $H_2(M^{(3)}; \mathbb{Z}_N) = \{ 0 \}$, we have $\left( \frac{\int_{X^+} \mathcal{D}b_1  \int_{X^-} \mathcal{D}b_2}{\int \mathcal{D}b} \right) = \frac{1}{N}$ and it is consistent with the above result. 
In general, there may be an additional local counterterm depending on the topology of $M$.
Even for such cases, we can redefine $\Bar{\mathscr{D}} (M^{(3)})$ so that this local counterterm of $M$ is absorbed.
Therefore, we conclude 
\begin{align}
    \mathscr{D} (M^{(3)}) \times \Bar{\mathscr{D}} (M^{(3)}) &= \frac{1}{N} \sum_{\Sigma \in H_2 (M^{(3)} ;\mathbb{Z}_N)} \eta(\Sigma) \rme^{-\frac{\im N}{4 \pi} \int \delta_\Sigma \wedge \delta_\Sigma}
\end{align}
At (\ref{eq:M3_discrete-theta}), we will show $\rme^{-\frac{\im N}{4\pi}\int \delta_\Sigma \wedge \delta_\Sigma} = (-1)^{Q(\Sigma)}$ and it gives the second fusion rule (\ref{eq:orientation_inv_fusion_rule}).

Similarly, we can derive the fusion rule for $\Bar{\mathscr{D}} (M^{(3)}) \times \mathscr{D} (M^{(3)})$.
In this case, the interval (B) is not gauged, so we have
\begin{align}
    &\mathrm{(A):}~ \calZ_{A} [b] \rme^{ \frac{\im N}{2\pi} \left(\int b_1 \wedge (B - b)\right)},\notag  \\
    &\mathrm{(B):}~ \calZ_{B} [b] \rme^{ \frac{\im N}{2\pi} \left(\int b \wedge B - \frac{1}{2} \int B^2 \right)}, \notag  \\
    &\mathrm{(C):}~ \calZ_{C} [b] \rme^{ \frac{\im N}{2\pi} \left(\int b_2 \wedge (B-b) \right)},
\end{align}
where $b_1$ is the gauge field on (A), $b_2$ is the gauge field on (C), and $b$ is the gauge field on $X$.
Since we can assume that $B$ vanishes around the interval to evaluate the fusion $\Bar{\mathscr{D}} (M^{(3)}) \times \mathscr{D} (M^{(3)})$, a parallel computation leads to
\begin{align}
    &\Bar{\mathscr{D}} (M^{(3)}) \times \mathscr{D} (M^{(3)})  = \frac{1}{N} \sum_{\Sigma \in H_2 (M^{(3)} ;\mathbb{Z}_N)} \eta(\Sigma).
\end{align}

\subsection{Derivation of (\ref{eq:ST-1_fusion_rule})}

\begin{figure}[t]
    \centering
    \includegraphics[width =  0.7 \linewidth]{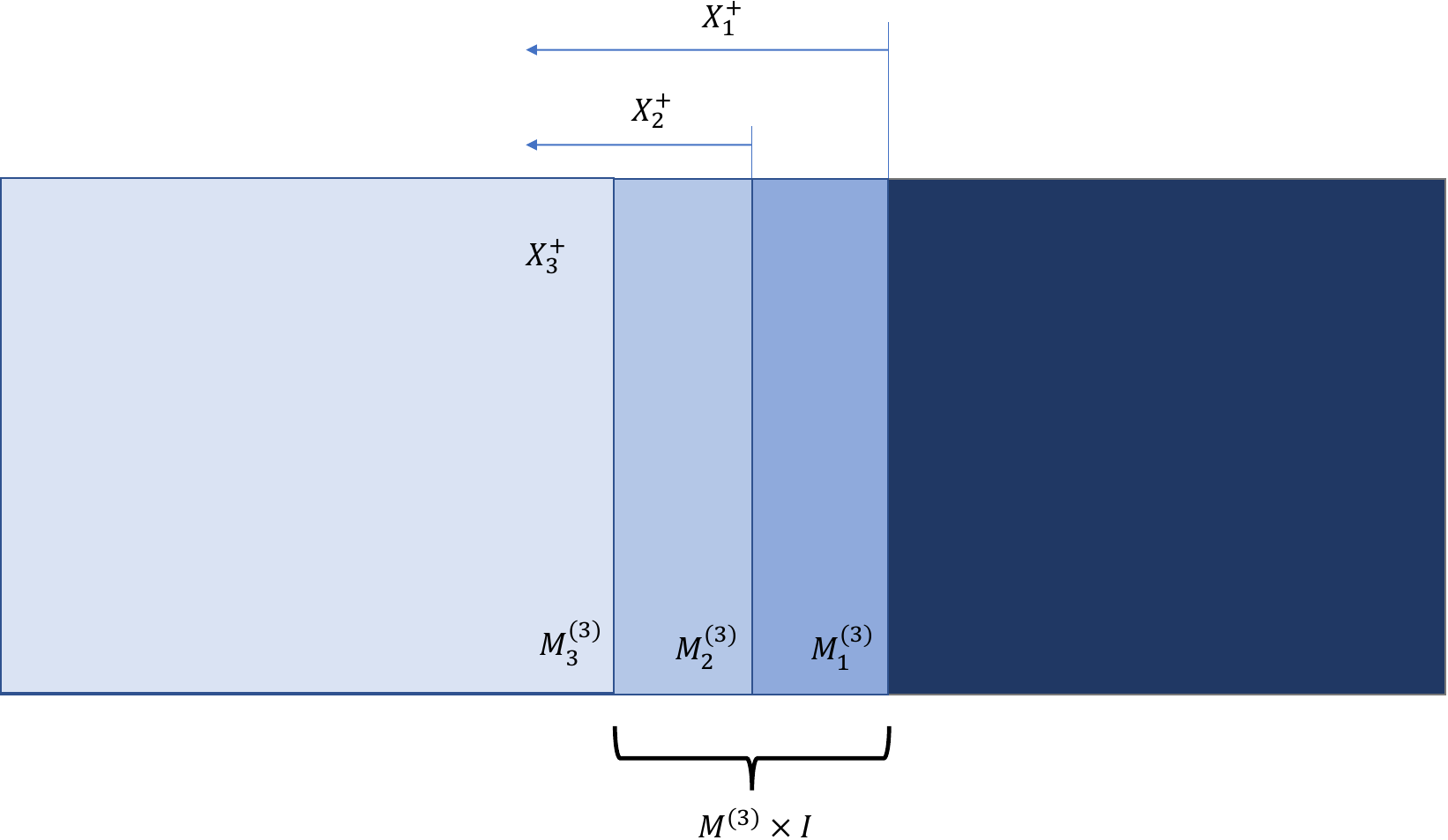}
    \caption{Fusion rule of three duality defects $\mathscr{D}(M^{(3)})$, and those hypersurfaces are infinitesimally displaced as $\mathscr{D}(M^{(3)}_1)$, $\mathscr{D}(M^{(3)}_2)$, and $\mathscr{D}(M^{(3)}_3)$. }
    \label{fig:calc_fusion_rule}
\end{figure}

As above, we split $M^{(3)}$ and compute
\begin{align}
    \mathscr{D} (M^{(3)}_1) \times \mathscr{D} (M^{(3)}_2) \times \mathscr{D} (M^{(3)}_3)
\end{align}
where $M^{(3)}_1, M^{(3)}_2,$ and $M^{(3)}_3$ are slightly shifted $M^{(3)}$, see Fig.~\ref{fig:calc_fusion_rule}. Similarly we define the regions $X_1^+,~X_2^+$, and $X_3^+$ so that $\partial X_1^+ = M^{(3)}_1, ~ \partial X_2^+ = M^{(3)}_2,~\partial X_3^+ = M^{(3)}_3$.

Let us evaluate the product $\mathscr{D} (M^{(3)}_1) \times \mathscr{D} (M^{(3)}_2) \times \mathscr{D} (M^{(3)}_3)$.
Since we know the action on line operators, we first consider the expectation value $\braket{\mathscr{D} (M^{(3)}_1) \times \mathscr{D} (M^{(3)}_2) \times \mathscr{D} (M^{(3)}_3)}$.

The product reads
\begin{align}
    & \braket{\mathscr{D} (M^{(3)}_1) \times \mathscr{D} (M^{(3)}_2) \times \mathscr{D} (M^{(3)}_3)} \notag \\
    &= \mathcal{N}_0 \frac{1}{\calZ_{\CR}^{\tau}} \int_{X_1^+} \Diff b_1 \int_{X_2^+} \Diff b_2 \int_{X_3^+} \Diff b_3\, \calZ_{\CR}^{\tau} [b_1]\rme^{\frac{\im N}{4\pi}\int (-b_1^2-b_2^2-b_3^2+2b_1\wedge b_2+2b_2\wedge b_3)} \notag \\
    &= \mathcal{N}_0 \frac{1}{\calZ_{\CR}^{\tau}} \int_{X_1^+} \Diff b_1 \int_{X_2^+} \Diff b_2 \int_{X_3^+} \Diff b_3\, \calZ_{\CR}^{\tau} [b_1] \rme^{\frac{\im N}{4\pi}\int (-b_1^2+2b_1\wedge b_2)} \rme^{-\frac{\im N}{4\pi}\int (b_2 - b_3)^2},
\end{align}
where $\mathcal{N}_0 $ represents the counterterm
\begin{align}
    \mathcal{N}_0  &:= N^{-\frac{\chi(X^+_1)}{2}-\frac{\chi(X^+_2)}{2}-\frac{\chi(X^+_3)}{2}}\rme^{\frac{\pi\im}{3}\sigma(X^+_1) + \frac{\pi\im}{3}\sigma(X^+_2) + \frac{\pi\im}{3}\sigma(X^+_3)} \notag \\
    &\rightarrow N^{-\frac{3\chi(X^+)}{2}}\rme^{\pi\im \sigma(X^+)}
\end{align}
Since $X_2^+$ includes $X_3^+$, we can shift $b_2$ by $b_3$:
\begin{align}
    &\braket{\mathscr{D} (M^{(3)}_1) \times \mathscr{D} (M^{(3)}_2) \times \mathscr{D} (M^{(3)}_3)} \notag \\
    &= \mathcal{N}_0 \frac{1}{\calZ_{\CR}^{\tau}} \int_{X_1^+} \Diff b_1 \int_{X_2^+} \Diff b_2  \, \calZ_{\CR}^{\tau} [b_1] \rme^{\frac{\im N}{4\pi}\int (-b_1^2+2b_1\wedge b_2)} \rme^{-\frac{\im N}{4\pi}\int (b_2)^2} \int_{X_3^+} \Diff b_3~\rme^{\frac{\im N}{4\pi}\int b_1 \wedge b_3}.
\end{align}
The last factor $\int \Diff b_3~\rme^{\frac{\im N}{4\pi}\int b_1 \wedge b_3}$ gives another boundary condition $\left. b_1 \right|_{M^{(3)}_3} = 0$.
From this condition, we can regard $b_1 \in H^2 (M^{(3)} \times I, \partial (M^{(3)} \times I); \mathbb{Z}_N)$, where $M^{(3)} \times I$ stands for the tubular region $X^-_3 \cap X^+_1$. Therefore,
\begin{align}
    &\braket{\mathscr{D} (M^{(3)}_1) \times \mathscr{D} (M^{(3)}_2) \times \mathscr{D} (M^{(3)}_3)} \notag \\
    &= \mathcal{N}_0 \frac{1}{\calZ_{\CR}^{\tau}} \left( \int_{X_3^+} \Diff b_3 \right)  \frac{|H^0 (X_1^+, \partial X_1^+; \mathbb{Z}_N)|}{|H^1 (X_1^+, \partial X_1^+; \mathbb{Z}_N)|} \notag \\
    & \times \sum_{b_1 \in H^2 (M^{(3)} \times I, \partial (M^{(3)} \times I); \mathbb{Z}_N)} \int_{X_2^+} \Diff b_2  \, \calZ_{\CR}^{\tau} [b_1] \rme^{\frac{\im N}{4\pi}\int (-b_1^2+2b_1\wedge b_2)} \rme^{-\frac{\im N}{4\pi}\int (b_2)^2}.
    \label{eq:evaluation-fusion-rule-splitted}
\end{align}

Note that the cross term between $b_1$ and $b_2$, namely $\rme^{\frac{\im N}{2\pi}\int b_1\wedge b_2}$, is written by the intersection between the Poincare dual of $b_1$ and that of $b_2$.
However, the dual surface of $b_2$ can be continuously deformed in $X_2^+$, from which we can ``separate'' $b_2$ from $b_1$ lying in the tubular region $M^{(3)} \times I$.
Thus, we have $\rme^{\frac{\im N}{2\pi}\int b_1\wedge b_2} = 1$ and
\begin{align}
    &\braket{\mathscr{D} (M^{(3)}_1) \times \mathscr{D} (M^{(3)}_2) \times \mathscr{D} (M^{(3)}_3)} \notag \\
    &=  \mathcal{N}_{ \mathscr{D}^3 }(M^{(3)}) \left( \frac{1}{\calZ_{\CR}^{\tau}} \sum_{b_1 \in H^2 (M^{(3)} \times I, \partial (M^{(3)} \times I); \mathbb{Z}_N)} \calZ_{\CR}^{\tau} [b_1] \rme^{-\frac{\im N}{4\pi}\int b_1^2} \right), \label{eq:evaluation-fusion-rule-b-1}
\end{align}
with
\begin{align}
    \mathcal{N}_{ \mathscr{D}^3 }(M^{(3)}) := & N^{-\frac{3\chi(X^+)}{2}}\rme^{\pi\im \sigma(X^+)} \frac{|H^0 (X^+, \partial X^+; \mathbb{Z}_N)|^3 |H^2 (X^+, \partial X^+; \mathbb{Z}_N)|}{|H^1 (X^+, \partial X^+; \mathbb{Z}_N)|^3} \notag \\
    &\times  \left( \sum_{b \in H^2 (X^+, \partial X^+; \mathbb{Z}_N)} \rme^{-\frac{\im N}{4 \pi} \int b \wedge b}\right). \tag{\ref{eq:fusion-rule-normalization}}
\end{align}

The right-hand side of (\ref{eq:evaluation-fusion-rule-b-1}) is, up to the normalization, the same as that discussed in \cite{Choi:2021kmx, Kaidi:2021xfk}. 
As a result, the $\mathbb{Z}_N^{[1]}$ gauge field on $M^{(3)} \times I$, $b_1$, can be labeled by a $\mathbb{Z}_N^{[0]}$ gauge field on $M^{(3)}$, $a_1$, and the discrete $\theta$ term becomes,
\begin{align}
    \rme^{-\frac{\im N}{4\pi}\int b_1 \wedge b_1} = \rme^{-\frac{\im N}{4\pi}\int a_1 \wedge d_M a_1}, \label{eq:M3_discrete-theta}
\end{align}
where $d_M$ denotes the derivative on $M^{(3)}$ \cite{Kaidi:2021xfk}.
For the sake of completeness, we briefly review its derivation.

The Poincar\'e--Lefschetz duality and shrinking the tiny interval lead to the isomorphism relating $\mathbb{Z}_N^{[1]}$ gauge field $b_1$ and $\mathbb{Z}_N^{[0]}$ gauge field $a_1$ (see also (\ref{eq:isomorphism-shrinking-interval-2})):
\begin{align}
    H^2 (M^{(3)} \times I, \partial (M^{(3)} \times I); \mathbb{Z}_N) \simeq H_2 (M^{(3)} \times I; \mathbb{Z}_N) \simeq H_2 (M^{(3)}; \mathbb{Z}_N) \simeq H^1 (M^{(3)}; \mathbb{Z}_N). \label{eq:isomorphism-shrinking-interval}
\end{align}
Since this map is essentially shrinking the interval, it will be locally represented by integration over the interval $\int_I$ reducing a 2-form to a 1-form.

Beforehand, we add some remarks on (\ref{eq:M3_discrete-theta}).
\begin{itemize}
    \item We have described an element of cohomology by a differential form.
Usually, we does not have to care about the lift from $\mathbb{Z}_N$.
Nevertheless, without care of the lift, the ``Bockstein operation'' $d_M a_1$ in (\ref{eq:M3_discrete-theta}) would vanish.
We need to pay attention to the lift in order to derive (\ref{eq:M3_discrete-theta}) precisely.
    \item The Bockstein operation $\frac{1}{N} d_M: H^1 (M^{(3)}; \mathbb{Z}_N) \rightarrow H^2 (M^{(3)}; \mathbb{Z})$ is trivial unless the cohomology $H^2 (M^{(3)}; \mathbb{Z})$ has a torsion part.
Since we assume that the manifold has no torsion in the main text, we may restrict ourselves to cases that the separating hypersurface $M^{(3)}$ also has no torsion.
In this case, the same result $\rme^{-\frac{\im N}{4\pi}\int b_1 \wedge b_1} = 1$ can be derived by a simple calculation without a care of the lift.
    \item The factor (\ref{eq:M3_discrete-theta}) is trivial for odd $N$.
Indeed, for odd $N$, there exists $2^{-1} \in \mathbb{Z}_N$, which leads to $\rme^{-\frac{\im N}{4\pi}\int b_1 \wedge b_1} = \rme^{-\frac{\im N}{2\pi} 2^{-1} \int b_1 \wedge b_1} = \rme^{-\im 2^{-1} \frac{ N^2}{2\pi}  \int a^{(1)}\wedge \frac{d_M a^{(1)}}{N}} = 1$.
This can also be seen from a computation of $\Omega^{\Spin}_3(B \mathbb{Z}_N)$:
SPT phases for odd $N$ are classified by $\mathbb{Z}_N$, which implies that the $\mathbb{Z}_N$ Chern-Simons phase at level $N$ (\ref{eq:M3_discrete-theta}) is trivial.
For even $N$, it can provide a nontrivial phase $\pm 1$.
In particular, for $N = 2$, where the Bockstein operation becomes the Steenrod square, this factor can be interpreted as the mod-2 triple intersection number $e^{i \pi \int_{M} a_1 \cup a_1 \cup a_1}$ \cite{Kaidi:2021xfk}.
\end{itemize}

Let us derive (\ref{eq:M3_discrete-theta}).
We express $b_1 \in H^2 (M^{(3)} \times I, \partial (M^{(3)} \times I); \mathbb{Z}_N)$ by a 2-form (with a lift to $\mathbb{Z}$) as,
\begin{align}
    \left. b_1 \right|_{\partial (M^{(3)} \times I)} = 0 ~~&(\operatorname{mod}~2 \pi),~~ \notag \\
    d b_1 = 0~~&(\operatorname{mod}~2 \pi),
\end{align}
where the right-hand side $(\operatorname{mod}~2 \pi)$ means $2 \pi \times$ (an integer coefficient cochain).
Remember that the lift of $b_1 \in H^2 (M^{(3)} \times I, \partial (M^{(3)} \times I); \mathbb{Z}_N)$ is normalized as $\int_\Sigma b_1 \in \frac{2 \pi}{N} \mathbb{Z}$ for a closed two-surface $\Sigma$ in our convention.

We decompose the two-form $b_1$ as,
\begin{align}
    b_1 = \omega^{(2)}(t) + \diff t \wedge \beta^{(1)}(t),
\end{align}
where $t$ is the coordinate of $I$, and $\omega^{(2)}(t)$ and $\beta^{(1)}(t)$ are a two-form and one-form on $M^{(3)}$, respectively.
As a representation of the isomorphism, the $\mathbb{Z}_N^{[0]}$ gauge field $a_1$ on $M^{(3)}$ is given by,
\begin{align}
    a_1 = \int_{-\epsilon}^\epsilon \diff t ~\beta^{(1)}(t). \label{eq:Z_n-gauge-field-on-M}
\end{align}
Then, the Pontryagin square is
\begin{align}
    \frac{N}{4\pi} \int b_1 \wedge b_1 = \frac{N}{2\pi} \int_{M^{(3)}} \int_{-\epsilon}^\epsilon \diff t ~ \omega^{(2)}(t) \wedge \beta^{(1)}(t).
\end{align}
Here, we can express $\omega^{(2)}$ as,
\begin{align}
    \omega^{(2)}(t) =  \int_{-\epsilon}^t \diff t' ~ \diff_M \beta^{(1)}(t') + \int_{[-\epsilon,t]} d b_1 + \omega^{(2)}(-\epsilon)
\end{align}
Since $\frac{1}{2\pi}db_1$ is a $\mathbb{Z}$ coefficient 3-cocycle, $\frac{1}{2\pi}\int_{[-\epsilon,t]} d b_1$ is also a $\mathbb{Z}$ coefficient 2-cocycle. Note also that the last term $\omega^{(2)}(-\epsilon)$ is a $\mathbb{Z}$ coefficient 3-cochain on $M^{(3)}$.

Using this expression, we have
\begin{align}
    \frac{N}{2\pi} \int_{M^{(3)}} \int_{-\epsilon}^\epsilon \diff t ~ \omega^{(2)}(t) \wedge \beta^{(1)}(t) &= \frac{N}{2\pi} \int_{M^{(3)}} \int_{-\epsilon}^\epsilon \diff t ~  \int_{-\epsilon}^t \diff t' ~ \diff_M \beta^{(1)}(t') \wedge \beta^{(1)}(t) \notag \\
    &~~~~ + 2\pi \int_{M^{(3)} \times I} \left( \int_{[-\epsilon,t]} \frac{db_1}{2\pi} \right) \wedge \left( \frac{N b_1}{2\pi}  \right) \notag \\
    &~~~~ + 2\pi \int_{M^{(3)}} \left( \frac{\omega^{(2)}(-\epsilon)}{2 \pi} \right) \wedge \left( \frac{N a_1}{2\pi}  \right).
\end{align}
The second term is the integral of the cup product of the $\mathbb{Z}$ coefficient 2-cocycle $ \int_{[-\epsilon,t]} \frac{db_1}{2\pi}$ and the $\mathbb{Z}$-coefficient cohomology element $ \frac{N b_1}{2\pi}$, which should be an integer.
Due to the same reason, the integral of the last term should be an integer.

Thus, we can omit the second and third terms to evaluate the discrete $\theta$-term $\rme^{-\frac{\im N}{4\pi}\int b_1 \wedge b_1}$.
From (\ref{eq:Z_n-gauge-field-on-M}), we obtain
\begin{align}
    \frac{N}{4\pi}\int b_1 \wedge b_1 &= \frac{N}{4\pi}\int a_1 \wedge d_M a_1 ~~(\operatorname{mod}~2\pi),
\end{align}
which indeed gives (\ref{eq:M3_discrete-theta}).

On the other hand, the external $\mathbb{Z}_N^{[1]}$ gauge field $b_1 \in H^2 (M^{(3)} \times I, \partial (M^{(3)} \times I); \mathbb{Z}_N)$ can be expressed by the insertion of $\mathbb{Z}_N^{[1]}$ symmetry operator $\eta (\Sigma)$ with the Poincar\'e--Lefschetz dual $\Sigma \in  H_2 (M^{(3)} \times I; \mathbb{Z}_N)$.
Therefore, the sum and insertion of $b_1$ field can be represented by
\begin{align}
    \frac{1}{\calZ_{\CR}^{\tau}} \sum_{b_1 \in H^2 (M^{(3)} \times I, \partial (M^{(3)} \times I); \mathbb{Z}_N)}\calZ_{\CR}^{\tau} [b_1] = \sum_{\Sigma \in  H_2 (M^{(3)} ; \mathbb{Z}_N)} \braket{\eta(\Sigma)}.
\end{align}
To sum up, we obtain
\begin{align}
    &\braket{\mathscr{D} (M^{(3)}_1) \times \mathscr{D} (M^{(3)}_2) \times \mathscr{D} (M^{(3)}_3)} \notag \\
    &= \mathcal{N}_{ \mathscr{D}^3 }(M^{(3)}) \left( \frac{1}{\calZ_{\CR}^{\tau}} \sum_{\Sigma \in H_2(M^{(3)}; \mathbb{Z}_N)} (-1)^{Q(\Sigma)} \braket{\eta(\Sigma )} \right),
\end{align}
where $(-1)^{Q(\Sigma)}$ is the phase factor $\pm 1$,
\begin{align}
    (-1)^{Q(\Sigma)} := \rme^{\frac{\im N}{4\pi}\int a_1(\Sigma) \wedge d_M a_1(\Sigma)}, 
\end{align}
and $a_1(\Sigma)$ is the Poincar\'e dual of $\Sigma \in H_2(M^{(3)}; \mathbb{Z}_N)$.

Now, we have derived the fusion rule in the absence of the operators,
\begin{align}
    \mathscr{D} (M^{(3)})^3
    &= \mathcal{N}_{ \mathscr{D}^3 }(M^{(3)}) ~\sum_{\Sigma \in H_2(M^{(3)}; \mathbb{Z}_N)} (-1)^{Q(\Sigma)} \eta(\Sigma )~~~(\mathrm{without~operators}).
\end{align}

As explained in the main text, the effect of $\mathscr{D} (M^{(3)})^3$ on line operators is the charge conjugation $\mathsf{C}(M^{(3)})$.
Therefore, we finally obtain
\begin{align}
    \mathscr{D} (M^{(3)})^3
    &= \mathcal{N}_{ \mathscr{D}^3 }(M^{(3)}) \mathsf{C}(M^{(3)})~\sum_{\Sigma \in H_2(M^{(3)}; \mathbb{Z}_N)} (-1)^{Q(\Sigma)} \eta(\Sigma),
\end{align}
which is the desired result (\ref{eq:ST-1_fusion_rule}).
As we have noted, $(-1)^{Q(\Sigma)}$ becomes trivial if $H^2 (M^{(3)} ; \mathbb{Z})$ has no torsion or if $N$ is odd.



As a side note, let us check the bulk independence of the normalization constant $\mathcal{N}_{ \mathscr{D}^3 }(M^{(3)})$ for a torsion free bulk.
Although this should hold from the definition of the defect $\mathscr{D}(M^{(3)})$, it is not apparent in the expression (\ref{eq:fusion-rule-normalization}).
We shall show
\begin{align}
    \mathcal{N}_{ \mathscr{D}^3 }(M^{(3)}; X^+) = \mathcal{N}_{ \mathscr{D}^3 }(M^{(3)}; Y^+)~\mathrm{~if~}\partial X^+ = \partial Y^+ = M^{(3)}, 
\end{align}
where $\mathcal{N}_{ \mathscr{D}^3 }(M^{(3)}; X^+)$ denotes the normalization constant (\ref{eq:fusion-rule-normalization}) to manifest $X^+$ dependence.
Since $X^+$ and $Y^+$ should be connected by cutting and gluing closed manifolds, let us see
\begin{align}
    \mathcal{N}_{ \mathscr{D}^3 }(M^{(3)}; X^+) = \mathcal{N}_{ \mathscr{D}^3 }(M^{(3)}; Y^+)~\mathrm{~if~}Y^+ = X^+ \sharp M^{(4)},
\end{align}
for any closed (spin) manifold $M^{(4)}$.
Here $\sharp$ denotes the connected sum (preserving the boundary).
Then, we have $H^2(Y^+, M^{(3)}; \mathbb{Z}_N) \simeq H^2(M^{(4)}; \mathbb{Z}_N) \oplus H^2(X^+, M^{(3)}; \mathbb{Z}_N) $, and correspondingly
\begin{align}
    \int b \wedge b = \int b_1 \wedge b_1 + \int b_2 \wedge b_2,
\end{align}
with $H^2(M^{(4)}; \mathbb{Z}_N)$-part $b_1$ and $H^2(X^+, M^{(3)}; \mathbb{Z}_N)$-part $b_2$.
Therefore, under the assumption that $H^*(M^{(4)}; \mathbb{Z})$ has no torsion, the ratio between $\mathcal{N}_{ \mathscr{D}^3 }(M^{(3)}; X^+ \sharp M^{(4)})$ and $\mathcal{N}_{ \mathscr{D}^3 }(M^{(3)}; X^+)$ is,
\begin{align}
    \mathcal{N}_{ \mathscr{D}^3 }&(M^{(3)}; X^+ \sharp M^{(4)}) / \mathcal{N}_{ \mathscr{D}^3 }(M^{(3)}; X^+) \notag \\
    &=  N^{-\frac{3\chi(M^{(4)})}{2}} \frac{|H^0 (M^{(4)}; \mathbb{Z}_N)|^3 |H^2 (M^{(4)}; \mathbb{Z}_N)|}{|H^1 (M^{(4)}; \mathbb{Z}_N)|^3}  \left( \sum_{b \in H^2 (M^{(4)}; \mathbb{Z}_N)} \rme^{-\frac{\im N}{4 \pi} \int b \wedge b}\right) \notag \\
    &= 1,
\end{align}
where we have used $e^{\im \pi \sigma(M^{(4)})} = 1$, $H^i (X^+\sharp M^{(4)}, \partial X^+; \mathbb{Z}_N) = H^i (M^{(4)}; \mathbb{Z}_N) \oplus H^i (X^+, \partial X^+; \mathbb{Z}_N)$ for $i \neq 0,4$, and  
\begin{align}
    \frac{|H^0 (M^{(4)}; \mathbb{Z}_N)|}{|H^1 (M^{(4)}; \mathbb{Z}_N)|} \sum_{b \in H^2 (M^{(4)}; \mathbb{Z}_N)} \rme^{-\frac{\im N}{4 \pi} \int b \wedge b} = N^{\chi(M^{(4)})/2},
\end{align}
from the locality and the calculation in (\ref{eq:proof_claim_gravitational}).
Therefore, although the normalization $\mathcal{N}_{ \mathscr{D}^3 }(M^{(3)})$ is expressed using $X^+$ in (\ref{eq:fusion-rule-normalization}), this indicates the bulk independence of $\mathcal{N}_{ \mathscr{D}^3 }(M^{(3)})$, as expected from the definition of the defect.

\section{Other self-dual parameters}\label{App:other_parameters}

In the main text, especially in Sec.~\ref{sec:constraints_dynamics}, we study the fixed point, $\tau_*=\frac{1+\sqrt{3}\im}{2}$, of the $ST^{-1}$ transformation, and constrain the possible ground states using the mixed gravitational anomaly. 
In this appendix, we study other self-dual parameters. 

\subsection{Fixed point of \texorpdfstring{$S$}{S} transformation: \texorpdfstring{$\tau=\im$}{tau=i}}

Here, we discuss the fixed point $\tau=\im$ of the $S$ transformation \eqref{eq:S_transformation}. 
Using \eqref{eq:self-duality-1} of Claim~\ref{claim:self-duality-1}, we obtain 
\begin{align}
    \calZ_{\CR/(\mathbb{Z}_N)_0}^{\tau=\im}[B]=N^{\frac{\chi}{2}} \calZ_{\CR}^{\tau=\im}[B]. 
    \label{eq:S_duality_relation}
\end{align}
Importantly, there is no signature-dependent phase factor. Because of the absence of the signature dependence, we cannot exclude the trivially gapped phase unlike the case of the $ST^{-1}$ transformation. 

Here, we note that $\tau=\im$ has $\theta=0$, and thus there is the $\mathsf{CP}$ invariance of the partition function, 
\begin{equation}
    \calZ_{\CR}^{\tau=\im}[\mathsf{CP}\cdot B]=\calZ_{\CR}^{\tau=\im}[B]. 
    \label{eq:CP_inv_Sdual}
\end{equation}
Because of this extra property, we can rule out the trivially gapped phase from the possible ground states for $N\ge 3$ (see also Ref.~\cite{Choi:2021kmx}).

In order to see this, let us assume that the ground state is trivially gapped, then the partition function should take the form of 
\begin{equation}
    \calZ_k[B]=\rme^{\frac{\im N k}{4\pi}\int B\wedge B}, 
\end{equation}
with some $k\sim k+N$. Under the $\mathsf{CP}$ transformation, $k$ flips its sign, and thus \eqref{eq:CP_inv_Sdual} requires that 
\begin{equation}
    k=-k \bmod N. 
\end{equation}
When $N$ is odd, $k=0$. When $N$ is even, $k=0$ or $k=\frac{N}{2}$. 
We here note that $k=0$ is chosen for the conjectured phase diagram of Fig.~\ref{fig:phase_diagram} since it gives the partition function of the monopole-induced confinement phase~\eqref{eq:Z_monopole}, $\calZ_{\mathrm{mon}}[B]=\calZ_{0}[B]=1$. 
However, under the $(\mathbb{Z}_N)_0$ gauging, it gives 
\begin{align}
    \int \Diff b\, \calZ_{\mathrm{mon}}[B]\rme^{\frac{\im N}{2\pi}\int b\wedge B}&=\int \Diff b\, \calZ_{\mathrm{mon}}[B]\rme^{\frac{\im N}{2\pi}\int b\wedge B}
    \propto \delta(B),  
\end{align}
and the partition function vanishes unless $\rme^{\im \int_{M_2}B}=1$ for any closed $2$-submanifolds $M_2\subset X$. 
This implies that all the Wilson loops are deconfined for the transformed theory, which can be identified as the Higgs phase. 
Indeed, Fig.~\ref{fig:phase_diagram} predicts that the $S$-dual point $\tau=\im$ consists of the monopole-induced confinement phase and the Higgs phase, and the above result is consistent with the observation. 
Especially for odd $N$, $k=0$ is the unique choice of the SPT state consistent with the $\mathsf{CP}$ invariance, and thus the $S$-duality relation~\eqref{eq:S_duality_relation} excludes it from the possible ground states. 

Let us then continue the discussion for the even $N$ case. Since the choice $k=0$ is inconsistent with the $S$-duality relation, we must choose $k=\frac{N}{2}$ as a possible SPT state. 
We need to check if \eqref{eq:S_duality_relation} can be satisfied:
\begin{align}
    \int \Diff b\, \calZ_{N/2}[b]\rme^{\frac{\im N}{2}\int b\wedge B}\stackrel{?}{=} N^{\frac{\chi}{2}}\calZ_{N/2}[B]. 
\end{align}
This equality is possible only if $\gcd(N,\frac{N}{2})=1$, i.e. $N=2$. 
That is, when $N\ge 3$, we can exclude the trivially gapped phase from possible ground states by using \eqref{eq:S_duality_relation} and \eqref{eq:CP_inv_Sdual}. 
When $N=2$, we may have a trivially gapped state that satisfies both anomaly matching conditions. If it is trivially gapped, the partition function is given by the level-$1$ SPT action with $\mathbb{Z}_2^{[1]}$ symmetry assuming that the UV regularization is consistent with the relations \eqref{eq:S_duality_relation} and \eqref{eq:CP_inv_Sdual}. 

\subsection{Fixed point of \texorpdfstring{$ST^{-1}ST^2S$}{ST[-1]ST[2]S} transformation: \texorpdfstring{$\tau_{**}=\frac{\sqrt{3}+\im}{2\sqrt{3}}$}{tau**=1/2+i/2sqrt(3)}}\label{sec:triple_point_oblique}

The $ST^{-1}ST^2S$ transformation gives 
\begin{equation}
    \tau\mapsto \frac{2\tau-1}{3\tau-1},\quad 
    \begin{pmatrix}
    n\\
    m
    \end{pmatrix}\mapsto
    \begin{pmatrix}
    2&1\\
    -3&-1
    \end{pmatrix}\begin{pmatrix}
    n\\
    m
    \end{pmatrix}=
    \begin{pmatrix}
    2n+m\\
    -3n-m
    \end{pmatrix}, 
\end{equation}
and the fixed point in the upper half-plane is given by
\begin{equation}
    \tau_{**}=\frac{\sqrt{3}+\im}{2\sqrt{3}}. 
\end{equation}
In the vicinity of the fixed point, this gives the $\frac{2\pi}{3}$ clockwise rotation, 
\begin{equation}
    \tau_{**}+\delta \tau\mapsto \tau_{**}+\rme^{-\frac{2\pi\im}{3}}\delta \tau. 
\end{equation}
We can also check that $(ST^{-1}ST^2 S)^3=\mathsf{C}$. 
According to Fig.~\ref{fig:phase_diagram}, this is the point where the monopole $(n,m)=(0,1)$ condensation, the dyon $(n,m)=(-1,1)$ condensation, and exotic dyon $(n,m)=(-1,2)$ condensation phases meet. 
This exotic dyon condensation phase is called the oblique confinement phase~\cite{tHooft:1981bkw}. 
When $N$ is even, it is described by the $\mathbb{Z}_2$ topological order since $W^{N/2}$ is deconfined. 
For odd $N$, the oblique confinement phase is an SPT state with $\mathbb{Z}_{N}^{[1]}$ symmetry~\cite{Honda:2020txe}. 

Let $\mathcal{T}$ be a $4$d QFT with $\mathbb{Z}_{N}^{[1]}$ symmetry, then we consider the following partition function,
\begin{align}
    &\quad \calZ_{((\mathcal{T}/(\mathbb{Z}_{N}^{[1]})_0)/(\mathbb{Z}_{N}^{[1]})_2)/(\mathbb{Z}_{N}^{[1]})_{-1}}[B]\notag\\
    &=\int\Diff b_1\Diff b_2\Diff b_3\,\calZ_{\mathcal{T}}[b_1]\rme^{\frac{\im N}{4\pi}\int(2b_2^2-b_3^2)+\frac{\im N}{2\pi}\int(b_1\wedge b_2+b_2\wedge b_3+b_3\wedge B)}. 
\end{align}
We can confirm that this is consistent with the naive relation $(ST^{-1}ST^2 S)^3=\mathsf{C}$ as follows:
\begin{align}
    &\quad\int \Diff b_{1,2,3} \Diff b'_{1,2,3} \Diff b''_{1,2,3}\calZ_{\mathcal{T}}[b_1]\rme^{\frac{\im N}{4\pi}\int(2b_2^2-b_3^2+2{b'_2}^2-{b'_3}^2+2{b''_2}^2-{b''_3}^2)}\notag\\
   &\qquad \times \rme^{\frac{\im N}{2\pi}\int(b_1\wedge b_2+b_2\wedge b_3+b_3\wedge {b'_1}+{b'_1}\wedge {b'_2}+{b'_2}\wedge {b'_3}+{b'_3}\wedge {b''_1}+{b''_1}\wedge {b''_2}+{b''_2}\wedge {b''_3}+{b''_3}\wedge B )}\notag\\
   &=N^{2\chi} \int \Diff b_{1,2,3}\Diff b'_3\Diff b''_3 \calZ_{\mathcal{T}}[b_1]\rme^{\frac{\im N}{4\pi}\int (2b_2^2+b_3^2+{b'_3}^2-{b''_3}^2)+\frac{\im N}{2\pi}\int(b_1\wedge b_2+b_2\wedge b_3-b_3\wedge b'_3 -b'_3\wedge b''_3+b''_3\wedge B)} \notag\\
   &=N^{2\chi}\int \Diff b_{1,2,3}\Diff b'_3\Diff b''_3 \calZ_{\mathcal{T}}[b_1]\rme^{\frac{\im N}{4\pi}\int (b_2^2+(b_3-{b'_3}+b_2)^2-{b''_3}^2)+\frac{\im N}{2\pi}\int(b_1\wedge b_2+b_2\wedge b'_3 -b'_3\wedge b''_3+b''_3\wedge B)}\notag\\
   &=N^{\frac{7\chi}{2}} \int \Diff b_1\Diff b_2 \calZ_{\mathcal{T}}[b_1]\rme^{\frac{\im N}{2\pi}\int (b_1\wedge b_2+b_2\wedge B)}\notag\\
   &=N^{\frac{9\chi}{2}} \calZ_{\mathcal{T}}[-B]. 
\end{align}
In the first line, we can perform the $b'_1$ and  $b''_1$ path integrals, which give the constraints $b'_2=-b_3$ and $b''_2=
-b'_3$, respectively. Therefore, the $b'_2$ and $b''_2$ path integrals becomes trivial, and we obtain the second line. 
By completing the square in terms of $b_3$ in the exponent, we obtain the third line, and thus the $b_3$ path integral just gives an overall constant. Moreover, since the quadratic term of $b'_3$ disappears in the above process, its path integral gives the constraint $b''_3=b_2$, which gives the fourth line, and we obtain the result. 

By applying \eqref{eq:self-duality-1} of Claim~\ref{claim:self-duality-1} repeatedly, we find
\begin{align}
    &\quad \calZ^{\tau}_{((\mathcal{CR}/(\mathbb{Z}_{N}^{[1]})_0)/(\mathbb{Z}_{N}^{[1]})_2)/(\mathbb{Z}_{N}^{[1]})_{-1}}[B]\notag\\
    &=N^{\frac{\chi}{2}}(S(\tau))^{\frac{\chi+\sigma}{4}}(S(\overline{\tau}))^{\frac{\chi-\sigma}{4}}
    \calZ^{S(\tau)}_{(\mathcal{CR}/(\mathbb{Z}_{N}^{[1]})_2)/(\mathbb{Z}_{N}^{[1]})_{-1}}[B]\notag\\
    &=N^{\chi}(S(\tau)\cdot ST^2S(\tau))^{\frac{\chi+\sigma}{4}} (S(\overline{\tau})\cdot ST^2S(\overline{\tau}))^{\frac{\chi-\sigma}{4}}
    \calZ^{ST^2S(\tau)}_{\mathcal{CR}/(\mathbb{Z}_{N}^{[1]})_{-1}}[B]\notag\\
    &=N^{\frac{3\chi}{2}}(S(\tau)\cdot ST^2S(\tau)\cdot ST^{-1}ST^2S(\tau))^{\frac{\chi+\sigma}{4}} (S(\overline{\tau})\cdot ST^2S(\overline{\tau})\cdot ST^{-1}ST^2S(\overline{\tau}))^{\frac{\chi-\sigma}{4}}\notag\\
    &\qquad \times \calZ^{ST^{-1}ST^2S(\tau)}_{\mathcal{CR}}[B]. 
\end{align}
We then obtain that 
\begin{align}
    \calZ^{\tau}_{((\mathcal{CR}/(\mathbb{Z}_{N}^{[1]})_0)/(\mathbb{Z}_{N}^{[1]})_2)/(\mathbb{Z}_{N}^{[1]})_{-1}}[B]
    &=N^{\frac{3\chi}{2}}(3\tau-1)^{-\frac{\chi+\sigma}{4}}(3\overline{\tau}-1)^{-\frac{\chi-\sigma}{4}}
    \calZ^{\frac{2\tau-1}{3\tau-1}}_{\mathcal{CR}}[B]. 
\end{align}
Substituting $\tau=\tau_{**}$, we find the relation~\eqref{eq:self_duality_ST-1ST2S} as $3\tau_{**}-1=\rme^{\frac{\pi\im}{3}}$, and let us recapitulate it here for convenience:
\begin{equation}
    \calZ^{\tau_{**}}_{((\CR/(\mathbb{Z}_N^{[1]})_0)/(\mathbb{Z}^{[1]}_{N})_{2})/(\mathbb{Z}^{[1]}_N)_{-1}}[B]
    =N^{\frac{3\chi(X)}{2}}\rme^{-\frac{\pi\im}{6}\sigma(X)} \calZ^{\tau_{**}}_{\CR}[B]. \label{eq:ST1ST2S-anomaly-constraint}
\end{equation}
Because of the phase factor that depends on the signature $\sigma(X)$, the SPT state can be ruled out from the possible ground states using the same logic for Claim~\ref{claim:gravitational_anomaly}.

Next, let us confirm if the phase diagram~\ref{fig:phase_diagram} is consistent with the above transformation. 
Let us assume that the partition function of the monopole-induced confinement phase is given by the level-$0$ SPT action \eqref{eq:Z_monopole}, $\calZ_{\mathrm{mon}}[B]=1$. 
Then, 
\begin{align}
    &\quad \int\Diff b_{1,2,3}\calZ_{\mathrm{mon}}[b_1]\rme^{\frac{\im N}{4\pi}\int(2b_2^2-b_3^2)+\frac{\im N}{2\pi}\int(b_1\wedge b_2+b_2\wedge b_3+b_3\wedge B)}\notag\\
    &=N^{\beta_0-\beta_1+\beta_2}\int \Diff b_{2,3}\delta(b_2)\rme^{\frac{\im N}{4\pi}\int(2b_2^2-b_3^2)+\frac{\im N}{2\pi}\int(b_2\wedge b_3+b_3\wedge B)}\notag\\
    &=N^{\chi}\int \Diff b_3 \rme^{\frac{\im N}{4\pi}\int(-b_3^2)+\frac{\im N}{2\pi}\int b_3\wedge B }\notag\\
    &= N^{\frac{3\chi}{2}}\rme^{\frac{\im N}{4\pi}\int B^2}=N^{\frac{3\chi}{2}}\calZ_{\mathrm{dyon}}[B].
\end{align}
Therefore, we obtain the level-$1$ SPT action, which is nothing but the partition function of the dyon-induced confinement phase~\eqref{eq:Z_dyon}. 
Applying the same prcedure to $\calZ_{\mathrm{dyon}}[B]$, we get 
\begin{align}
    &\quad \int \Diff b_{1,2,3}\calZ_{\mathrm{dyon}}[b_1] \rme^{\frac{\im N}{4\pi}\int(2b_2^2-b_3^2)+\frac{\im N}{2\pi}\int(b_1\wedge b_2+b_2\wedge b_3+b_3\wedge B)}\notag\\
    &=\int \Diff b_{1,2,3} \rme^{\frac{\im N}{4\pi}\int ((b_1+b_2)^2+(b_2+b_3)^2-2b_3^2) +\frac{\im N}{2\pi}\int(b_3\wedge B)}\notag\\
    &=N^{\chi}\int \Diff b_3 \rme^{\frac{\im N}{4\pi}\int (-2b_3^2) +\frac{\im N}{2\pi}\int(b_3\wedge B)}. 
\end{align}
The result of this path integral depends on whether $N$ is even or odd. 
When $N$ is even, this path integral vanishes unless $\rme^{\im \frac{N}{2}\int_{M_2}B}=1$ for any closed $2$-submanifolds $M_2\subset X$, which means that the charge $\frac{N}{2}$ Wilson loop is deconfined. 
When $N$ is odd, this gives the level-$\frac{N+1}{2}$ SPT action. 
The result is consistent with the observation in Ref.~\cite{Honda:2020txe} that the oblique confinement phase is the $\mathbb{Z}_2$ topological order for even $N$ and the SPT state for odd $N$. 
The level-$\frac{N+1}{2}$ SPT action for odd $N$ has been derived from the global inconsistency argument~\cite{Honda:2020txe}.
Therefore, we can define\footnote{We can also understand this expression as the $S$-duality transformation from the $(2,1)$-dyon-condensed state $\rme^{\frac{\im N}{4\pi}\int (-2B^2)}$.
Hence, it is natural that the resulting phase is the $(-1,2)$-dyon-condensed state.}
\begin{align}
    Z_{\mathrm{oblique}}[B] &= N^{- \chi/2} \int \Diff b~ \rme^{\frac{\im N}{4\pi}\int (-2b^2) +\frac{\im N}{2\pi}\int(b \wedge B)} \notag \\
    &= \begin{cases}
    \rme^{\frac{\im N}{4\pi} \frac{N+1}{2} \int B \wedge B} & (\mathrm{odd~}N) \\
    2^{\beta_2/2} \rme^{\frac{\im N}{4\pi} \frac{N}{2} \int B \wedge B} \delta(\frac{NB}{2})& (\mathrm{even~}N)
    \end{cases}
\end{align}
as a natural candidate for the partition function of the $(-1,2)$-dyon-condensed state.

Based on the transformation shown above, the following linear combination of partition functions of these phases
\begin{align}
   \calZ^{\tau_{**}}_{\CR}[B] = Z_{\mathrm{mon}}[B] + \rme^{\frac{2\pi\im}{3}\sigma(X)} Z_{\mathrm{dyon}}[B] + \rme^{-\frac{2\pi\im}{3}\sigma(X)} Z_{\mathrm{oblique}}[B] \label{eq:monopole-dyon-oblique}
\end{align}
satisfies the constraint (\ref{eq:ST1ST2S-anomaly-constraint}).
Here we have used $\sigma(X) \in 16 \mathbb{Z}$ (Rokhlin’s theorem), which yields $\rme^{-\frac{\pi\im}{6}\sigma(X)} = \rme^{-\frac{2\pi\im}{3}\sigma(X)}$.
Again, this indicates that the new anomaly matching constraint (\ref{eq:ST1ST2S-anomaly-constraint}) is consistent with the conjectured phase diagram.
One can also check that the combination (\ref{eq:monopole-dyon-oblique}) reproduces the mixed anomaly between $\mathbb{Z}_N^{[1]}$ and $\mathsf{CP}$, up to a pure gravitational counterterm:
\begin{equation}
    \mathsf{CP}:\calZ^{\tau_{**}}_{\CR}[B]\mapsto \rme^{-\frac{2 \pi\im}{3}\sigma(X)}\rme^{-\im \frac{N}{4\pi}\int_{X} B\wedge B}\calZ^{\tau_{**}}_{\CR}[B]. 
\end{equation}

\section{Computation of the bordism group \texorpdfstring{$\Omega^{\Spin}_4(B^2 \mathbb{Z}_N)$}{}}
\label{App:computation_bordism}

In this Appendix, we derive
\begin{align}
\Omega^{\Spin}_4(B^2 \mathbb{Z}_N) \simeq \mathbb{Z} \oplus \mathbb{Z}_N,
\label{eq:spin-bordism}
\end{align}
where $B^2\mathbb{Z}_N=K(\mathbb{Z}_N,2)$ is the second Eilenberg-MacLane space. 
This shows that the SPT action with $\mathbb{Z}_{N}^{[1]}$ symmetry is classified by $\mathbb{Z}_N$, and thus we can confirm that the topological action, 
\begin{equation}
    \frac{\im N k}{4\pi }\int B\wedge B
\end{equation}
with $k\sim k+N$, characterizes those SPT phases. 
This result was used to derive Claim \ref{claim:gravitational_anomaly} at (\ref{eq:sec-4-classification-SPTs}) and for similar discussions in Appendix \ref{App:other_parameters}.

We can approximate the spin bordism group by using the Atiyah–Hirzebruch spectral sequence\footnote{Rigorously speaking, the Atiyah–Hirzebruch spectral sequence only determines the bordism group up to extension.
In the calculation below, we will see that the nontrivial elements of the $E^\infty$ page are $E^\infty_{0,4} \simeq \mathbb{Z}$ and $E^\infty_{4,0} \simeq \mathbb{Z}_N$, which implies $0 \rightarrow \mathbb{Z} \rightarrow \Omega^{\Spin}_4(B^2 \mathbb{Z}_N) \rightarrow \mathbb{Z}_N \rightarrow 0$. 
In general, a possible extension is $\mathbb{Z} \oplus \mathbb{Z}_k$ with $k|N$.
However, we know that there are independent $\mathbb{Z}$ corresponding to the signature $\sigma(X)$ and $\mathbb{Z}_N$ corresponding to $\frac{N}{4 \pi }\int B \wedge B$. This guarantees (\ref{eq:spin-bordism}).}, 
\begin{equation}
    E^2_{p,q}=H_p(B^2 \mathbb{Z}_N; \Omega^{\Spin}_q(pt.))\Rightarrow \Omega^{\Spin}_{p+q}(B^2\mathbb{Z}_N). 
    \label{eq:AHSS}
\end{equation}
The bordism group of a point $\Omega^{\Spin}_q(pt.)$ is well-known, e.g. \cite{Kapustin:2014dxa, Garcia-Etxebarria:2018ajm},
\begin{align}
\begin{array}{c|cccccc}
    q & ~0 & ~1 & ~2 & ~3 & ~4 & 5 \\ \hline
    \Omega^{\Spin}_q(pt.) & ~\mathbb{Z} & ~\mathbb{Z}_{2} & ~\mathbb{Z}_{2} & ~0 & ~\mathbb{Z} & ~0
    \end{array}
    \label{eq:bordism-point}
\end{align}
The homology $H_d (B^2 \mathbb{Z}_N; \mathbb{Z})$ is given by \footnote{In Sec. 3.5 of \cite{Garcia-Etxebarria:2018ajm}, 
$H_d (B^2 \mathbb{Z}_{p^k}, \mathbb{Z})$ for prime $p$ is provided. For a general integer $N = p_1^{k_1} \cdots p_\ell ^{k_\ell}$, we have $B^2\mathbb{Z}_{N} \simeq B^2\mathbb{Z}_{p_1^{k_1}} \times \cdots \times B^2\mathbb{Z}_{p_\ell^{k_\ell}}$ (from the definition of Eilenberg-MacLane space) and (\ref{eq:B2ZN-homology}) from the K\"unneth formula.}
\begin{align}
    &\begin{array}{c|cccccc}
    d & ~0 & ~1 & ~2 & ~3 & ~4 & ~5 \\ \hline
    H_d(B^2\mathbb{Z}_N;\mathbb{Z}) & ~\mathbb{Z} & ~0 & ~\mathbb{Z}_N & ~0 & ~\mathbb{Z}_{N} & ~0
    \end{array}\quad (\mbox{odd}\, N),\\
    &\begin{array}{c|cccccc}
    d & ~0 & ~1 & ~2 & ~3 & ~4 & 5 \\ \hline
    H_d(B^2\mathbb{Z}_N;\mathbb{Z}) & ~\mathbb{Z} & ~0 & ~\mathbb{Z}_N & ~0 & \mathbb{Z}_{2N} & \mathbb{Z}_2
    \end{array}\quad (\mbox{even}\, N).
    \label{eq:B2ZN-homology}
\end{align}
We then have the $E_2$ pages for odd $N$ (Table~\ref{tab:E2page_odd}) and even $N$ (Table~\ref{tab:E2page_even}).
For odd $N$, there are no differentials for low degrees, and we immediately obtain \eqref{eq:spin-bordism} for odd $N$. 



    \begin{table}[t]
        \centering
        \begin{tabular}{ |c|cccccc| } 
 \hline
 4 & \hl{$\mathbb{Z}$} & 0 & $\mathbb{Z}_N$ & 0 & $\mathbb{Z}_N$ & 0\\ 
 3 & 0 & \hl{0} & 0 & 0 & 0 & 0\\ 
 2 & $\mathbb{Z}_2$ & 0 & \hl{0} & 0 & 0 & 0\\ 
 1 & $\mathbb{Z}_2$ & 0 & 0 & \hl{0} & 0 & 0\\
 0 & $\mathbb{Z}$ & 0 & $\mathbb{Z}_N$ & 0 & \hl{$\mathbb{Z}_{N}$} & 0\\ 
 \hline
 $q/p$ & 0 & 1 & 2 & 3 & 4 & 5\\
 \hline
\end{tabular}
        \caption{$E_2$ page for odd $N$.}
        \label{tab:E2page_odd}
\end{table}

    \begin{table}[t]
        \centering
        \begin{tabular}{ |c|cccccc| } 
 \hline
 4 & \hl{$\mathbb{Z}$} & 0 & $\mathbb{Z}_N$ & 0 & $\mathbb{Z}_{2N}$ & $\mathbb{Z}_{2}$\\ 
 3 & 0 & \hl{0} & 0 & 0 & 0 & 0\\ 
 2 & $\mathbb{Z}_2$ & 0 & \hl{$\mathbb{Z}_2$} & $\mathbb{Z}_2$ & $\mathbb{Z}_2$ & $\mathbb{Z}_2 \oplus \mathbb{Z}_2$\\ 
 1 & $\mathbb{Z}_2$ & 0 & $\mathbb{Z}_2$ & \hl{$\mathbb{Z}_2$} & $\mathbb{Z}_2$ & $\mathbb{Z}_2 \oplus \mathbb{Z}_2$\\
 0 & $\mathbb{Z}$ & 0 & $\mathbb{Z}_N$ & 0 & \hl{$\mathbb{Z}_{2N}$} & $\mathbb{Z}_2$\\ 
 \hline
 $q/p$ & 0 & 1 & 2 & 3 & 4 & 5\\
 \hline
\end{tabular}
        \caption{$E_2$ page of even $N$.}
        \label{tab:E2page_even}
\end{table}

For even $N$, the computation is more complicated.
The nontrivial differentials of the $E_2$ page are:
\begin{align}
    d_{4,0}^2: E^2_{4,0} = \mathbb{Z}_{2N} \rightarrow E^2_{2,1} = \mathbb{Z}_{2}, \notag \\
    d_{4,1}^2: E^2_{4,1} = \mathbb{Z}_{2} \rightarrow E^2_{2,2} = \mathbb{Z}_{2}, \notag \\
    d_{5,0}^2: E^2_{5,0} = \mathbb{Z}_{2} \rightarrow E^2_{3,1} = \mathbb{Z}_{2}.
\end{align}
We introduce several technicalities to evaluate these differentials.

First, we note the ring structure of the mod-2 cohomology.
For $N = 2^s$, there is a well-known description by Serre \cite{Serre:mod2cohomology}.
This Serre's description yields $H^*(B^2\mathbb{Z}_{2^s}; \mathbb{Z}_{2}) = \mathbb{Z}_{2}[u_2,\delta_s \iota_2, Sq^{2} \delta_s \iota_2]$, where $\iota_2 \in H^2 (B^2\mathbb{Z}_{2^s}; \mathbb{Z}_{2^s})$ is the fundamental class of the Eilenberg-MacLane space, $u_2 \in H^2 (B^2\mathbb{Z}_{2^s}; \mathbb{Z}_{2})$ is its mod 2 reduction, $\delta_s: H^2 (B^2\mathbb{Z}_{2^s}; \mathbb{Z}_{2^s}) \rightarrow H^3 (B^2\mathbb{Z}_{2^s}; \mathbb{Z}_{2})$ is the connecting homomorphism associated to $0 \rightarrow \mathbb{Z}_{2^s} \rightarrow \mathbb{Z}_{2^{s+1}} \rightarrow \mathbb{Z}_{2} \rightarrow 0$, $Sq^{2}: H^i (B^2\mathbb{Z}_{2^s}; \mathbb{Z}_{2}) \rightarrow H^{i+2} (B^2\mathbb{Z}_{2^s}; \mathbb{Z}_{2})$ is the second Steenrod square, and $\mathbb{Z}_{2}[u_2,\delta_s \iota_2, Sq^{2} \delta_s \iota_2]$ stands for the $\mathbb{Z}_{2}$ polynomial algebra on generators $(u_2,\delta_s \iota_2, Sq^{2} \delta_s \iota_2)$.
For a general integer $N$, we factorizes $N$ into prime factors as $N = 2^s \times (\mathrm{odd})$.
Then, from (\ref{eq:B2ZN-homology}) and the K\"unneth formula, the mod-2 cohomology essentially stems from the $2^s$ factor: $H^d (B^2\mathbb{Z}_{N}; \mathbb{Z}_{2}) \simeq H^d (B^2\mathbb{Z}_{2^s}; \mathbb{Z}_{2})$ at least for $d = 0,1,2,3,4,5$.
This determines the cohomology structure which we will use.

Second, the differentials $d^2_{p,0}$ and $d^2_{p,1}$ are given by~\cite{Teichner:MR1214960}
\begin{align}
    d^2_{p,0}: H_p (B^2\mathbb{Z}_{N}; \mathbb{Z}) &\rightarrow H_{p-2} (B^2\mathbb{Z}_{N}; \mathbb{Z}_{2}) \notag \\
     \xi  &\mapsto \operatorname{Sq}^2_* \circ \rho (\xi), \\
    d^2_{p,1}: H_p (B^2\mathbb{Z}_{N}; \mathbb{Z}_{2}) &\rightarrow H_{p-2} (B^2\mathbb{Z}_{N}; \mathbb{Z}_{2}) \notag \\
     \xi  &\mapsto \operatorname{Sq}^2_* (\xi), 
\end{align}
where $\rho: H_p (B^2\mathbb{Z}_{N}; \mathbb{Z}) \rightarrow H_p (B^2\mathbb{Z}_{N}; \mathbb{Z}_2)$ is the mod-2 reduction, and $\operatorname{Sq}^2_*: H_p (B^2\mathbb{Z}_{N}; \mathbb{Z}_{2}) \rightarrow H_{p-2} (B^2\mathbb{Z}_{N}; \mathbb{Z}_{2})$ denotes the dual second Steenrod square\footnote{The dual Steenrod square is defined as follows.
The Steenrod square is defined in terms of the cohomology $H^* (B^2\mathbb{Z}_{N}; \mathbb{Z}_2)$.
Note the relation $H^d (M; \mathbb{Z}_2) \simeq \operatorname{Hom}(H_d (M; \mathbb{Z}_2), \mathbb{Z}_2) \simeq H_d (M; \mathbb{Z}_2)$ from the universal coefficient theorem, which yields the Kronecker pairing $(\cdot, \cdot):H^d (M; \mathbb{Z}_2) \times H_d (M; \mathbb{Z}_2) \rightarrow \mathbb{Z}_2$. The dual Steenrod square is characterized by
\begin{align}
    (\alpha, \operatorname{Sq}^2_* \beta) = (\operatorname{Sq}^2 \alpha,  \beta) ~~~\mathrm{for~} \alpha \in H^{d-2} (M; \mathbb{Z}_2),~\beta \in H_{d} (M; \mathbb{Z}_2).
\end{align}
}.

Now, let us look into each differential:
\begin{itemize}
    \item $d_{4,0}^2: E^2_{4,0} = \mathbb{Z}_{2N} \rightarrow E^2_{2,1} = \mathbb{Z}_{2}$.
    
    We note the following properties:
    \begin{itemize}
        \item The generator of $E^2_{2,1} = H_2 (B^2\mathbb{Z}_{N}; \mathbb{Z}_{2}) \simeq \mathbb{Z}_{2}$ is the dual of $u_2 \in H^2 (B^2\mathbb{Z}_{N}; \mathbb{Z}_{2})$. The generator of $H_4 (B^2\mathbb{Z}_{N}; \mathbb{Z}_{2}) \simeq \mathbb{Z}_{2}$ is the dual of $u_2 \cup u_2 = \operatorname{Sq}^2 u_2 \in H^4 (B^2\mathbb{Z}_{N}; \mathbb{Z}_{2})$.
        Therefore, the dual Steenrod square $\operatorname{Sq}^2_*: H_4 (B^2\mathbb{Z}_{N}; \mathbb{Z}_{2}) \rightarrow H_2 (B^2\mathbb{Z}_{N}; \mathbb{Z}_{2})$ is a bijection.
        \item Since $H_3 (B^2\mathbb{Z}_{N}; \mathbb{Z}) = \{ 0\}$, $H_4 (B^2\mathbb{Z}_{N}; \mathbb{Z}_{2}) \simeq \mathbb{Z}_{2}$ consists of the mod 2 reduction of $H_4 (B^2\mathbb{Z}_{N}; \mathbb{Z})$. Thus, the mod 2 reduction $\rho: H_4 (B^2\mathbb{Z}_{N}; \mathbb{Z}) \rightarrow H_4 (B^2\mathbb{Z}_{N}; \mathbb{Z}_2)$ is nontrivial and surjective.
    \end{itemize} 
    
    These properties imply that the map $d_{4,0}^2: E^2_{4,0} = \mathbb{Z}_{2N} \rightarrow E^2_{2,1} = \mathbb{Z}_{2}$ is the nontrivial one.
    We obtain the element of the $E_3$ page as $E^3_{4,0} = \operatorname{ker} d_{4,0}^2 \simeq \mathbb{Z}_{N}$.
    
    \item $d_{4,1}^2: E^2_{4,1} = \mathbb{Z}_{2} \rightarrow E^2_{2,2} = \mathbb{Z}_{2}$.
    
    As above, this map is bijective. These elements are ``killed'' by the differential. We have $E^3_{2,2} = E^2_{2,2}/ \operatorname{im} d_{4,1}^2 \simeq \{ 0 \}$.

    \item $d_{5,0}^2: E^2_{5,0} = \mathbb{Z}_{2} \rightarrow E^2_{3,1} = \mathbb{Z}_{2}$.
    
    We can evaluate this differential similar to $d_{4,0}^2$, but $H_5 (B^2\mathbb{Z}_{N}; \mathbb{Z}_{2}) \simeq \mathbb{Z}_{2} \oplus \mathbb{Z}_{2}$ makes the situation more complicated. Let us consider $\rho$ and $\operatorname{Sq}^2_*$ separately.
    
    \begin{itemize}
        \item The mod 2 reduction $\rho: H_5 (B^2\mathbb{Z}_{N}; \mathbb{Z}) \rightarrow H_5 (B^2\mathbb{Z}_{N}; \mathbb{Z}_2)$.
        The generators of $H_5 (B^2\mathbb{Z}_{N}; \mathbb{Z}_2) \simeq \mathbb{Z}_{2} \oplus \mathbb{Z}_{2}$ are the duals of $Sq^{2} \delta_s \iota_2$ and $u_2 \cup \delta_s \iota_2$.
        We need to determine which generator $\operatorname{im} \rho$ corresponds.

        To this end, let us consider the exact sequence associated to $0 \rightarrow \mathbb{Z}_{2^s} \xrightarrow[]{\times 2} \mathbb{Z}_{2^{s+1}} \xrightarrow[]{\operatorname{mod~} 2} \mathbb{Z}_{2} \rightarrow 0$:
\begin{align}
    H_5(B^2\mathbb{Z}_{N}; \mathbb{Z}_{2^{s+1}}) \xrightarrow[]{\rho^{(s)}} H_5(B^2\mathbb{Z}_{N}; \mathbb{Z}_{2}) \xrightarrow[]{\delta^{(s)}} H_4(B^2\mathbb{Z}_{N}; \mathbb{Z}_{2^{s}}), \label{eq:2s-exact-sequence}
\end{align}
where $\rho^{(s)}$ is the mod 2 reduction and $\delta^{(s)}$ is the connecting homomorphism.
By considering the mod-$2^{s+1}$ reduction and the mod-2 reduction from $H_5(B^2\mathbb{Z}_{N}; \mathbb{Z}) \simeq \mathbb{Z}_2$ to $H_5(B^2\mathbb{Z}_{N}; \mathbb{Z}_{2^{s+1}}) \simeq \mathbb{Z}_2 \oplus \mathbb{Z}_{2^{s+1}}$ and $H_5(B^2\mathbb{Z}_{N}; \mathbb{Z}_{2}) \simeq \mathbb{Z}_2 \oplus \mathbb{Z}_{2}$, we notice that $\rho = \rho^{(s)} \circ \tilde{\rho}^{(s)}$, where $\tilde{\rho}^{(s)}: H_5(B^2\mathbb{Z}_{N}; \mathbb{Z}) \rightarrow H_5(B^2\mathbb{Z}_{N}; \mathbb{Z}_{2^{s+1}})$ is the mod-$2^{s+1}$ reduction.
Thus, we have $\operatorname{im} \rho \subset \operatorname{im} \rho^{(s)}$.
On the other hand, the connecting homomorphism $\delta^{(s)}$ is nontrivial.
Indeed, we know there is an element $\xi_5 \in H_5(B^2\mathbb{Z}_{N}; \mathbb{Z}_{2})$ dual to $u_2 \cup \delta_s \iota_2$.
We can rewrite this element as\footnote{Note that $ \frac{\iota_2 \cup \iota_2}{2} \in H^4(B^2\mathbb{Z}_{N}; \mathbb{Z}_{2^s})$ because of $2^s \iota_2 \cup \iota_2 = 0$ and that the connecting homomorphism can be written as $\delta_s \omega = \left[ \frac{1}{2} d \omega \right]$.} $u_2 \cup \delta_s \iota_2 = \delta_s \left( \frac{\iota_2 \cup \iota_2}{2} \right)$ with the connecting homomorphism $\delta_s: H^p(B^2\mathbb{Z}_{N}; \mathbb{Z}_{2^s}) \rightarrow H^{p+1}(B^2\mathbb{Z}_{N}; \mathbb{Z}_{2})$.
Therefore, by taking the Kronecker pairing, we have
\begin{align}
    1= (u_2 \cup \delta_s \iota_2, \xi_5) = \left( \frac{\iota_2 \cup \iota_2}{2} , \delta^{(s)}\xi_5 \right),
\end{align}
which yields $\delta^{(s)}\xi_5  \neq 0$. From the exactness of the sequence (\ref{eq:2s-exact-sequence}), the element $\xi_5$ dual to $u_2 \cup \delta_s \iota_2$ is perpendicular to $\operatorname{im} \rho^{(s)}$.
Therefore, we have $\operatorname{im} \rho = \operatorname{im} \rho^{(s)} \simeq \mathbb{Z}_2$, and its generator is dual to $Sq^{2} \delta_s \iota_2$.

        \item The (restricted) dual Steenrod square $\operatorname{Sq}^2_*: \operatorname{im} \rho \rightarrow H_3(B^2\mathbb{Z}_{N}; \mathbb{Z}_{2})$ is bijective. The generator dual to $Sq^{2} \delta_s \iota_2$ is mapped to an element dual to $\delta_s \iota_2$, which is the generator of $H^3(B^2\mathbb{Z}_{N}; \mathbb{Z}_{2})$.
    \end{itemize}
    
    We have seen that the differential $d_{5,0}^2 = \operatorname{Sq}^2_* \circ \rho: E^2_{5,0} = \mathbb{Z}_{2} \rightarrow E^2_{3,1} = \mathbb{Z}_{2}$ is bijective.
    Therefore, the corresponding element of the $E_3$ page is $E^3_{3,1} = E^2_{3,1}/\operatorname{im} d_{5,0}^2 \simeq \{ 0 \}$.
\end{itemize}

We can also see that $d_{2,0}^2 = \operatorname{Sq}^2_* \circ \rho: E^2_{2,0} \rightarrow E^2_{0,1}$ is trivial and that $d_{5,1}^2= \operatorname{Sq}^2_*: E^2_{5,1} \rightarrow E^2_{3,2}$ is surjective.
Now, we can ``turn the page'' and have the $E_3$ page as Table~\ref{tab:E3page_even}. 
It would be clear that no higher differentials are relevant to $\Omega^{Spin}_4(B^2 \mathbb{Z}_N) $.
This completes the derivation of \eqref{eq:spin-bordism} for even $N$.

    \begin{table}[t]
        \centering
        \begin{tabular}{ |c|cccccc| } 
 \hline
 4 & \hl{$\mathbb{Z}$} & 0 & $\mathbb{Z}_N$ & 0 & * & *\\ 
 3 & 0 & \hl{0} & 0 & 0 & 0 & 0\\ 
 2 & $\mathbb{Z}_2$ & 0 & \hl{0} & 0 & * & *\\ 
 1 & $\mathbb{Z}_2$ & 0 & 0 & \hl{0} & 0 & *\\
 0 & $\mathbb{Z}$ & 0 & $\mathbb{Z}_N$ & 0 & \hl{$\mathbb{Z}_{N}$} & 0\\ 
 \hline
 $q/p$ & 0 & 1 & 2 & 3 & 4 & 5\\
 \hline
\end{tabular}
        \caption{$E_3$ page for even $N$. Only elements related to $\Omega^{Spin}_4(B^2 \mathbb{Z}_N) $ are shown.}
        \label{tab:E3page_even}
\end{table}




\bibliographystyle{JHEP}
\bibliography{./QFT.bib,./refs.bib} 

\end{document}